\documentclass[12pt,letter]{article}
\pdfoutput=1
\usepackage{graphicx, epsfig, color,cite}
\usepackage{amsmath,amsthm,amsfonts,geometry,etoolbox}

\usepackage{amssymb}
\usepackage{caption,subcaption,graphicx,appendix}
\usepackage[hidelinks]{hyperref}
\def\to{\rightarrow}

\def\bi{\begin{itemize}}
\def\ei{\end{itemize}}

\def\tchi{\tilde\chi}

\def\DeltaBG{\Delta_{\rm BG}}
\def\DeltaHS{\Delta_{\rm HS}}
\def\DeltaEW{\Delta_{\rm EW}}

\def\tf{\tilde f}

\def\tst{\tilde t}

\def\tg{\tilde g}

\def\alt{\lesssim}
\def\agt{\gtrsim}
\def\be{\begin{equation}}  
\def\ee{\end{equation}}  
\def\bea{\begin{eqnarray}}  
\def\eea{\end{eqnarray}}

\newcommand{\myeq}{\begin{small}\begin{equation}\begin{aligned}}
\newcommand{\myeqend}{\end{aligned}\end{equation}\end{small}}

\begin{document}
\begin{titlepage}
\begin{flushright}
OU-HEP-230501
\end{flushright}

\vspace{0.5cm}
\begin{center}
  {\Large \bf On practical naturalness and its implications\\
    for weak scale supersymmetry}\\

\vspace{1.2cm} \renewcommand{\thefootnote}{\fnsymbol{footnote}}
{\large Howard Baer$^{1}$\footnote[1]{Email: baer@ou.edu },
Vernon Barger$^2$\footnote[2]{Email: barger@pheno.wisc.edu},
Dakotah Martinez$^1$\footnote[3]{Email: dakotah.s.martinez-1@ou.edu} and
Shadman Salam$^3$\footnote[4]{Email: ext.shadman.salam@bracu.ac.bd}
}\\ 
\vspace{1.2cm} \renewcommand{\thefootnote}{\arabic{footnote}}
{\it 
$^1$Homer L. Dodge Department of Physics and Astronomy,\\
University of Oklahoma, Norman, OK 73019, USA \\[3pt]
}
{\it 
$^2$Department of Physics,
University of Wisconsin, Madison, WI 53706 USA \\[3pt]
}
{\it 
$^3$Center for Mathematical Sciences, Department of Mathematics and Natural Sciences,
Brac University, Dhaka 1212, Bangladesh \\[3pt]
}

\end{center}

\vspace{0.5cm}
\begin{abstract}
\noindent
We revisit the various measures of naturalness for models of
weak scale supersymmetry including 1. electroweak (EW) naturalness,
2. naturalness via sensitivity to high scale parameters (EENZ/BG),
3. sensitivity of Higgs soft term due to high scale (HS) radiative corrections and
4. stringy naturalness (SN) from the landscape.
The EW measure is most conservative and seems unavoidable;
it is also model independent in that its value is fixed only by the weak scale
spectra which ensues, no matter which model is used to generate it.
The EENZ/BG measure is ambiguous depending on which ``parameters of ignorance''
one includes in the low energy effective field theory (LE-EFT).
For models with calculable soft breaking terms, then the EENZ/BG measure
reduces to the tree-level EW measure.
The HS measure began life as a figurative expression and probably
shouldn't be taken more seriously than that.
SN is closely related to EW naturalness via the atomic principle,
although it is also sensitive to the distribution of soft terms on
the landscape.
If the landscape favors large soft terms, as in a power law distribution,
then it favors $m_h\simeq 125$ GeV along with sparticles beyond
present LHC reach.
In this context, SN appears as a
probability measure where more natural models are expected to be more
prevalent on the landscape than finetuned models.
We evaluate by how much the different measures vary against one another with
an eye to determining by how much they may overestimate finetuning;
we find overestimates can range up to a factor of over 1000.
In contrast to much of the literature, we expect the string landscape
to favor EW natural SUSY models over finetuned models so that the landscape
is {\it not} an alternative to naturalness.
\end{abstract}
\end{titlepage}
\tableofcontents
\pagebreak

\section{Introduction}
\label{sec:intro}

Weak scale supersymmetry provides a very clean solution to the 
hierarchy of scales problem\cite{Witten:1981nf,Kaul:1981wp} of
particle physics and is actually supported by data from four different virtual
effects:\footnote{Radiative corrections have historically been a reliable guide to new physics and (as just a few examples)
  indeed have presaged the
  discovery of the $W$ and $Z$ vector bosons, the top quark and the Higgs bosons.}
\begin{enumerate}
\item the successful running of gauge couplings to unified values within the
Minimal Supersymmetric Standard Model (MSSM)\cite{Dimopoulos:1981yj,Amaldi:1991cn,Ellis:1990wk,Langacker:1991an},
\item the predicted large value of the top quark mass needed for
radiatively-induced electroweak symmetry breaking (REWSB)\cite{Ibanez:1982fr,Ibanez:1983wi,Alvarez-Gaume:1983drc},
\item the match between the narrow theory-predicted window
  of $m_h$ values within the MSSM and the subsequent Higgs boson
  discovery\cite{Slavich:2020zjv} and
\item precision electroweak corrections which, in the $m_t$ vs. $m_W$ plane,
actually favor heavy SUSY over the SM\cite{Heinemeyer:2013dia}.
\end{enumerate}

In addition, some remnant SUSY is expected to survive
superstring compactification from 10/11 to four spacetime dimensions
on a Calabi-Yau manifold\cite{Candelas:1985en}.
In fact, it is conjectured that the landscape of all geometric, stable,
string/M theory compactifications to Minkowski spacetime (at leading order)
are supersymmetric\cite{Acharya:2019mcu};
manifolds which do not respect these conditions typically lead to
Witten bubble-of-nothing instabilities.
Also, in contrast to the SM, SUSY leads to EW vacuum stability at
ultra-high energies owing to gauge sector contributions
(D-terms) to Higgs quartic couplings\cite{Ellis:2000ig}. 
Plus, highly motivated SM extensions which introduce a new high energy scale--
such as the inclusion of see-saw neutrinos or a Peccei-Quinn sector to solve
the strong CP problem-- avoid the Higgs mass blow-up
due to the introduced new high mass scales\cite{Vissani:1997ys,Farina:2013mla,Klein:2019jxa}
unless the underlying model is supersymmetric.
And with respect to the axion solution to the strong CP problem, intrinsically
supersymmetric discrete $R$-symmetries\cite{Lee:2011dya},
which are expected to emerge from string compactifications\cite{Nilles:2017heg},
provide an avenue for emergence of the required global $U(1)_{PQ}$ symmetry
with sufficient precision as to solve the axion
quality problem\cite{Baer:2018avn,Bhattiprolu:2021rrj}.
In such models, the PQ scale $f_a$ is related to the hidden sector SUSY
breaking scale $m_{hidden}\sim f_a\sim 10^{11}$ GeV so that $f_a$ lies
within the cosmological sweet spot  for axion production via coherent
oscillations in the early universe\cite{Baer:2018avn}.
String instanton effects on the axion
quality are also ameliorated within the MSSM\cite{Demirtas:2021gsq}.
And while there are few compelling mechanisms for successful baryogenesis
left within the rubric of the SM, the introduction of SUSY leads to
several new and/or improved mechanisms to address the matter-antimatter
asymmetry\cite{Dine:2003ax,Bae:2015efa}.

In spite of this impressive litany of successes, it is common nowadays to
dismiss weak scale supersymmetry (WSS)\cite{Baer:2006rs} as a
viable beyond-the-Standard Model (BSM) theory due to the apparent lack of new
physics signals at the CERN Large Hadron Collider (LHC)\cite{Craig:2022cef}.
The data from LHC, which is by-and-large in accord with SM
expectations\cite{Narain:2022qud},
is in contrast to early theoretical expectations for WSS based upon naturalness
arguments that superpartners would emerge with mass values not far from
the weak scale $m_{weak}\simeq m_{W,Z,h}\sim 100$ GeV\cite{Ellis:1986yg,Barbieri:1987fn,deCarlos:1993rbr,Anderson:1994tr,Dimopoulos:1995mi,Chankowski:1997zh,Chankowski:1998xv,Bastero-Gil:1999jqv,Abe:2007kf}.
At present, such arguments are being used to set policy and guide
future facilities for the High Energy Physics (HEP)
frontier\cite{EuropeanStrategyforParticlePhysicsPreparatoryGroup:2019qin,Butler:2023glv}.
Given the stakes involved, it is essential to go back and review
the naturalness-based arguments to assess when and where and if they present
a reliable guide to the search for new physics.
After all, the original naturalness arguments are over 30 years old,
and the HEP community has hopefully learned a lot since then.

In this paper, we revisit several proposed naturalness measures which have
been applied to various supersymmetric models.
As opposed to 't Hooft naturalness, these measures determine the degree of what is defined in 
Sec. \ref{sec:Pnat} as {\it practical naturalness}: that all independent contributions to some 
observable $\cal{O}$ are comparable to or less than $\cal{O}$.
Historically, the first of these is the EENZ/BG\cite{Ellis:1986yg,Barbieri:1987fn} measure (labeled here as $\DeltaBG$)
which determines the sensitivity of the measured value of the weak scale
to variation in model parameters $p_i$ ($i$ labels the various parameters
under consideration). Typically the $p_i$ have been taken to be the
various soft SUSY breaking terms starting at a high
effective field theory (EFT) cutoff scale
$\Lambda =m_{GUT}\simeq 2\times 10^{16}$ GeV:
\be
\DeltaBG \equiv max_i|\frac{\partial\log m_Z^2}{\partial\log p_i}|=
max_i |\frac{p_i}{m_Z^2}\frac{\partial m_Z^2}{\partial p_i}|.
\label{eq:DBG}
\ee
For small $\DeltaBG\alt 30$, then sparticle masses are expected below the
several hundred GeV range although in some special regions of model parameter
space, such as the focus point region\cite{Feng:1999mn,Feng:1999zg}
of the minimal supergravity\cite{Arnowitt:1993qp} (mSUGRA)
or constrained MSSM\cite{Kane:1993td} (CMSSM)  model, multi-TeV scale top squarks can be allowed.
Despite its popularity, this measure has been argued to {\it overestimate}
finetuning in SUSY models by large factors and to give ambiguous answers
depending on exactly which parameters are chosen to be the
fundamental $p_i$\cite{Baer:2013gva,Baer:2014ica}.

A second measure, which we label here as $\DeltaHS$ (for high scale sensitivity
of the up-Higgs soft mass $m_{H_u}^2$), starts with the approximate
SUSY Higgs mass relation $m_h^2\sim \mu^2+m_{H_u}^2(weak)$
where $m_{H_u}^2(weak) =m_{H_u}^2(\Lambda ) +\delta m_{H_u}^2$.
One then requires
  \be
  \DeltaHS=\delta m_{H_u}^2/m_h^2
\label{eq:DHS}
  \ee
to be small.
(As mentioned earlier, it is the large top-quark Yukawa coupling $f_t$
which radiatively drives $m_{H_u}^2$ from its large SUGRA value at the
high scale to negative values at the weak scale so that EW symmetry is
spontaneously broken.)
This measure, which is inconsistent with $\DeltaBG$ in that it doesn't
allow for multi-TeV top squarks even in the FP region,
has lead to intense scrutiny of LHC top squark searches since it is
expected that
$\delta m_{H_u}^2\sim \frac{6 f_t^2}{(4\pi)^2}m_{\tst}^2 \log\frac{\Lambda^2}{m_{\tst}^2}$\cite{Murayama:2000dw,Harnik:2003rs,Chacko:2005ra,Kitano:2005wc,Kitano:2006gv,Papucci:2011wy,Brust:2011tb}.
$\DeltaHS$ was found to lead to violations of the finetuning rule\cite{Baer:2013gva}:
that it is not allowed to claim finetuning amongst {\it dependent} terms
which contribute to some observable ${\cal O}$. In this case, $\delta m_{H_u}^2$
and $m_{H_u}^2(\Lambda )$ are dependent, leading to overestimates in finetuning.

A third measure is the {\it electroweak} measure $\DeltaEW$\cite{Baer:2012up,Baer:2021tta} which is touted
to be more conservative and model independent than the others,
and also unavoidable (within the context of the MSSM).
It is based on the SUSY Higgs potential minimization condition
\be
m_Z^2/2=\frac{m_{H_d}^2+\Sigma_d^d-(m_{H_u}^2+\Sigma_u^u)\tan^2\beta}{\tan^2\beta -1}-\mu^2\simeq -m_{H_u}^2-\mu^2-\Sigma_u^u(\tst_{1,2})
\label{eq:mzs}
\ee
where all right-hand-side (RHS) entries are taken as their weak scale values and
\be
\DeltaEW\equiv max_i|entries\ on\ RHS\ of\ Eq.\ \ref{eq:mzs}|/(m_Z^2/2) .
\label{eq:DEW}
\ee
This measure was preceded by Chan {\it et al.}\cite{Chan:1997bi}
who suggested that the magnitude of the SUSY conserving $\mu$ parameter
could serve as a finetuning measure all by itself.
This measure is sometimes criticized in that it apparently lacks
sensitivity to high scale parameters (more on this later).

A fourth entry is not at present a quantifiable measure, but known nonetheless
as {\it stringy naturalness} (SN), and arises from Douglas' consideration of the
string landscape picture\cite{Douglas:2004zg}:
\begin{quotation} {\bf Stringy naturalness:}
  An observable ${\cal O}_1$ is more (stringy) natural
than observable ${\cal O}_2$ if more {\it phenomenologically viable} string vacua
lead to ${\cal O}_1$ than to ${\cal O}_2$.
\end{quotation}
To quantify stringy naturalness, at least two ingredients are needed:
1. the expected distribution of some quantity within the landscape of vacua possibilities and 2. an anthropic selection ansatz for which many choices would lead to universes that are unable to support observers.
For the case of SUSY models, the first of these is usually how soft terms are
distributed in the landscape while the second of these is the magnitude of
the weak scale itself: if the predicted value of $m_{weak}$
within each pocket universe is too far displaced from its measured value
in our universe, then nuclear physics goes astray,
and atoms as we know them fail to appear-- leading to no complex chemistry
as seems to be needed for life as we know it (atomic principle)\cite{Agrawal:1997gf}.
An attempt to compute and display stringy naturalness via density of dots
in model parameter space has been made in Ref. \cite{Baer:2019cae}.

In the present work, we reexamine these several measures of naturalness,
filling in some of the many gaps of understanding that exist in the literature.
Part of our work is based on a new computation of $\DeltaBG$
naturalness based on evaluating numerically the derivatives in Eq. \ref{eq:DBG}.
This new computation is embedded in the publicly available code
DEW4SLHA\cite{Baer:2021tta} so that the updated code can provide
values of each of the measures $\DeltaBG$, $\DeltaHS$ and $\DeltaEW$
given an input SUSY Les Houches Accord (SLHA) file\cite{Skands:2003cj}.\footnote{The code
DEW4SLHA, written by D. Martinez, is available at  \it{https://www.dew4slha.com}.}
We also compute ratios of naturalness measures to determine the extent
of which some measures can overestimate finetuning in SUSY models.
For instance, in the SUSY theory review contained in
the Particle Data Book\cite{ParticleDataGroup:2022pth},
it is suggested that the overestimates may range up to a factor 10;
in contrast, we find overestimates ranging up to factors of over 1000.

\section{Some models of weak scale SUSY}
\label{sec:models}

In our deliberations, we make reference to several SUSY models which we
briefly review here for the reader.

\subsection{mSUGRA/CMSSM model}
\label{ssec:cmssm}

Some of our numerical work refers to the mSUGRA\cite{Arnowitt:1993qp}
or CMSSM model\cite{Kane:1993td} with parameter space
\be
m_0,\ m_{1/2},\ A_0,\ \tan\beta\ \ {\rm and}\ \ {\rm sign}(\mu )\ \ \ ({\rm mSUGRA/CMSSM}).
\ee
In this model, $m_0$ refers to a unified high scale scalar soft breaking mass,
usually defined at the scale $m_{GUT}$ where the three gauge couplings unify.
While unified gaugino masses can be developed in many supersymmetric models
with a simple choice for gauge kinetic function, the scalar mass unification
is an (unmotivated) simplifying assumption that violates expectations from
gravity-mediated SUSY breaking models where non-universal scalar masses are
expected unless imposed by some symmetry\cite{Soni:1983rm,Kaplunovsky:1993rd,Brignole:1993dj}.
For instance, scalar masses within a matter generation may be expected to
unify to $m_0(i)$ (for generation index $i=1-3$) due to the fact that all
matter fills out a complete 16-dimensional spinor representation of $SO(10)$.
However, generational universality $m_0(1)=m_0(2)=m_0(3)$ is not expected
and leads to the famous SUSY flavor and CP problems\cite{Gabbiani:1996hi}. Furthermore,
the Higgs fields $H_u$ and $H_d$ live in different $SU(5)$
(or general $SO(10)$)
representations from matter scalars, and hence are also not expected to unify.

\subsection{Non-universal Higgs models (NUHM)}
\label{ssec:nuhmi}

These models come in several different guises and meet the expectation that 
$m_{H_u}\ne m_{H_d}\ne m_0(i)$.
The simplest case, NUHM1\cite{Baer:2004fu} assumes
$m_{H_u}=m_{H_d}\ne m_0(i)$ as expected in simple $SO(10)$ GUTs,
while NUHM2\cite{Ellis:2002wv,Ellis:2002iu,Baer:2005bu}
with $m_{H_u}\ne m_{H_d}\ne m_0(i)$ occurs in $SU(5)$ or general $SO(10)$ GUTs.
In all cases, when we speak of SUSY GUT models we have an eye towards
{\it local} grand unification\cite{Buchmuller:2005sh} wherein the geography
of fields on a string compactified manifold determines the GUT symmetry
properties\cite{Nilles:2009yd,Nilles:2014owa}.
Sometimes an NUHM3 model with $m_0(1)=m_0(2)\ne m_0(3)$ is used\cite{Pomarol:1995xc}
and sometimes NUHM4 with $m_0(1)\ne m_0(2)\ne m_0(3)$ is used especially
for discussion of the SUSY flavor and CP problems\cite{Baer:2019zfl}.
For these NUHMi models ($i=1-4$), frequently the GUT values of $m_{H_u}$
and $m_{H_d}$ are traded for the weak scale
values of the superpotential $\mu$ parameter and the pseudoscalar Higgs mass
$m_A$ via the scalar potential minimization conditions.
Thus, the NUHMi parameter space is given by
\be
m_0(i),\ m_{1/2},\ A_0,\ \tan\beta ,\ \mu ,\ m_A .
\ee
The NUHMi models are particularly convenient to explore {\it natural}
supersymmetry since one can directly restrict oneself to
natural values of the $\mu$ parameter: $\mu\sim 100-350$ GeV.

\begin{figure}[!htbp]
\begin{center}
\includegraphics[height=0.4\textheight]{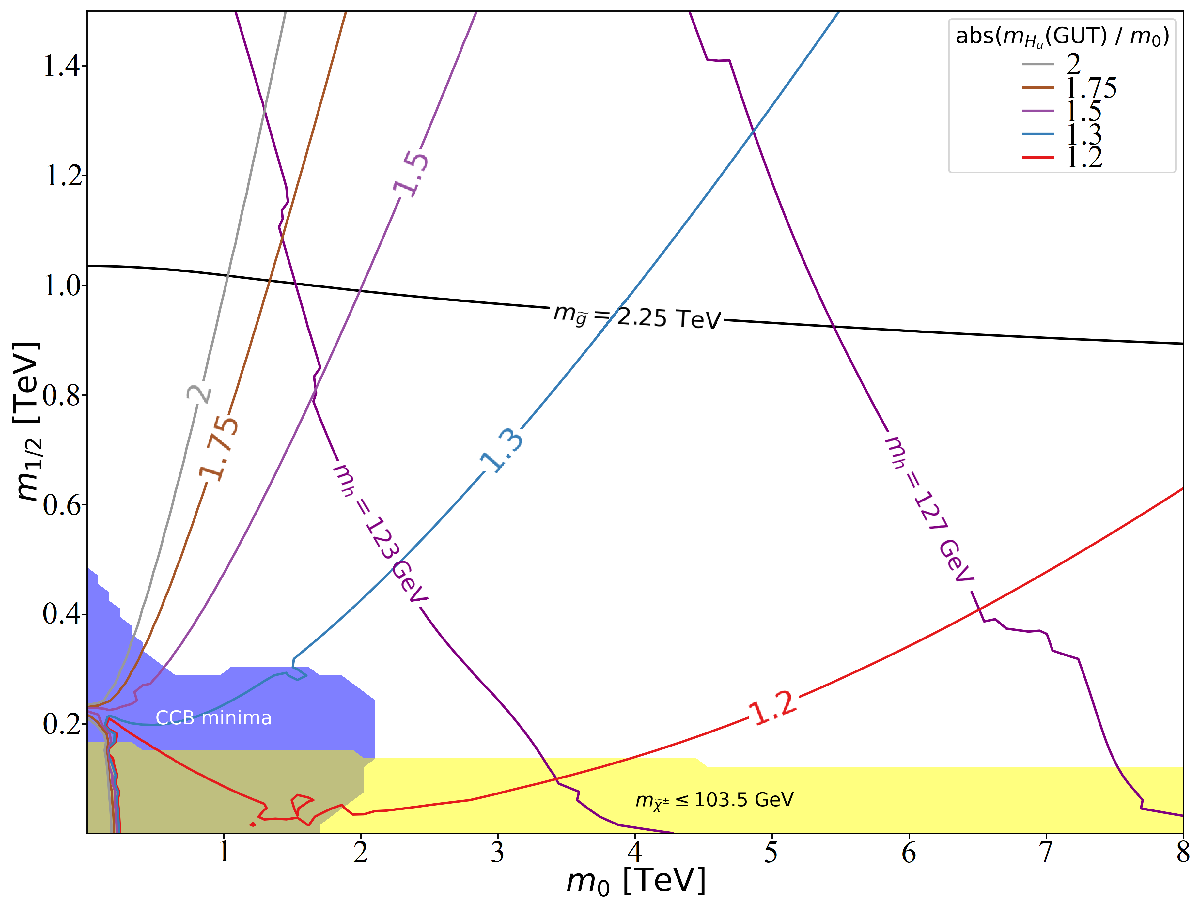}
\caption{Ratio of $m_{H_u}(m_{GUT}) /m_0$ in the NUHM2 model
  which is needed to ensure that $\mu =200$ GeV.
  We also take $A_0=-1.6 m_0$ and $\tan\beta =10$ with $m_A=2$ TeV. The blue shaded region is excluded, as these points lead to CCB minima. The yellow shaded region near the bottom has the lightest chargino below LEP2 limits, $m_{\widetilde{\chi}_{1}^{\pm}}<103.5$ GeV.
The spectra is calculated using SoftSUSY.
\label{fig:mHu_ratio}}
\end{center}
\end{figure}

In Fig. \ref{fig:mHu_ratio}, we plot in the $m_0$ vs. $m_{1/2}$ plane
of the NUHM2 model the ratio of $m_{H_u}(m_{GUT})/m_0$
which is needed to ensure that the SUSY $\mu$ parameter is fixed
at a natural value of $\mu =200$ GeV. We also adopt $A_0=-1.6 m_0$
and $\tan\beta =10$ with $m_A=2$ TeV. From the plot, we see that
$m_{H_u}(m_{GUT})\sim 2m_0$ along the left-hand-side of the plot, but dips
to a ratio of about 1.2-1.5 for the bulk of the plane which respects
the $m_h\sim 123-127$ GeV range (assuming about a 2 GeV error bar on the
$m_h$ theory calculation). Thus, only modest deviations of order 20-50\%
are required in order ensure one of the most fundamental requirements of
naturalness, namely $|\mu |\sim m_{weak}$.

\subsection{Natural anomaly and mirage mediation models}
\label{ssec:others}

Later in this paper we shall explore naturalness in the guise of
anomaly-mediated SUSY breaking (AMSB)\cite{Randall:1998uk,Giudice:1998xp,Baer:2018hwa}
and mixed modulus-AMSB SUSY models, also known as mirage mediation (MM)\cite{Choi:2005ge,Baer:2016hfa}.

\section{Naturalness and practical naturalness}
\label{sec:Pnat}

Supersymmetry offers a 't Hooft technically natural solution\cite{tHooft:1979rat} to the hierarchy of scales problem in that, as the hidden sector SUSY
breaking scale $m_{hidden}$ (which determines the magnitude of the soft terms
via $m_{soft}\sim m_{3/2}\sim m_{hidden}^2/m_P$ in gravity-mediation and hence
of the weak scale via the scalar potential minimization conditions)
is taken to zero, the model becomes more (super)symmetric.
The SUSY solution to this {\it big} hierarchy problem (BHP)--
stabilizing the weak scale so that it doesn't blow up to the Planck or GUT scale--
is {\it not} the naturalness issue which concerns many
contemporary SUSY theorists. Indeed, 't Hooft naturalness remains a valid solution
to the BHP even for very large gaps $m_{soft}\gg m_{weak}$.
Instead, it is the so-called {\it little hierarchy problem} (LHP) which is
of concern\cite{Barbieri:2000gf,Birkedal:2004xi}:
\begin{quotation}
{\hskip-1.5em}how can it be that $m_{weak}\sim m_{W,Z,h}\sim 100$ GeV is so much
smaller than the soft SUSY breaking terms, which-- according to LHC data--
are $m_{soft}\agt 1$ TeV (owing to LHC bounds $m_{\tg}\agt 2.2$ TeV,
$m_{\tst_1}\agt 1.1$ TeV, $\cdots$)\cite{Canepa:2019hph}.
\end{quotation}

In addressing the LHP, what is of concern is what we call the notion
of {\bf practical naturalness} (PN)\cite{Baer:2015rja}\footnote{This is in
accord with Veltman's notion of naturalness as presented in Ref. \cite{Veltman:1980mj}.
See also Susskind\cite{Susskind:1978ms}.}:
\begin{quotation}
  {\hskip-1.5em}An observable ${\cal O}=o_1+\cdots +o_n$  is practically natural if all {\it independent}
  contributions $o_i$ to ${\cal O}$ are comparable to or less than ${\cal O}$. 
\end{quotation}
(Here, {\it comparable to} means within a factor of several from the measured value.)
Practical naturalness embodies the notion of naturalness that is most often
used in successful applications of naturalness.
For instance, by requiring the charm quark mass contribution
\be
\Delta m_K(c)\simeq \frac{G_F}{\sqrt{2}}\frac{\alpha}{6\pi}\frac{f_K^2m_K}{\sin^2\theta_W}\cos^2\theta_C\sin^2 \theta_C\frac{m_c^2}{m_W^2}
\ee
to be comparable to or less than the measured $K_L-K_S$ mass difference $\Delta m_K$,
Gaillard and Lee\cite{Gaillard:1974hs} were able to predict $m_c\sim 1.5$ GeV several months
before the charm quark was discovered\footnote{It is still a breathtaking
  exercise to plug in the numbers and see the charm quark mass emerge.}.
An essential element of practical naturalness is that the contributions $o_i$
should be independent of one another in the sense that if one of the
$o_i$ is varied, then the others don't necessarily vary.
For instance, Dirac was bothered by various divergent contributions to
perturbative QED observables. However, these were {\it dependent}
contributions in that if the regulator was varied, the different divergent
terms would also vary. One should always first {\it combine dependent terms}
before evaluating naturalness. Once dependent terms are combined, then
a measure of naturalness emerges:
\be
\Delta\equiv max_i|o_i|/|{\cal O}|.
\label{eq:Delta}
\ee
Using PN, we see that QED perturbation theory is practically natural
in that the leading terms are comparable to the measured observables
whilst higher order terms (once dependent terms are combined) are
typically much smaller.

\section{Some measures of naturalness}

\subsection{Sensitivity to high scale parameters: EENZ/BG naturalness}
\label{ssec:DBG}

Historically, the first measure of SUSY model naturalness was proposed
by Ellis {\it et al.} in Ref. \cite{Ellis:1986yg} and subsequently used by
Barbieri and Giudice\cite{Barbieri:1987fn} to compute sparticle mass
upper bounds in the mSUGRA/CMSSM model: Eq. \ref{eq:DBG}.
The measure purports to compute sensitivity of the measured value of the 
weak scale to variation in high scale parameters $p_i$.
The $\DeltaBG$ measure is actually a measure of practical naturalness
of the weak scale in the case where $m_Z^2=a_1 p_1+\cdots a_n p_n$.
Let's suppose the $j$th contribution to $m_Z^2$ is largest. Then
$\DeltaBG =max_i|(p_i/m_Z^2)\partial m_Z^2/\partial p_i|=|a_j p_j/m_Z^2|$
in accord with Eq. \ref{eq:Delta}.
The various $|a_i p_i/m_Z^2|\equiv c_i$ terms are labeled
sensitivity coefficients\cite{Feng:2013pwa}.
The rub here is what choice to take as to the free parameters $p_i$.
\footnote{Giudice remarks in Ref. \cite{Giudice:2008bi}: ``It may well be that,
  in some cases, Eq. \ref{eq:DBG} overestimates the amount of tuning.
  Indeed, Eq. \ref{eq:DBG} measures the sensitivity of the prediction of
  $m_Z$ as we vary parameters in {\it theory space}.
  However, we have no idea how this {\it theory space} looks like,
  and the procedure of independently varying all parameters may be too
  simple-minded'' . See also discussion in Ref. \cite{Chankowski:1998xv}.}

The starting point is to express $m_Z^2$ in terms of weak scale SUSY parameters
as in Eq. \ref{eq:mzs}:
\be
m_Z^2 \simeq -2m_{H_u}^2-2\mu^2
\label{eq:mZsapprox}
\ee
where the partial equality obtains for moderate-to-large $\tan\beta$ 
values and where we assume for now that the radiative corrections are small.
To evaluate $\Delta_{BG}$, one needs to know the explicit dependence of 
$m_{H_u}^2$ and $\mu^2$ on the fundamental parameters.
Semi-analytic solutions to the one-loop renormalization group equations
for $m_{H_u}^2$ and $\mu^2$ can be found for instance in Refs. \cite{Ibanez:1984vq,Lleyda:1993xf}. For the case of $\tan\beta =10$,
then\cite{Abe:2007kf,Martin:2007gf,Feng:2013pwa}
\bea
m_Z^2& \simeq & -2.18\mu^2 + 3.84 M_3^2+0.32M_3M_2+0.047 M_1M_3 \nonumber \\
& & -0.42 M_2^2+0.011 M_2M_1-0.012M_1^2-0.65 M_3A_t \nonumber \\
& & -0.15 M_2A_t-0.025M_1 A_t+0.22A_t^2+0.004 M_3A_b\nonumber \\
& &-1.27 m_{H_u}^2 -0.053 m_{H_d}^2\nonumber \\
& &+0.73 m_{Q_3}^2+0.57 m_{U_3}^2+0.049 m_{D_3}^2 -0.052 m_{L_3}^2+0.053 m_{E_3}^2\nonumber \\
& &+0.051 m_{Q_2}^2-0.11 m_{U_2}^2+0.051 m_{D_2}^2 -0.052 m_{L_2}^2+0.053 m_{E_2}^2\nonumber \\
& &+0.051 m_{Q_1}^2-0.11 m_{U_1}^2+0.051 m_{D_1}^2 -0.052 m_{L_1}^2+0.053 m_{E_1}^2 ,
\label{eq:mZsparam}
\eea
where all terms on the right-hand-side are understood to be
$GUT$ scale parameters.
As an example, if we adopt $m_{Q_3}^2$ as a fundamental parameter, then the sensitivity
coefficient $c_{m_{Q_3}^2}=0.73 m_{Q_3}^2/m_Z^2$ and for $m_{Q_3}=3$ TeV, then
one finds $c_{m_{Q_3}^2}\simeq 800$ so that $\DeltaBG>800$ and the model is
certainly finetuned. If instead we declare all scalar masses unified to $m_0$,
then there are large cancellations and instead one finds
$c_{m_0^2}=0.013 m_0^2/m_Z^2\sim 14.2$: a reduction in finetuning by over a
factor 50!
Clearly, whether or not soft terms are correlated or not makes a big
difference in the evaluation of $\DeltaBG$!

\subsubsection{Numerical routine to compute $\DeltaBG$}

The evaluation of $\Delta_{BG}$ can be done by approximating the partial derivatives with the method of two-point finite central difference quotients. That is, for finding the partial derivative with respect to a parameter $p_{1}$ of $m_{Z}^{2}(p_{1},p_{2},\ldots,p_{n})$, where $p_{i}$ are the fundamental parameters of the model chosen for evaluating $\Delta_{BG}$, then

\begin{equation}
    \frac{\partial m_{Z}^{2}\left(p_{1},p_{2},\ldots,p_{n}\right)}{\partial p_{1}}\approx\frac{m_{Z}^{2}(p_{1}+h_{1},p_{2},\ldots,p_{n}) - m_{Z}^{2}(p_{1}-h_{1},p_{2},\ldots,p_{n})}{2h_{1}}.
\label{eq:2pt_deriv_approx}
\end{equation}
$h_{1}$ is the size of the variation of the differentiation parameter $p_{1}$, which is then used to determine the resulting change in $m_{Z}^{2}$. Since this is a partial derivative, all other input parameters are left fixed at their original values prior to differentiation. 

To compute this derivative, $m_{Z}^{2}$ must be evaluated in the right-hand side of Eq. \ref{eq:2pt_deriv_approx} as an \emph{output} of the $m_{Z}^{2}$ Higgs minimization condition, Eq. \ref{eq:mzs}, at the weak renormalization scale $Q_{\text{SUSY}}=\sqrt{m_{\widetilde{t}_{1}}m_{\widetilde{t}_{2}}}$ to minimize radiative corrections in the Higgs minimization condition. For the partial derivative of $m_{Z}^{2}$ with respect to $p_{i}$, the GUT-scale parameter $p_{i}$ defined at the renormalization scale $Q_{\text{GUT}}$ is varied to $p_{i}+h_{i}$, with $h_{i}\ll p_{i}$. Then the \emph{new} set of GUT-scale parameters $\{p_{1},p_{2},\ldots, p_{i}+h_{i},\ldots,p_{n-1},p_{n}\}$ are evolved from $Q_{\text{GUT}}$ down to $Q_{\text{SUSY}}$ using the full two-loop MSSM renormalization group equations (RGEs). Lastly, the varied value $m_{Z}^{2}(p_{1},p_{2},\ldots, p_{i}+h_{i},\ldots, p_{n-1},p_{n})$ is computed from the tree-level Higgs minimization condition for $m_{Z}^{2}$, giving a value slightly deviated from $91.2^{2}$. This value is then used in Eq. \ref{eq:2pt_deriv_approx} and the process is repeated for the other direction of variation.

In this numerical derivative approach, two sources of error can enter and skew the results: truncation error and roundoff error. Below are some descriptions of these errors and how we minimize them.
\begin{itemize}
    \item Truncation error is the error of approximating the true, analytical derivative of $m_{Z}^{2}$, a tangent line to the $m_{Z}^{2}$ curve, with our numerical two-point method, producing a secant line to the $m_{Z}^{2}$ curve. For a given derivative variation size of $h$, the truncation error for this two-point method is suppressed by a term of $\mathcal{O}(h^{2})$. This error remains relatively small so long as the step size $h<1$ and the higher-order derivatives of $m_{Z}^{2}$ are reasonably bounded.
    \item Roundoff error comes from representing the values $p_{1},p_{2},\ldots,p_{n},$ and $h_{1}$ in Eq. \ref{eq:2pt_deriv_approx} as floating point numbers, where the computer must ``round off" most decimal values after a certain number of digits due to storage limitations in binary. Because of this, there is a non-zero spacing between two consecutive floating point numbers $x$ and $y$, and this spacing is called the unit of least precision (denoted ULP($x$)). Careful error analysis reveals that the roundoff error is proportional to the step size used in the evaluation. This roundoff error is then minimized when, for a two-point central difference, the step size $h_{i}$ for the derivative with respect to some $p_{i}$ is chosen as $h_{i}\approx\left[\text{ULP}\left(p_{i}\right)\right]^{1/3}$. In order for $h_{i}<1$ to occur, the ULP$\left(p_{i}\right)$ must then also be less than unity.
\end{itemize}
Numerical error may also enter through the numerical solution of the RGEs, though similar numerical considerations can help control these errors as well. With these sources of error in mind, the error in evaluating this derivative will remain small, i.e., $\mathcal{O}(<1)$, so long as $|p_{i}|\lesssim10^{15}$ in magnitude for all $i$. This leads to $h_{i}<1$ for double-precision floating point numbers. DEW4SLHA offers the option of performing this calculation with even higher accuracy derivative approximations, such as a four-point or eight-point central difference quotient to further minimize truncation error.

The numerical evaluation of $\DeltaBG$ has several advantages over the
semi-analytic formulae using expansions such as Eq. \ref{eq:mZsparam}.
\bi
\item The numeric routine uses full two-loop RGEs including all
  third generation Yukawa couplings\cite{Martin:1993zk} and one loop threshold
  effects while semi-analytic expansions use one-loop RGEs without
  threshold effects.
\item The semi-analytic expansions were formulated to compute the
  Higgs potential at a scale $Q\sim m_Z$ whilst the numeric routine uses
  an optimized scale choice $Q^2=m_{\tst_1}m_{\tst_2}$ which matches the
  higher scales for MSSM/SM decoupling that are expected from LHC data.
\item Usually the semi-analytic expansions are computed for a particular
  $\tan\beta$ value while the numeric evaluation is valid for all $\tan\beta$.
  \ei

\begin{figure}[!htbp]
\begin{center}
  \includegraphics[height=0.25\textheight]{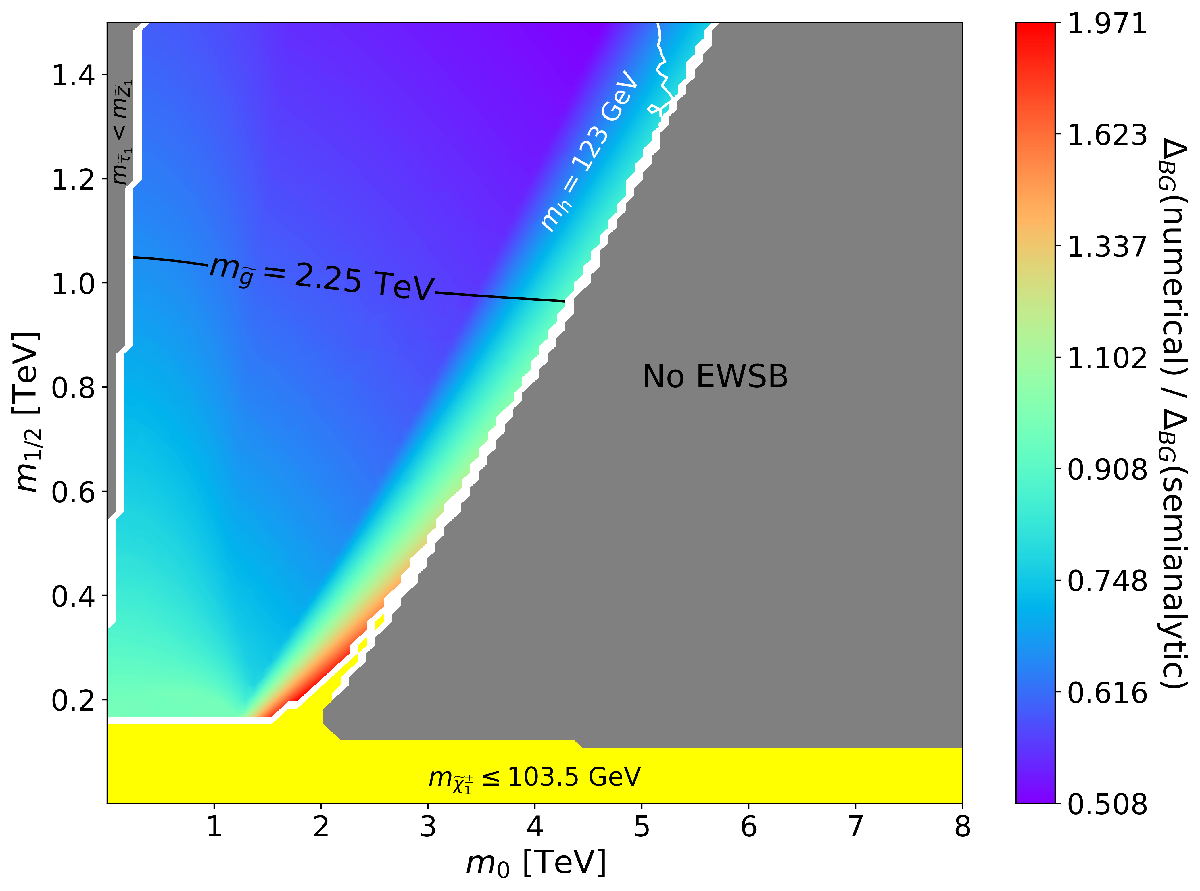}\\
  \includegraphics[height=0.25\textheight]{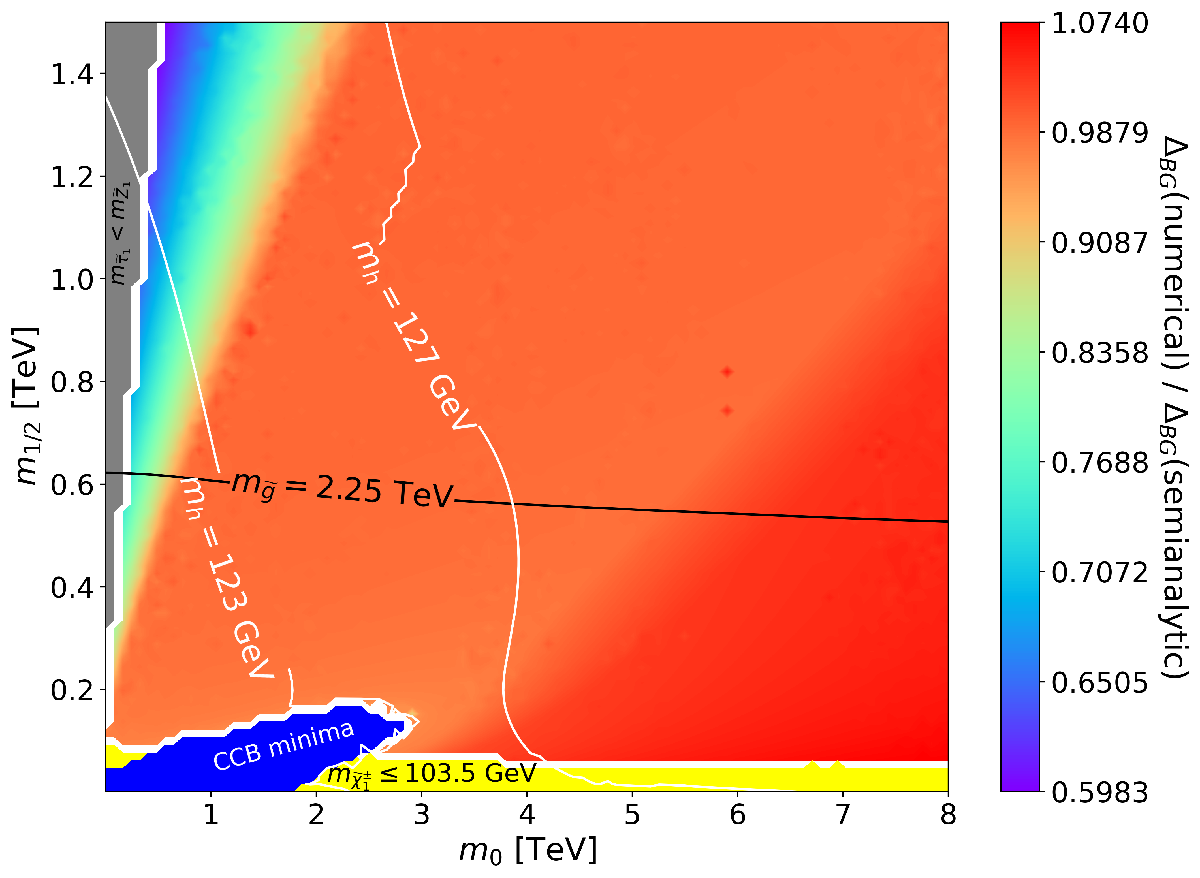}\\
  \includegraphics[height=0.25\textheight]{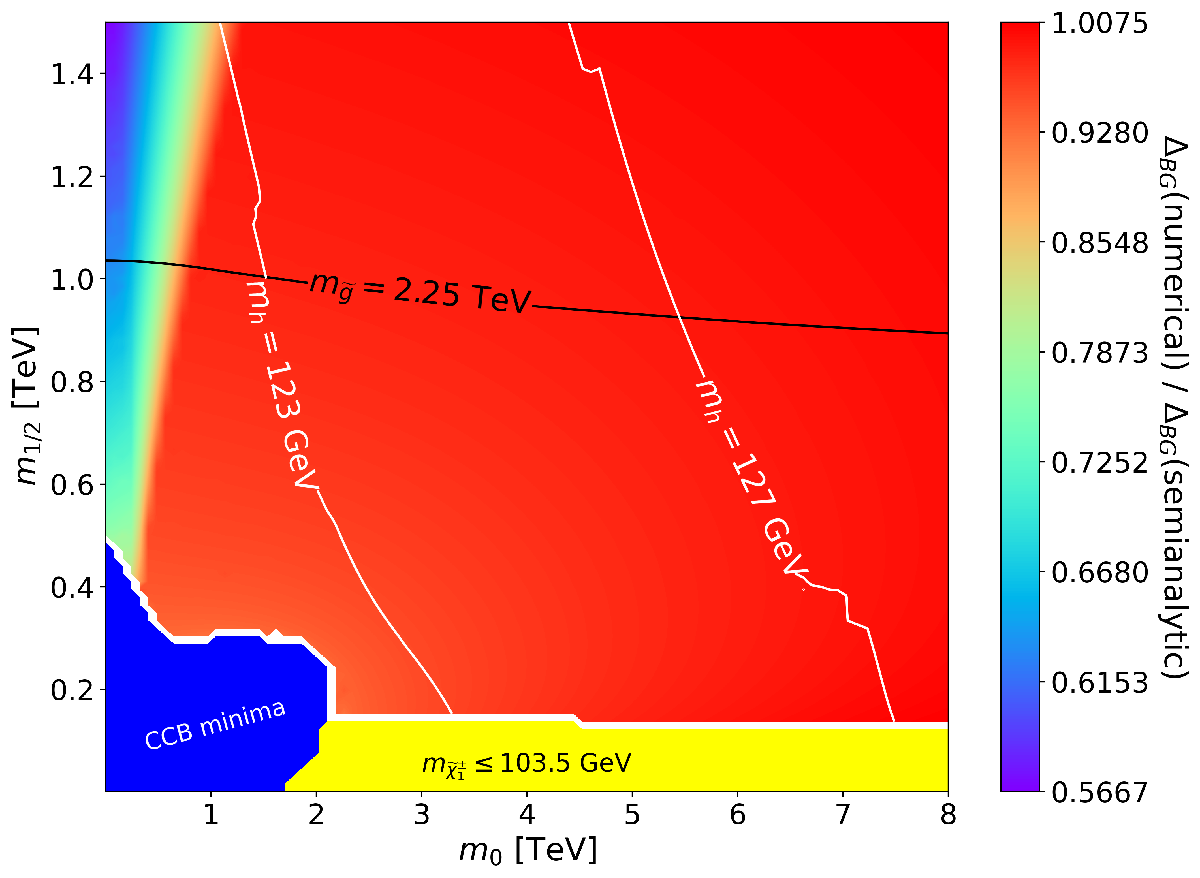}
\caption{Plot of $\Delta_{BG}(\text{numerical})/\DeltaBG (\text{semianalytic})$ in the $m_0$ vs. $m_{1/2}$ plane of {\it a}) the CMSSM/mSUGRA model with $A_0=0$, $\tan\beta =10$ and $\mu >0$, {\it b}) the CMSSM/mSUGRA model with $A_0=-2m_0$ and {\it c}) the NUHM2 model with $\mu =200$ GeV and $A_{0}=-1.6m_{0}$ with $m_A=2$ TeV. We use the code DEW4SLHA to compute $\Delta_{BG}(\text{numerical})$ using a numerical algorithm for the sensitivity coefficients and SoftSUSY v4.1.17 for the spectrum.
\label{fig:DBG_numeric}}
\end{center}
\end{figure}

To illustrate the comparison between the two methods, in
  Fig. \ref{fig:DBG_numeric}{\it a}) we compute the ratio
  $\Delta_{BG}(\text{numerical})/\DeltaBG (\text{semianalytic})$ in the $m_0$ vs. $m_{1/2}$
  plane of the mSUGRA/CMSSM plane for $A_0=0$ and $\tan\beta =10$ with $\mu >0$.
  The blue region corresponds to a ratio $\sim 0.5$ while for small $m_0$
  we find $\Delta_{BG}(\text{numerical})/\DeltaBG (\text{semianalytic})\alt 1$ and
  for large $m_0$ then we find $\Delta_{BG}(\text{numerical})/\DeltaBG (\text{semianalytic})\agt 1$
  with the ratio reaching as high as $\sim 2$ near the lower focus point region.
  
In Fig. \ref{fig:DBG_numeric}{\it b}) we again compute the ratio 
  $\DeltaBG(\text{numerical})/\DeltaBG (\text{semianalytic})$ in the
  $m_{0}$ vs $m_{1/2}$ plane of the mSUGRA/CMSSM, but now for $A_0=-2m_{0}$
  and $\tan\beta=10$ with $\mu > 0$. The large value of $A_{0}$ here
  permits the Higgs mass to be within
  the allowed range of $125\pm2$ GeV. The broad orange and red regions
  throughout the RHS of the plane correspond to where
  $\DeltaBG(\text{numerical})\sim\DeltaBG(\text{semianalytic})$.
  The largest discrepancy between the evaluation methods occurs on the
  LHS of the plane near the stau LSP region, where
  $\DeltaBG(\text{numerical})\sim0.6\DeltaBG(\text{semianalytic})$.
  Fig. \ref{fig:DBG_numeric}{\it c}) instead shows the ratio comparing
  the numerical method to the semianalytic method in the $m_{0}$ vs $m_{1/2}$
  plane of the \emph{NUHM2} model with $\mu =200$ GeV, $m_A=2$ TeV and $A_0=-1.6 m_0$. 
Again, the broad orange and red region on
  the RHS of this plane shows very good agreement between the two methods:
  $\DeltaBG(\text{numerical})\sim\DeltaBG(\text{semianalytic})$. On the LHS
  above the CCB minima region, where $m_{1/2}>m_{0}$, then the semianalytic method
  result becomes somewhat larger than the numerical method result, leading to a
  minimal ratio $\DeltaBG(\text{numerical})\sim0.57\DeltaBG(\text{semianalytic})$.

\subsubsection{Numerical results for $\DeltaBG$}

\begin{figure}[!htbp]
\begin{center}
\includegraphics[height=0.4\textheight]{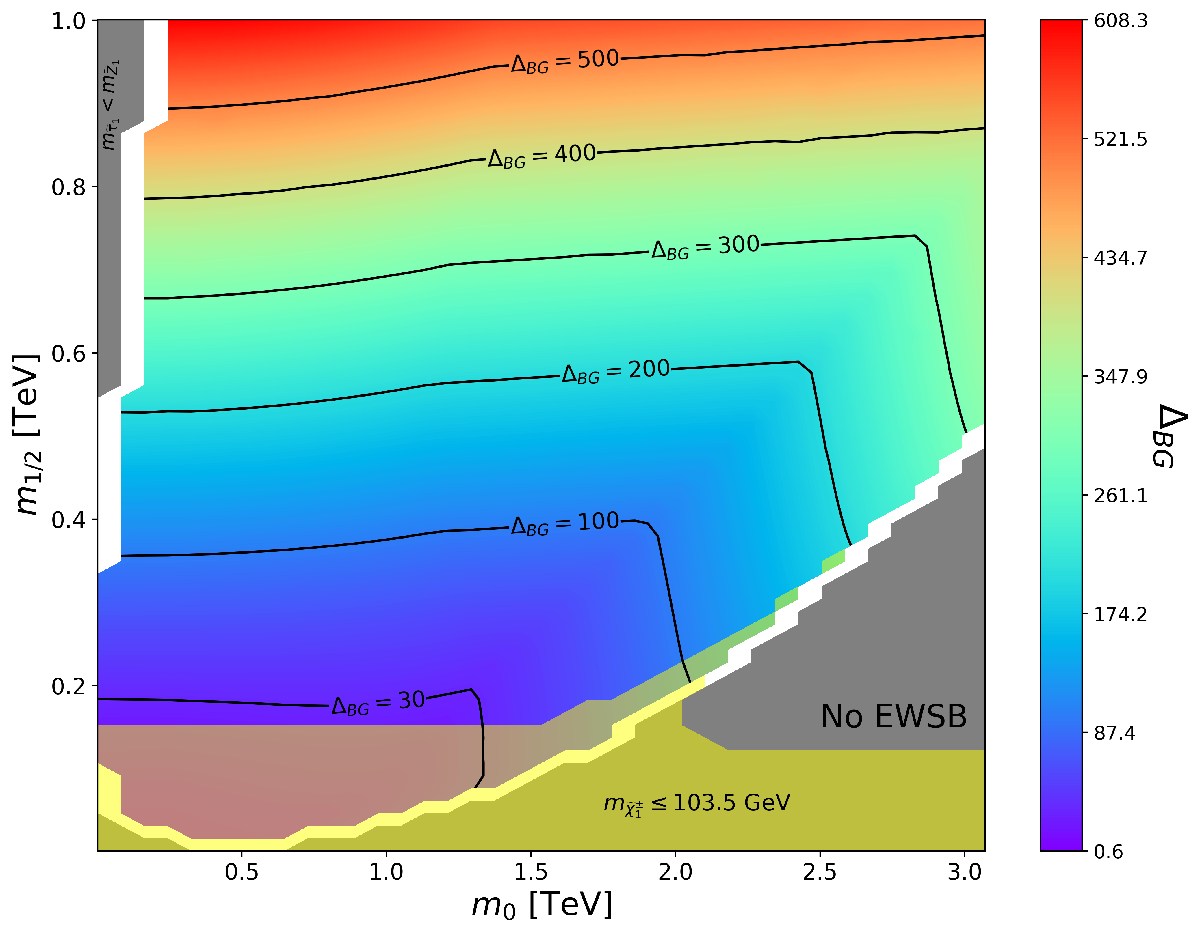}
\caption{Plot of naturalness contours $\Delta_{BG}$ in the $m_0$ vs. $m_{1/2}$
  plane of the CMSSM/mSUGRA model with $A_0=0$, $\tan\beta =10$ and $\mu >0$.
  We use the code DEW4SLHA to compute $\Delta_{BG}$ using a numerical
  algorithm for the sensitivity coefficients and SoftSUSY for the spectrum.
\label{fig:DBG_cmssm}}
\end{center}
\end{figure}

In Fig. \ref{fig:DBG_cmssm}, we compute contours and color-coded regions
of $\DeltaBG$ in the mSUGRA/CMSSM model using a numerical routine
to evaluate the sensitivity coefficients. This routine is embedded in the
publicly available computer code DEW4SLHA which computes the three
measures of naturalness $\DeltaBG$, $\DeltaHS$ and $\DeltaEW$ for
any model based on its Les Houches Accord spectrum generator output file. The results in
Fig. \ref{fig:DBG_cmssm} agree well with those presented by Allanach
{\it et al.} in Ref. \cite{Allanach:2000ii}.

In truth, the various supposedly independent high scale soft terms are
introduced by hand in the mSUGRA/CMSSM model
as a {\it parametrization of our ignorance} as to the SUSY breaking mechanism.
Indeed, in the case of gravity-mediation, if we specify a specific SUSY breaking
mechanism, then all soft terms are calculable in terms of the gravitino mass
$m_{3/2}$.
An example is the famous dilaton-dominated SUSY breaking model\cite{Brignole:1993dj}:
in this case
\be
m_0=m_{3/2}\ \  {\rm with}\ \  m_{1/2}=-A_0=\sqrt{3}m_{3/2}.
\ee
In such a case, then it doesn't make sense that the soft terms are independent:
invoking PN, we should combine dependent terms in Eq. \ref{eq:mZsparam}.
Then $m_Z^2\simeq -2.18\mu^2+14.494 m_{3/2}^2$. Adopting $m_{3/2}=3$ TeV
as in the previous example, then we find $\mu =7735$ GeV and $\DeltaBG =c_{m_{3/2}^2}=15683$.

\begin{figure}[!htbp]
\begin{center}
\includegraphics[height=0.4\textheight]{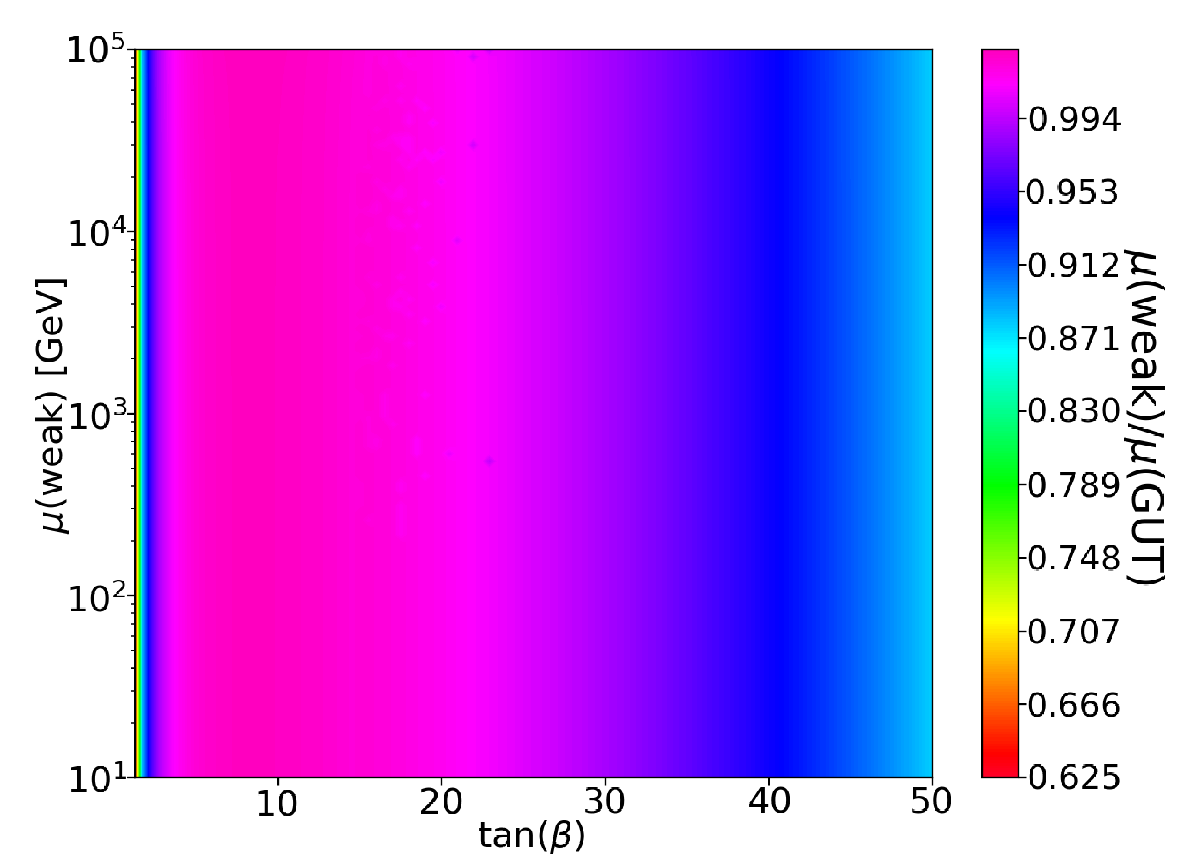}
\caption{Ratio of $\mu /\mu_0$ in the $\tan\beta$ vs. $\mu (\text{weak})$ plane, where $\mu_{0}$ is the GUT-scale value of the $\mu$ parameter.
\label{fig:mumu0}}
\end{center}
\end{figure}

The SUSY $\mu$ parameter evolves very little from the GUT scale
to the weak scale, due to the supersymmetric non-renormalization theorems\cite{Grisaru:1979rn}. The ratio of $\mu (m_{weak})/\mu(m_{GUT})$ is shown in Fig. \ref{fig:mumu0} for the $\tan\beta$ vs. $\mu (m_{weak})$ plane in the mSUGRA/CMSSM model. The deviation between $\mu (m_{weak})/\mu (m_{GUT})$
is typically a few percent, climbing to $\sim 10\%$ at very large $\tan\beta$.

Now, in the case where all soft terms are determined in terms of $m_{3/2}$
(such as gravity-mediation, anomaly-mediation and mirage-mediation),
then we expect roughly that
\be
m_Z^2\simeq -2\mu^2 +a\cdot m_{3/2}^2
\ee
and since $\mu$ hardly evolves, then $a\cdot m_{3/2}^2\simeq -2 m_{H_u}^2(weak)$.
In this case -- with all correlated soft terms
(which we may dub as the SUGRA1 model) --
then $\DeltaBG\sim c_{m_{3/2}^2}= a m_{3/2}^2/m_Z^2\simeq max [2\mu^2, 2 m_{H_u}^2(weak)] /m_Z^2$.
This latter case we will find is
nearly the same as $\DeltaEW$ aside from the inclusion of the radiative
corrections to the weak scale scalar potential.

\begin{figure}[!t]
\begin{center}
\includegraphics[height=0.4\textheight]{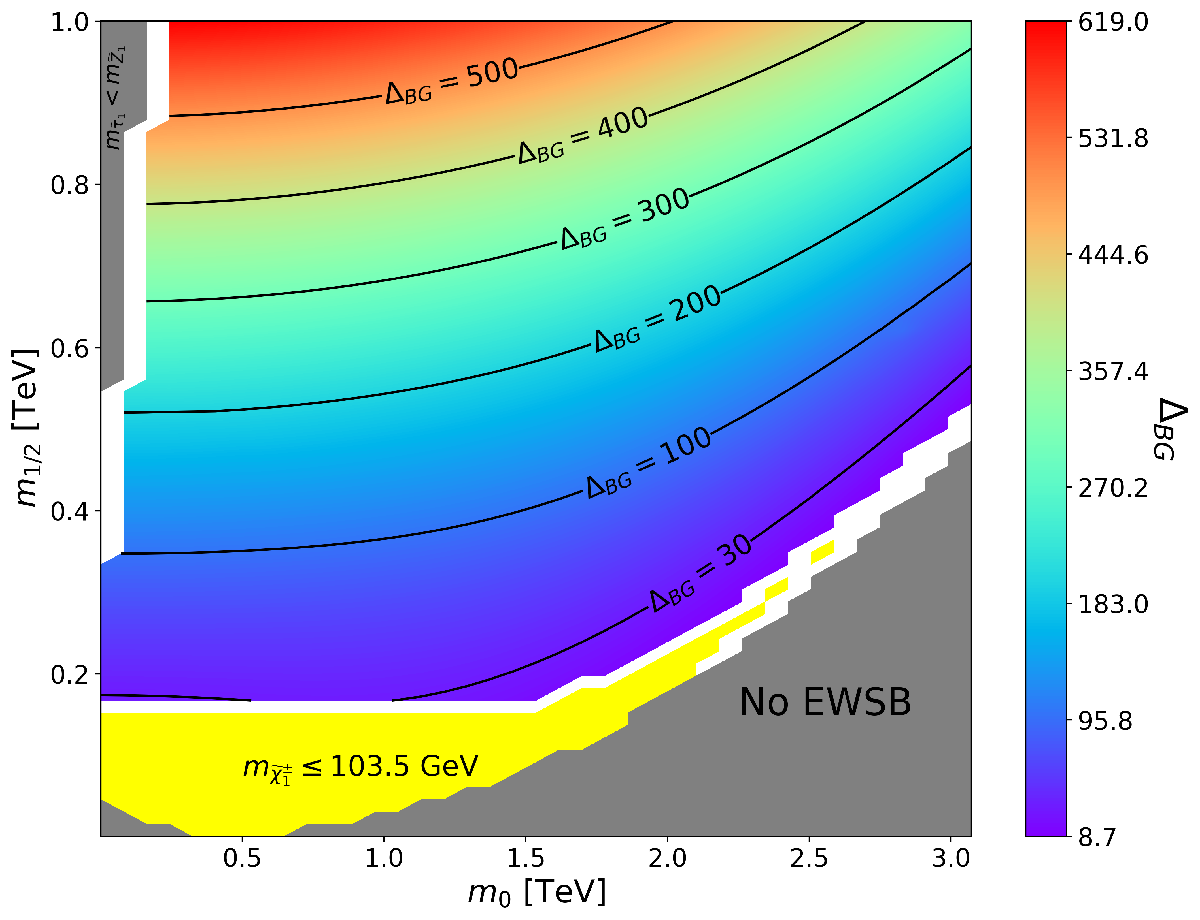}
\caption{Plot of naturalness contours $\Delta_{BG}$ in the $m_0$ vs. $m_{1/2}$
  plane of the one-soft-parameter SUGRA1 model with $A_0=0$, $\tan\beta =10$ and $\mu >0$. We use SoftSUSY to generate the spectra.
  \label{fig:DBG_SUG1}}
 
\end{center}
\end{figure}

In Fig. \ref{fig:DBG_SUG1}, we plot naturalness contours in the same parameter plane as in Fig. \ref{fig:DBG_cmssm}, but now assuming instead the one-soft-parameter SUGRA1 model, For SUGRA1, we have \be
m_Z^2=-2.18\mu_0^2+a\cdot m_0^2\ \ \ \text{(SUGRA1)}
\label{eq:SUG1}
\ee
where the constant $a$ can be determined via $a=(m_Z^2+2.18\mu_0^2)/m_0^2$.
In this case, the naturalness contours roughly follow the contours
of constant $\mu$ value. (The $\mu$ term all by itself has been advocated
as a measure of naturalness by Chan {\it et al.}\cite{Chan:1997bi}.) For the case of SUGRA1, the naturalness contours are very different from the case of independent high scale soft terms assumed in the mSUGRA/CMSSM model.

One may also define a SUGRA2 model. Here, we assume that since gaugino masses arise from the gauge kinetic function, this soft term is independent of the others which are determined instead by the K\"ahler function, but where $A_0$ is determined in terms of $m_0$ (such as $A_0=-2m_0$) so that
\be
m_Z^2=-2.18\mu_0^2+3.786 m_{1/2}^2-0.427 m_0^2+1.642 m_{1/2}m_0\ \ \ (SUGRA2).
\label{eq:SUG2}
\ee
Finally, SUGRA3 allows that $A_0$ is somehow independent from $m_0$ (or $m_{3/2}$)
so that
\be
m_Z^2=-2.18\mu_0^2+3.786 m_{1/2}^2+0.013 m_0^2+1.642 m_{1/2}m_0+0.22 A_0^2\ \ \ (SUGRA3).
\label{eq:SUG3}
\ee
For the three cases, we find that the $\DeltaBG$ values are very different
in the SUGRA1, SUGRA2 or SUGRA3 models just depending on which parameters
are assumed to be truly independent.

\begin{figure}[!htbp]
\begin{center}
  \includegraphics[height=0.23\textheight]{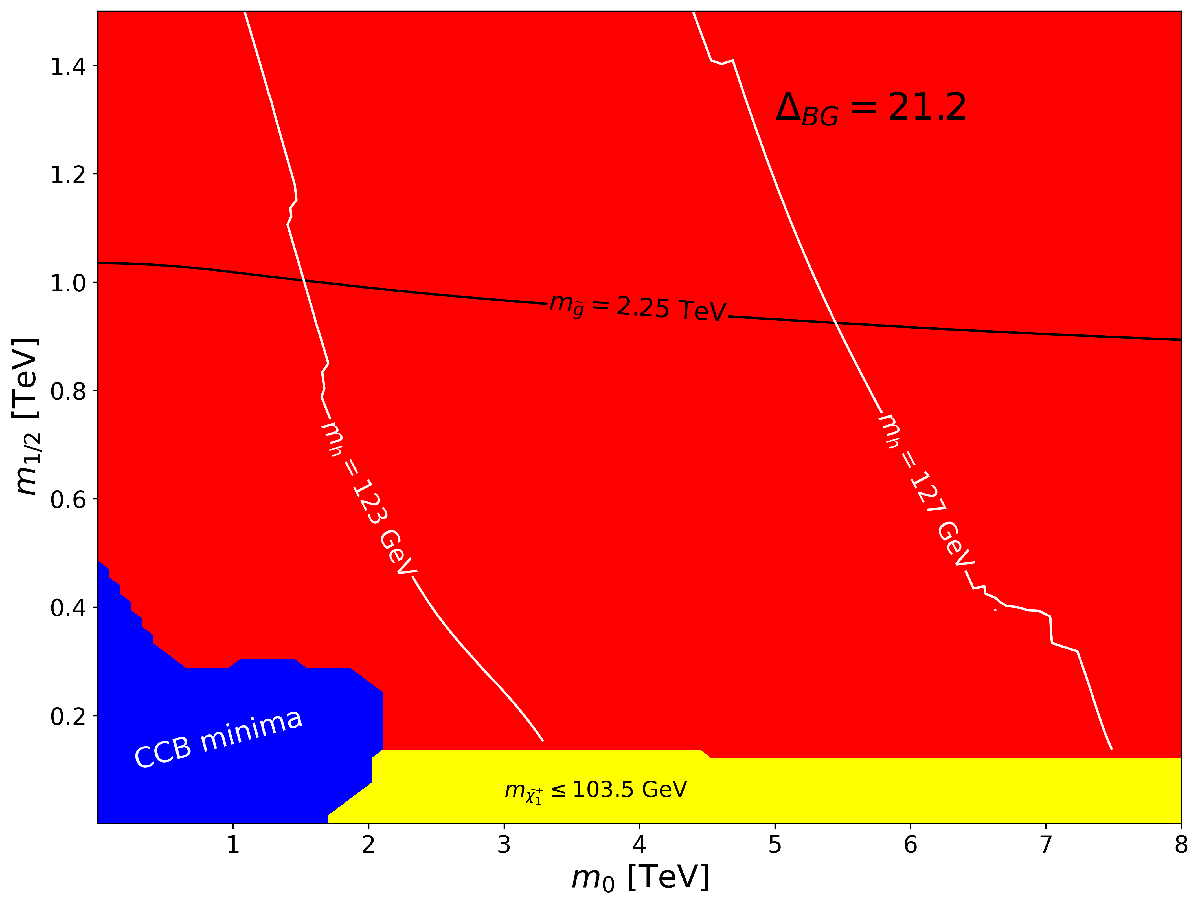}\\
  \includegraphics[height=0.25\textheight]{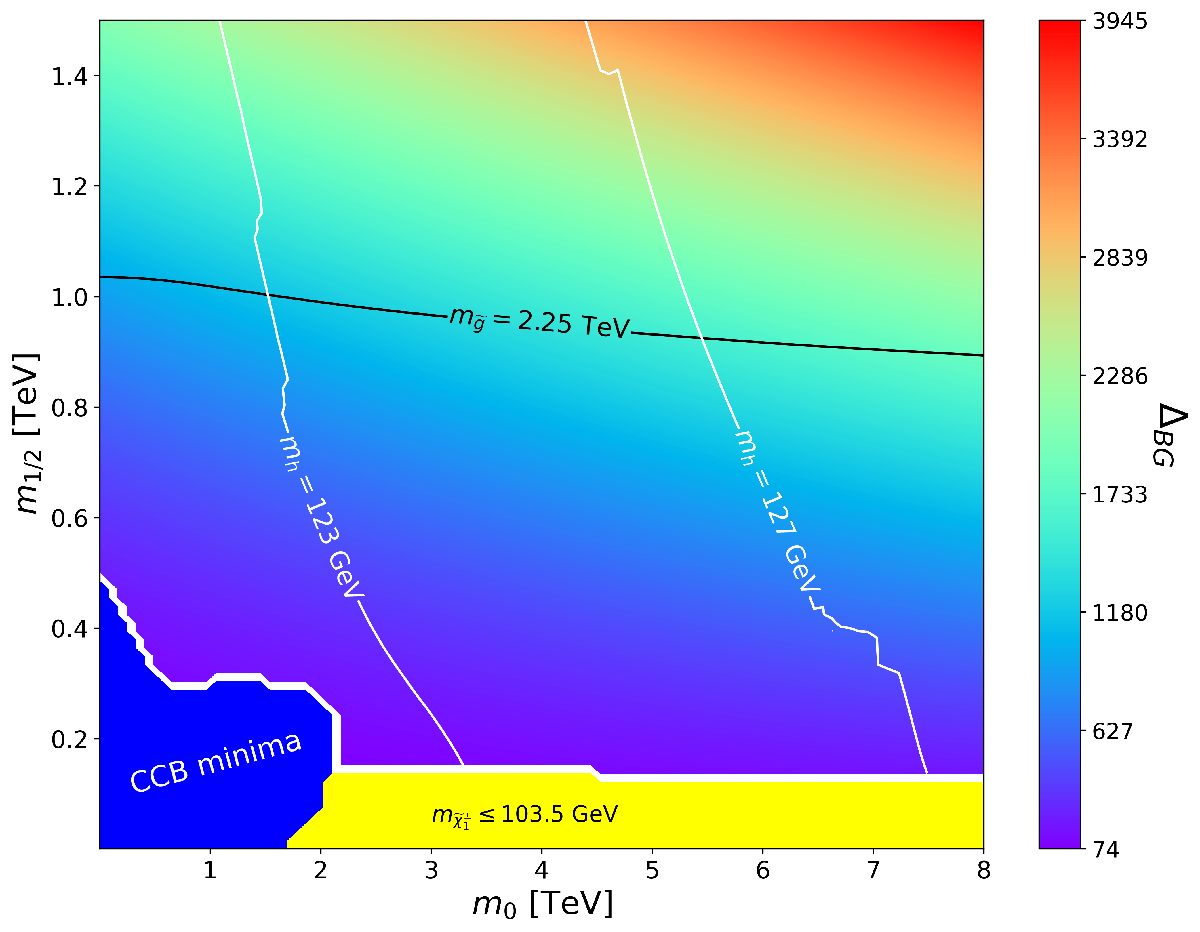}\\
  \includegraphics[height=0.25\textheight]{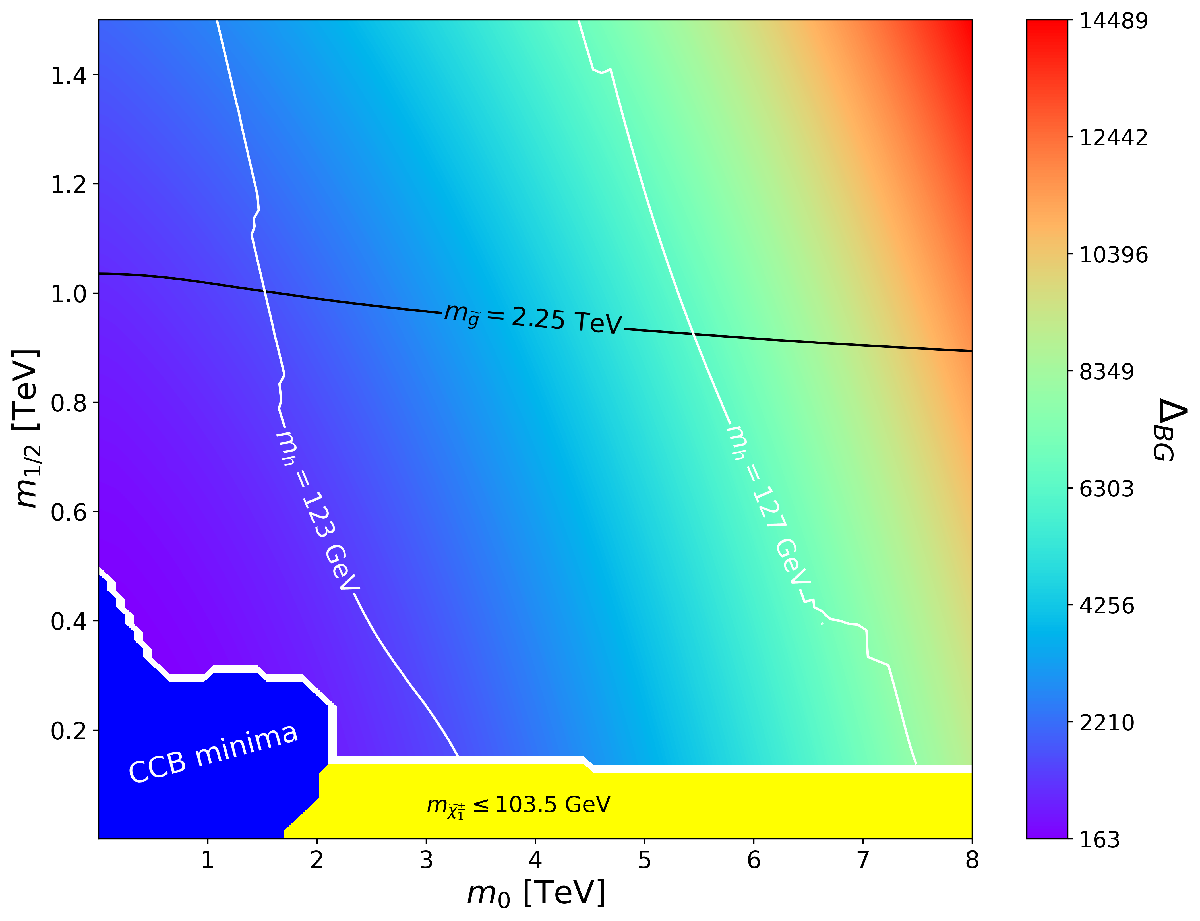}
  \caption{Plot of $\DeltaBG$ values in the $m_0$ vs. $m_{1/2}$ plane
    for the NUHM2 model for $A_0=-1.6 m_0$, $\tan\beta =10$ with $\mu =200$ GeV
    and $m_A=2$ TeV.
    In {\it a}), we plot $\DeltaBG$ assuming a single
    independent soft parameter $m_{3/2}$ while in {\it b}) we
    plot $\DeltaBG$ for assumed two independent soft parameters
    $m_0$ and $m_{1/2}$ while in {\it c}) we plot assuming
    all three of $m_0$, $m_{1/2}$ and $A_0$ are independent.
      The spectrum is calculated using SoftSUSY
  and the naturalness measures with DEW4SLHA.
\label{fig:SUGiinNUHM2}}
\end{center}
\end{figure}

In Fig. \ref{fig:SUGiinNUHM2}, we show color coded regions of $\DeltaBG$
as computed in the $m_0$ vs. $m_{1/2}$ plane of the NUHM2 model where $\tan\beta =10$, $A_0=-1.6 m_0$ with $\mu =200$ GeV and $m_A=2$ TeV. In frame {\it a}), we assume all soft terms are correlated as in Eq. \ref{eq:SUG1}. In this case, since $\mu$ is fixed, there is a constant value of $\DeltaBG=21.2$ throughout the plane. 

In frame {\it b}), we instead assume two independent soft parameters
$m_0$ and $m_{1/2}$ (but with $A_0$ fixed in terms of $m_0$) so that
we are in the SUGRA2 model, Eq. \ref{eq:SUG2}.
Here, the value of $\DeltaBG$ is vastly different from frame {\it a}),
reaching up to values of $\sim 3900$ in the upper-right corner:
a factor of $\sim 180$ times greater than the frame {\it a}) value.
Here, the $\DeltaBG$ finetuning is dominated by the $m_{1/2}$ value
but not so much by $m_0$. In frame {\it c}), instead we show values
of $\DeltaBG$ assuming three independent soft parameters as in
Eq. \ref{eq:SUG3}. In this case, with $A_0$ fixed as $A_0=-1.6 m_0$
but nonetheless declared as independent, we see a greater dependence
on $m_0$, so $\DeltaBG$ increases as $m_0$ increases, mainly because
$A_0$ increases with increasing $m_0$. Here, $\DeltaBG$ reaches maximal
values of $\sim 14500$ in the upper-right corner, a factor $\sim 680$
larger than the frame {\it a}) value!

In summary, from the discussion of this Section, we see that the
measure $\DeltaBG$ could be a legitimate finetuning measure if there
could be consensus on what constitutes {\it independent parameters} of
the model. The plots also illustrate the extreme model-dependence of
$\DeltaBG$, where $\DeltaBG$ can obtain values differing by several
orders of magnitude depending on which parameters $p_i$ are assumed
fundamental or independent.

\subsection{High scale finetuning}
\label{ssec:DHS}

An alternative to EENZ/BG naturalness which we label as high scale finetuning
(HS) emerged early on in the 21st century.
It may have been intended originally as a figurative bullet point
indicator to argue for sparticle masses near the weak scale\cite{Murayama:2000dw}, but later was taken more seriously\cite{Harnik:2003rs,Kitano:2006gv,Papucci:2011wy,Brust:2011tb}.
This measure seeks to
apply PN to the Higgs boson mass relation (see {\it e.g.}
Eq. 10 of \cite{Bae:2015nva})
\be
m_h^2\simeq \mu^2 +m_{H_u}^2(weak) +{\rm EW} +{\rm mixing}
\label{eq:mhs}
\ee
where the EW corrections and mixings are already $\alt m_h^2$.
The idea then is to break $m_{H_u}^2(weak)$ into
$m_{H_u}^2(m_{GUT})+\delta m_{H_u}^2$ and require $\delta m_{H_u}^2\alt m_h^2$.
The full one-loop expression for $\delta m_{H_u}^2$ may be obtained by
integrating its one-loop RGE from $m_{GUT}$ to $m_{weak}$:
\be
\frac{dm_{H_u}^2}{dt}=\frac{2}{16\pi^2}\left(-\frac{3}{5}g_1^2M_1^2-3g_2^2M_2^2
+\frac{3}{10}g_1^2 S+3f_t^2X_t\right)
\label{eq:dmHu_approx}
\ee
where $t=\log Q$, $S=m_{H_u}^2-m_{H_d}^2+Tr\left[ {\bf m}_Q^2-{\bf m}_L^2-2{\bf m}_U^2+{\bf m}_D^2+{\bf m}_E^2\right]$ and $X_t=m_{Q_3}^2+m_{D_3}^2+m_{H_u}^2+A_t^2$.
In the literature\cite{Kitano:2006gv,Papucci:2011wy,Brust:2011tb},
to gain a simple expression, the terms with gauge couplings are ignored and $X_t$ is approximated as $X_t\sim m_{Q_3}^2+m_{D_3}^2+A_t^2$, where $m_{Q_3}^{2},m_{D_3}^{2},$ and $A_t^2$ here are GUT-scale values.
Then a single step integration leads to
\be
\delta m_{H_u}^2\sim -\frac{3}{8\pi^2}f_t^2\left(m_{Q_3}^2+m_{U_3}^2+A_t^2\right)\log\left( \Lambda /m_{weak}\right)
\label{eq:dmHu2_approx}
\ee
where the high scale $\Lambda$ is usually assumed $\sim m_{GUT}$.
The $\DeltaHS$ measure famously promoted three light third generation
squarks below the 500 GeV scale\cite{Papucci:2011wy}, and motivated
intensive searches by the LHC collaborations to root out light top-squark signals.

In order to compare $\DeltaHS$ more appropriately with $\DeltaBG$ and
$\DeltaEW$, we slightly redefine $\DeltaHS$ in terms of $m_Z^2/2$\cite{Baer:2012mv} where
in this case we take
\be
m_Z^2/2=\frac{(m_{H_d}^2(\Lambda )+\delta m_{H_d}^2+\Sigma_d^d)-(m_{H_u}^2(\Lambda )+\delta m_{H_u}^2+\Sigma_u^u )\tan^2\beta}{\tan^2\beta -1}-(\mu^2(\Lambda )+\delta\mu^2)
\label{eq:mzsHS}
\ee
and $\Lambda$ is some input high scale, perhaps $m_P$ or $m_{GUT}$.
Then
\be
\DeltaHS =max|{\rm largest}\ {\rm term}\ {\rm on}\ {\rm RHS}\ {\rm of}\
{\rm Eq.}\ \ref{eq:mzsHS}|.
\ee
In this way, the three measures become equal in certain limiting cases.

\begin{figure}[!htbp]
\begin{center}
\includegraphics[height=0.4\textheight]{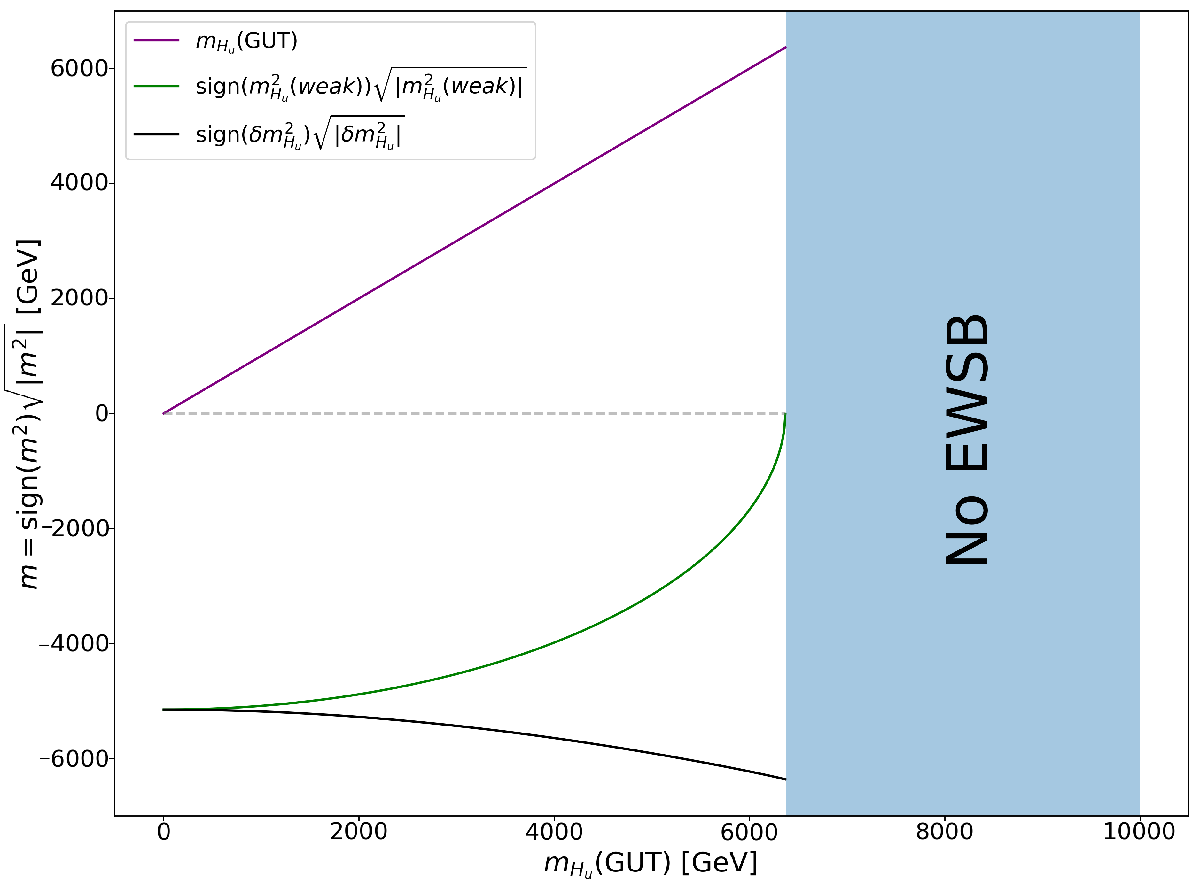}
\caption{Plot of $sign(\delta m_{H_u}^2)\cdot\sqrt{|\delta m_{H_u}^2 |}$
  vs. $m_{H_u}(GUT)$ for the NUHM2 model with $m_0=5$ TeV,
  $m_{1/2}=1.2$ TeV, $A_0=-1.6m_0$, $\tan\beta =10$ and $m_{H_d}=5$ TeV.
\label{fig:dmHus}}
\end{center}
\end{figure}

\begin{figure}[!htbp]
\begin{center}
\includegraphics[height=0.3\textheight]{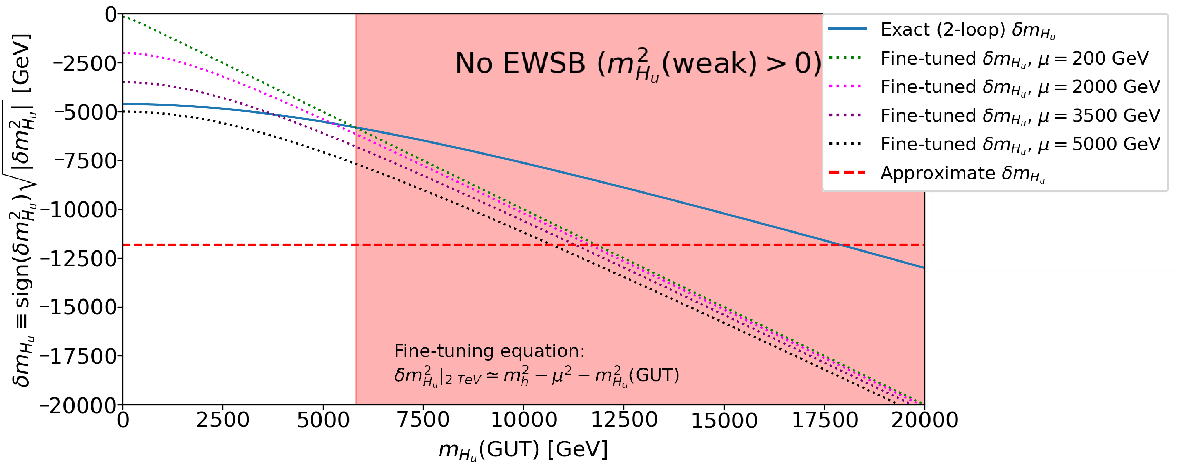}
\caption{Plot of $sign(\delta m_{H_u}^2)\cdot\sqrt{|\delta m_{H_u}^2 |}$
  vs. $m_{H_u}(GUT)$ for the NUHM2 model with $m_0=4.5$ TeV,
  $m_{1/2}=1$ TeV, $A_0=-7.2$ TeV and $\tan\beta =10$ with $m_{A}=2$ TeV.
  We show the approximate expression Eq. \ref{eq:dmHu_approx} (red-dashed curve)
  along with exact 2-loop expression (blue solid) along with the value gleaned from finetuning for various values of $\mu$.
  \label{fig:dmHus_approx}}
\end{center}
\end{figure}

The $\DeltaHS$ measure is problematic on several counts
\begin{enumerate}
  \item It violates the PN precept in that, in simplifying $\delta m_{H_u}^2$,
all dependence on $m_{H_u}^2(\Lambda )$ is lost, which hides the fact that
$\delta m_{H_u}^2$ is actually dependent on $m_{H_u}^2 (\Lambda )$. In fact,
the bigger the assumed value for $m_{H_u}^2(\Lambda )$, then the bigger is the
cancelling correction $\delta m_{H_u}^2$\cite{Baer:2015fsa}.
This is shown in Fig. \ref{fig:dmHus} where we show the exact two-loop
value of $\delta m_{H_u}^2$ vs. $m_{H_u}^2(GUT)$, where the clear
dependence of $\delta m_{H_u}^2$ on $m_{H_u}^2(GUT)$ is shown.
The plot also shows that the bigger $m_{H_u}(GUT)$ becomes, then the
more EW-natural the model becomes in that $m_{H_u}^2(weak)$ becomes comparable
to $m_Z^2$ on the right-hand-side shortly before EWSB is no longer broken.
The splitting up of $m_{H_u}^2(weak)$
into $m_{H_u}^2(\Lambda )+\delta m_{H_u}^2$ turns $\DeltaHS$ into contradiction
with $\DeltaBG$, where $m_{H_u}^2(weak)$ is expanded into high scale parameters
in Eq. \ref{eq:mZsparam} but not split into
$m_{H_u}^2(\Lambda )+\delta m_{H_u}^2$.
This splitting of $m_{H_u}^2(weak)$ into dependent parts destroys the
cancellations needed for focus point SUSY\cite{Feng:1999mn,Feng:1999zg} which is
promoted as allowing for TeV-scale top-squarks.
\item Electroweak symmetry breaking in SUSY models is accomplished by
  driving $m_{H_u}^2$ to negative values owing to the large top-quark Yukawa coupling $f_t$. Indeed, the REWSB mechanism is touted as one of the triumphs of WSS
  since it required $m_t\sim 100-200$ GeV\cite{Alvarez-Gaume:1983drc}
  at a time when experiments seemed to indicate $m_t\sim 40$ GeV.
  By requiring $\delta m_{H_u}^2$ to be small, then often $m_{H_u}^2(weak)$
  will not be
  large-negative enough to cause EWSB.
  In the context of vacua selection in the string landscape,
  such models without EWSB would likely not lead to inhabitable universes
  and would be vetoed. This can be viewed as a selection mechanism to favor
  models with large enough $\delta m_{H_u}^2$ such that EW symmetry is
  properly broken (see {\it e.g.} Fig. 3 of Ref. \cite{Baer:2016lpj}.)
\item There is also substantial ambiguity in evaluating $\DeltaHS$.
  In Fig. \ref{fig:dmHus_approx} we show the value of
  $\delta m_{H_u}\equiv sign(\delta m_{H_u}^2)\sqrt{|\delta m_{H_u}^2|}$
  vs. $m_{H_u}(m_{GUT})$ for a NUHM2 benchmark point with $m_0=4.5$ TeV,
  $m_{1/2}=1$ TeV, $A_0=-7.2$ TeV with $\tan\beta =10$ and $m_A=2$ TeV.
  The approximate expression Eq. \ref{eq:dmHu2_approx} is shown as the
  flat red-dashed line which of course doesn't depend on $m_{H_u}^2(m_{GUT})$.
  The solid blue curve is the exact two-loop RG expression for
  $\delta m_{H_u}$ and is shown to deviate from the approximate result
  by well over a factor of 2 at low $m_{H_u}(m_{GUT})$ and only agrees
  with the approximation far into the excluded region where the
  electroweak symmetry isn't properly broken.
  Alternatively, one may use the $m_h^2\simeq \mu^2+m_{H_u}^2+\delta m_{H_u}^2$
  equation for a particular set of input parameters including
  $m_{H_u}^2(m_{GUT})$ ({\it e.g.} in the NUHM2 model) to compute the
  value of $\delta m_{H_u}^2$ and then try to finetune $m_{H_u}^2(m_{GUT})$
  to enforce $m_h=125$ GeV. But as one tunes the value of $m_{H_u}^2(m_{GUT})$,
  then the value of $\delta m_{H_u}^2$ changes accordingly (as indicated by the
  varius dotted lines for different input $\mu$ values), so that instead
  of finetuning, one must adopt an iterative procedure to try and find a
  solution. Sometimes the solution will migrate into the noEWSB region
  while other times the iterations can find a viable solution.
\end{enumerate}

\subsection{Electroweak naturalness}
\label{ssec:DEW}

As mentioned before, the electroweak naturalness measure $\DeltaEW$
measures the largest contribution on the right-hand-side of Eq. \ref{eq:mzs}
and compares that to $m_Z^2/2$. This is the most conservative, unavoidable
measure of naturalness since it is independent of any high scale model.
Even when high scale parameters are correlated in some way, those correlations
are typically lost under RG running and subsequent computation of the
physical sparticle mass eigenstates. The interpretation of $\DeltaEW$ is
clear: if any one of the RHS contributions to Eq. \ref{eq:mzs} is far larger
than $m_Z^2/2$, then it is highly implausible (but not impossible) that some
other contribution would accidentally be large, opposite-sign such that
the two conspire to give an $m_Z$ value of just $91.2$ GeV.
In this sense, natural models correspond to plausible models; models
with large $\DeltaEW$ are logically possible, but highly implausible.
We'll see later that this manifests itself as a {\it probability}, or
likelihood, to emerge from scans over the string landscape.

The tree-level contributions to $\DeltaEW$ are instructive:
\begin{itemize}
\item the SUSY conserving $\mu$ parameter, which sets the mass scale for
  the $W$, $Z$, $h$ and $higgsinos$ enters the weak scale directly.
  We already know that $m_{W,Z,h}\sim 100$ GeV; the higgsinos should lie
  within a factor of several of the measured value of the weak scale.
  In light of LHC constraints, the SUSY LSP is likely a higgsino-like
  lightest neutralino, or at worst a gaugino-higgsino admixture.
\item The value of $m_{H_u}^2$, where $H_u$ acts as the SM Higgs doublet,
  should be driven to small, negative values since it also sets the mass of
  the $W$, $Z$ and $h$ bosons.
\item The value of $m_{H_d}$-- which sets the mass scale for the heavier Higgs
  bosons $A$, $H$ and $H^\pm$-- can be much larger since its contribution to the
  weak scale is suppressed by a factor $\tan\beta$.
  \end{itemize}

The loop-level contributions $\Sigma_u^u$ and $\Sigma_d^d$ are proportional
to the individual particle/sparticle masses but since the $\Sigma_d^d$ terms are
suppressed by $\tan\beta$, the $\Sigma_u^u$ terms are usually dominant.
Of the $\Sigma_u^u$ terms, usually $\Sigma_u^u(\tst_{1,2})$ are largest
owing to the large top-quark Yukawa coupling.
Since these terms are all
suppressed by loop factors, the particle/sparticle masses which enter the
$\Sigma_u^u$ terms can be at the TeV or beyond scale before becoming
comparable to the weak scale. Explicit expressions for the $\Sigma_u^u$
and $\Sigma_d^d$ are given in the Appendices to Ref's \cite{Baer:2012cf}
and \cite{Baer:2021tta}.
The dominant terms are given by
\be
\Sigma_u^u(\tst_{1,2})=\frac{3}{16\pi^2}F(m_{\tst_{1,2}}^2)
\left[ f_t^2-g_Z^2\mp\frac{f_t^2A_t^2-8g_Z^2(\frac{1}{4}-\frac{2}{3}x_W)
 \Delta_t}{m_{\tst_2}^2-m_{\tst_1}^2}\right]
\label{eq:Sigmauu}
 \ee
where $F(m^2)=m^2\left(\log\frac{m^2}{Q^2}-1\right)$ and the optimized
scale choice is taken as $Q^2=m_{\tst_1}m_{\tst_2}$.  Also,
$\Delta_t=(m_{\tst_L}^2-m_{\tst_R}^2)/2+m_Z^2\cos 2\beta (\frac{1}{4}-\frac{2}{3}x_W)$ with $g_Z^2=(g^2+g^{\prime 2})/8$ and $x_W=\sin^2\theta_W$;
in the denominator of Eq. \ref{eq:Sigmauu}, the tree-level masses should be used. 

\begin{figure}[!htbp]
\begin{center}
\includegraphics[height=0.4\textheight]{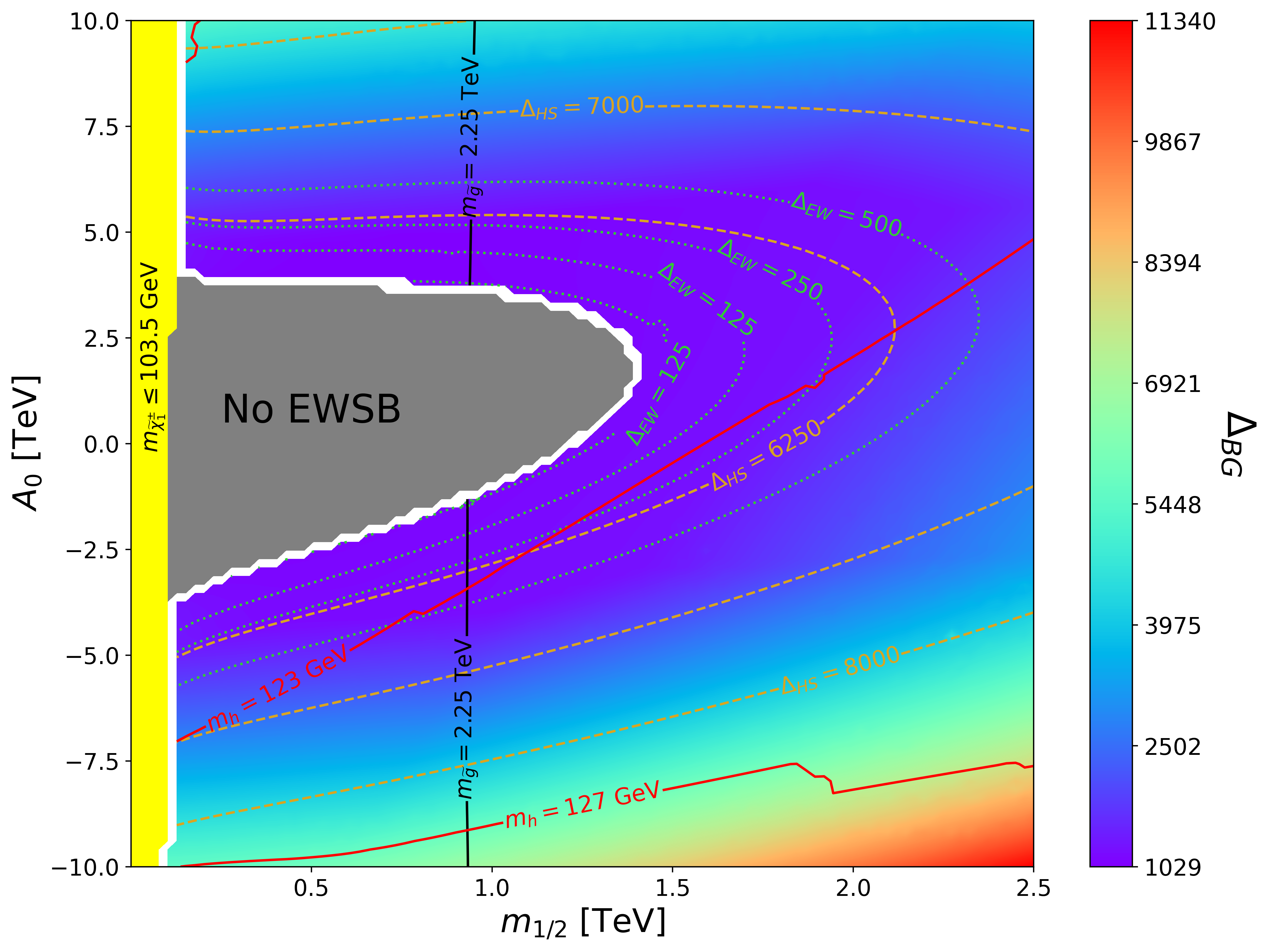}
\caption{Plot of color-coded values of $\DeltaBG$ in the $m_{1/2}$
  vs. $A_0$ plane of the mSUGRA/CMSSM   model for $m_0=5 $ TeV,
  $\tan\beta =10$ and $\mu >0$. We also show contours of Higgs mass $m_h=123$ and $127$ GeV, and contours of $\DeltaEW$ and $\DeltaHS$.
\label{fig:mhfA0}}
\end{center}
\end{figure}

Some highlights of the $\Sigma_u^u$ terms include the following.
\begin{itemize}
\item For $\DeltaEW\alt 30$, the top-squark contributions
  $\Sigma_u^u(\tst_{1,2})$ allow for
  top-squarks up to $m_{\tst_1}\alt 3$ TeV and $m_{\tst_2}\alt 8$ TeV.
  The explicit expressions contain
  large cancellations for large $A_t$ both for $\Sigma_u^u(\tst_1)$ and
  $\Sigma_u^u(\tst_2)$. The large $A_t$ helps to lift $m_h$ into the 125 GeV
  range since $m_h$ is maximal for $x_t\sim \sqrt{6}m_{\tst}$\cite{Slavich:2020zjv}. This is in contrast to $\DeltaBG$ and $\DeltaHS$ which both prefer
  small trilinear soft terms.
  In Fig. \ref{fig:mhfA0} we show color-coded regions of $\DeltaBG$
  in the $m_{1/2}$ vs. $A_0$ plane of the mSUGRA/CMSSM model for
  $m_0=5$ TeV, $\tan\beta =10$ and $\mu >0$.
  We also show contours of Higgs mass $m_h=123$ and $127$ GeV, and contours
  of $\DeltaEW$ and $\DeltaHS$. The grey region around $A_0\sim 0$ is the
  focus point region. From the plot, we see  that $\DeltaHS$ is always large,
  $\DeltaHS\agt 6000$, due to the large value of $m_0$.
  Meanwhile, $\DeltaBG$ reaches as low as $\sim 1000$, also in the FP region.
  $\DeltaEW$ can reach as low as 62 in between the two $\DeltaEW =125$
  contours. As expected from the mSUGRA/CMSSM model, no points allow for
  both low finetuning and $m_h\sim 125$ GeV.
\item Since first/second generation Yukawa couplings are tiny,
  then these sparticle masses can be much larger than the third generation,
  with first/second generation squarks and sleptons ranging up to $30-50$ TeV.
  In the context of the string landscape, this leads to a
  quasi-degeneracy/decoupling solution to the
  SUSY flavor and CP problems\cite{Baer:2019zfl}.
\item Gluinos affect the $\Sigma_u^u$ via RG running and
  directly at the two-loop level\cite{Dedes:2002dy}.
  They can range up to $m_{\tg}\alt 6$ TeV for
  $\DeltaEW\alt 30$, well beyond present LHC bounds\cite{Baer:2017pba}.
  \end{itemize}

A positive feature of $\DeltaEW$ is its model independence (within the
context of models for which the MSSM is the weak scale EFT).
The amount of finetuning only depends on the weak scale spectrum which
is generated, but not on how it is obtained. Thus, if one generates a
certain weak scale spectrum via some high scale model, or just the pMSSM,
then one gets the same value of $\DeltaEW$. This of course isn't true for
the measures $\DeltaHS$ or $\DeltaBG$.

A common criticism of $\DeltaEW$ is that it doesn't account for high scale
parameter choices and correlations.
This is not exactly true as discussed earlier.
The $\mu$ parameter evolves only slightly from $m_{GUT}$ to $m_{weak}$,
as shown in Fig. \ref{fig:mumu0}.
With $\mu (m_{GUT})\simeq \mu (m_{weak})$, and in the context of all soft terms
correlated (as should be the case in a well specified SUSY breaking model),
then $\DeltaBG\simeq \DeltaEW$, sans the radiative corrections $\Sigma_u^u$
and $\Sigma_d^d$. Also, if the dependent terms $m_{H_u}^2(\Lambda )$ and $\delta m_{H_u}^2$ are combined, as required by PN, then $\DeltaHS \simeq \DeltaEW$,
sans radiative corrections. Furthermore, the specific choices of high scale
parameters can lead to more or less finetuning via Eq. \ref{eq:mzs}.
In fact, a string landscape selection for larger soft terms often results in
smaller values of $\DeltaEW$ as compared to any selection for small or
weak scale soft terms\cite{Baer:2021uxe}.

\subsection{Stringy naturalness: anthropic origin of the weak scale}
\label{ssec:stringy}

A fourth entry into the naturalness debate comes from Douglas\cite{Douglas:2004qg} with regards to
the string landscape: stringy naturalness, as remarked above.
An advantage of stringy naturalness is that it actually provides an
explanation for the magnitude of the weak scale, and not just naturalness
of the weak scale.
The distribution of vacua in the multiverse as a function of $m_{soft}$
is expected to be
\be
dN_{vac}\sim f_{SUSY}(m_{soft})\cdot f_{EWSB}(m_{soft}) dm_{soft}.
\ee
Douglas\cite{Douglas:2004qg} advocates for a power-law draw to large soft terms based on the supposition 
that there is no favored value for SUSY breaking fields on the landscape: 
$f_{SUSY}\sim m_{soft}^{2n_F+n_D-1}$ where $n_F$ is the number of (complex-valued)
$F$-breaking fields and $n_D$ is the number of (real-valued) $D$-breaking fields giving
rise to the ultimate SUSY breaking scale. The distribution $f_{EWSB}$ is suggested as
$f_{EWSB}=\Theta (30-\DeltaEW )$\cite{Baer:2017uvn} such that the value of the weak scale in 
each pocket universe lies within the ABDS window\cite{Agrawal:1997gf}, the so-called 
{\it atomic principle}. 
At present, SN does not admit a clear numerical measure\cite{Baer:2019cae}.

\section{Comparison of measures}
\label{sec:compare}

\begin{figure}[!htbp]
\begin{center}
\includegraphics[height=0.35\textheight]{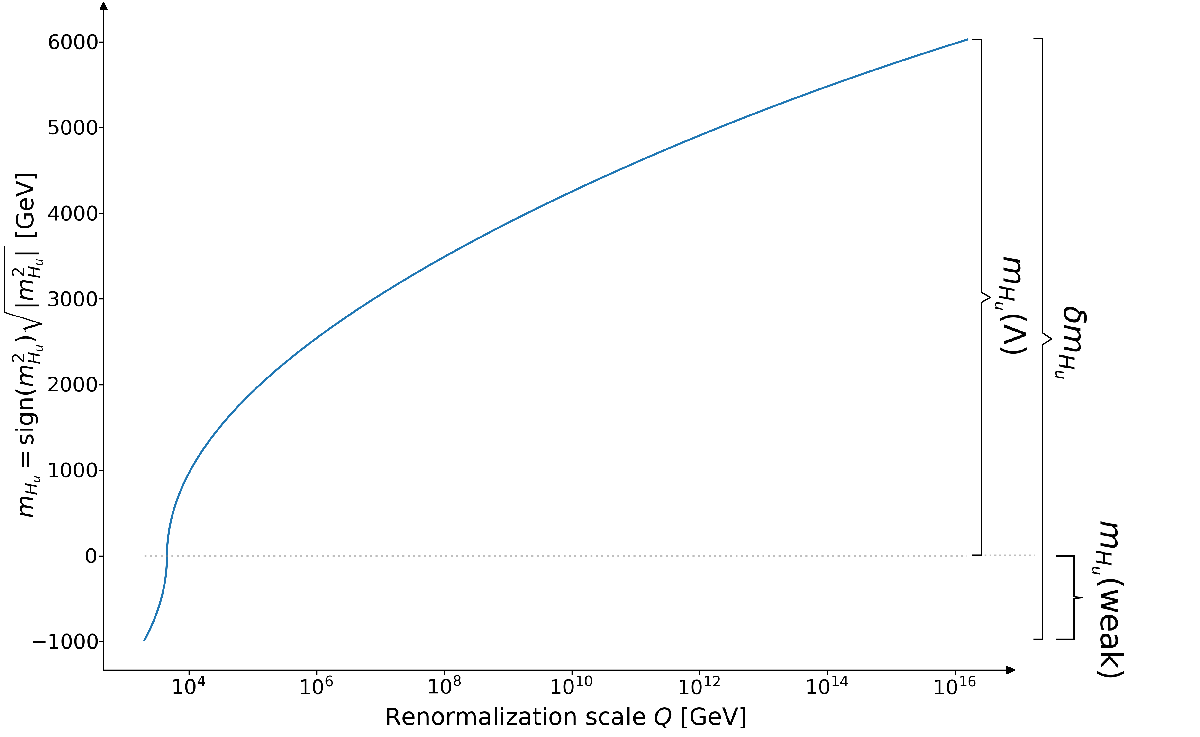}
\caption{Evolution of the the $m_{H_u}^2$ soft SUSY breaking up-Higgs mass
  from $Q=m_{GUT}$ to $Q=m_{weak}$ for the NUHM2 model with $m_0=4.5$ TeV,
  $m_{1/2}=1$ TeV, $A_0=-1.6 m_0$, $\tan\beta =10$ and $m_A=2$ TeV.
\label{fig:mHu}}
\end{center}
\end{figure}

From the previous discussion, it becomes clear that the various
naturalness measures are defined very differently and hence we expect them
to favor different regions of model parameter space.
A figurative view of how the different measures compare can be gleaned from
Fig. \ref{fig:mHu}, which plots the evolution of the soft Higgs mass-squared
parameter from $Q=m_{GUT}$ to $Q=m_{weak}$ in the NUHM2 model for
$m_0=4.5$ TeV, $m_{1/2}=1$ TeV, $A_0=-1.6 m_0$, $\tan\beta =10$ and $m_A=2$ TeV.
The right-side brackets correspond roughly to the different naturalness
measures. In NUHM2, the BG measure contains a sensitivity coefficient
$c_{m_{H_u}^2}\sim 1.27 |m_{H_u}(\Lambda )/m_Z|^2$ (or $c_{m_{H_u}}\sim 2.54 |m_{H_u}(\Lambda )/m_Z|^2$ if $m_{H_{u}}(\Lambda)$ is the fundamental parameter, instead of $m_{H_{u}}^{2}(\Lambda)$). 
Where this coefficient is the maximal contribution to $\DeltaBG$, then this ``distance'' is
the approximate measure. For $\DeltaHS\simeq \delta m_{H_u}^2/m_Z^2$,
the relevant measure is instead the bracketed distance $\delta m_{H_u}^2$, and so
$\DeltaHS$ is usually (but not always) larger than $\DeltaBG$, since in order for EW
symmetry to break, $m_{H_u}^2$ must be driven to negative values.
Notice then from the plot that the only way for $\delta m_{H_u}^2$ to
be small is also if $m_{H_u}^2(\Lambda )$ is small: this is why
$\DeltaHS$ favors only the low $m_0$ region when mSUGRA/CMSSM
universality with $m_{H_u}= m_0$ is required. In contrast, low values
of $\DeltaEW$ require low $m_{H_u}^2(weak)$, and so low $\DeltaEW$ can be
found for any value of $m_{H_u}^2(\Lambda )$ such that $m_{H_u}^2$ is barely
driven to negative values.
This is some times called criticality\cite{Giudice:2006sn,Baer:2016lpj}, or
barely broken electroweak symmetry\cite{Arkani-Hamed:2005zuc}.
This latter quality is favored
by the string landscape where as large as possible values of
$m_{H_u}^2(\Lambda )$ are statistically favored so long as $m_{Hu}^2$ is
just barely driven to negative values\cite{Baer:2017uvn}.

\begin{figure}[!htbp]
\begin{center}
  \includegraphics[height=0.365\textwidth]{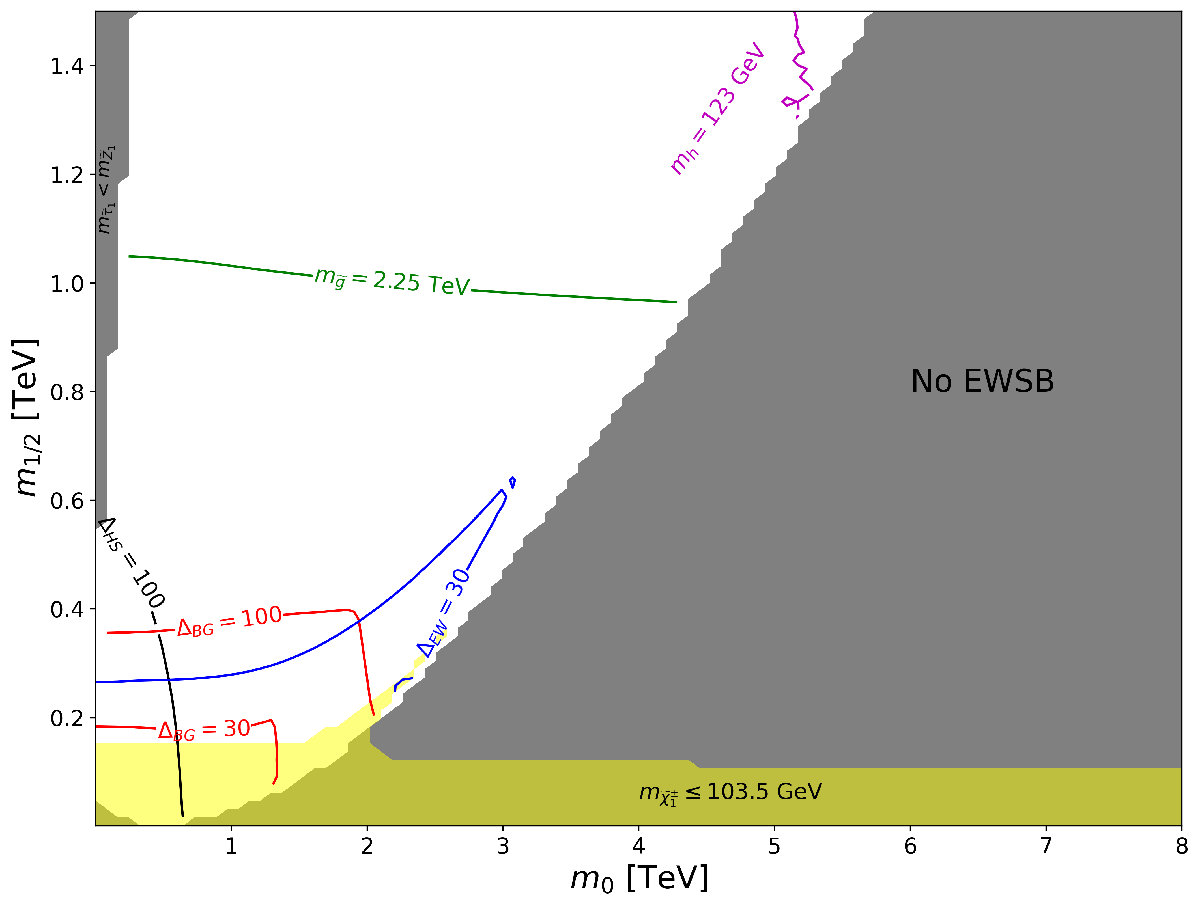}
  \includegraphics[height=0.365\textwidth]{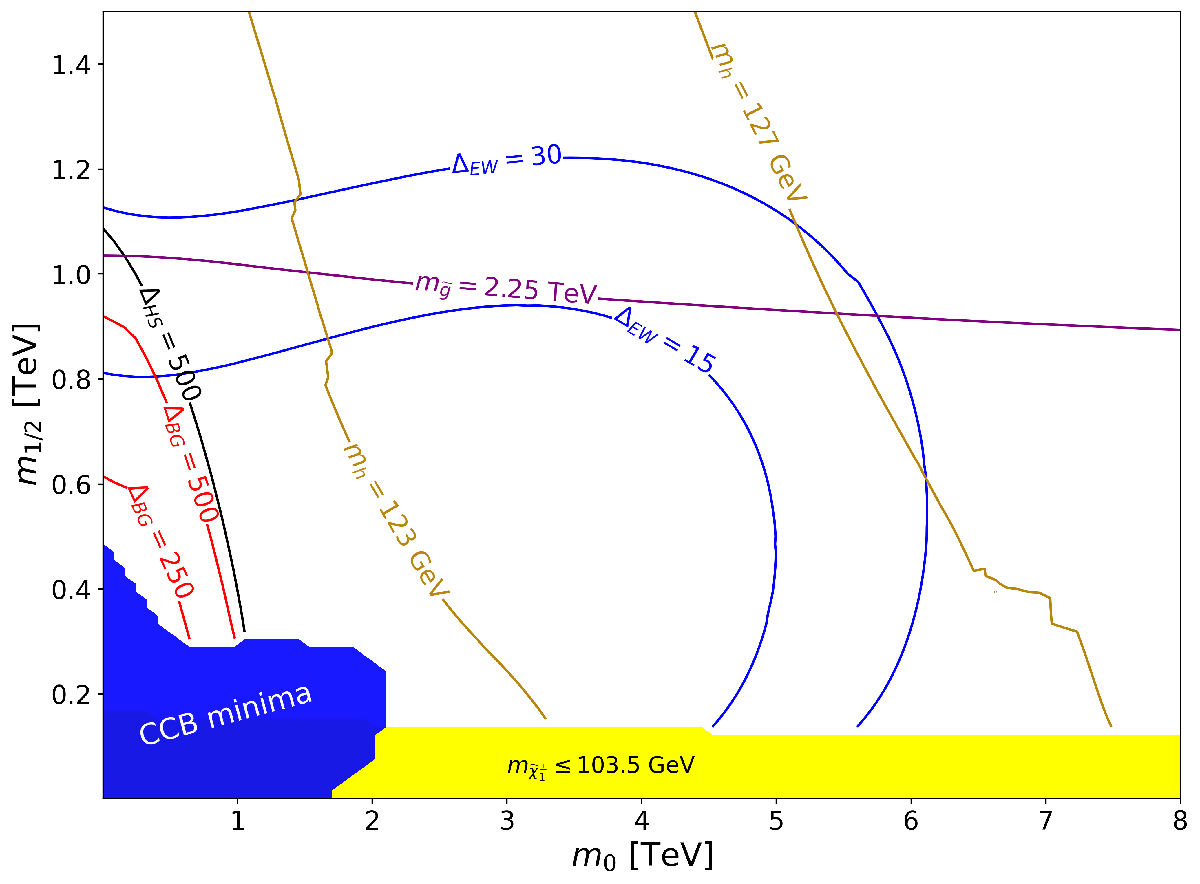}
\caption{Plot of various naturalness contours in {\it a}) the
  mSUGRA/CMSSM $m_0$ vs. $m_{1/2}$ plane for $A_0=0$ and $\tan\beta =10$ and
  $\mu >0$. In {\it b}), we plot naturalness contours in the
  $m_0$ vs. $m_{1/2}$ plane of the NUHM2 model with $A_0=-1.6 m_0$,
  $\tan\beta =10$, $\mu=200$ GeV and $m_A=2$ TeV.
  The spectrum is calculated using SoftSUSY
  and the naturalness measures with DEW4SLHA.
\label{fig:nat_planes}}
\end{center}
\end{figure}

\begin{figure}[!htbp]
\begin{center}
  \includegraphics[height=0.365\textwidth]{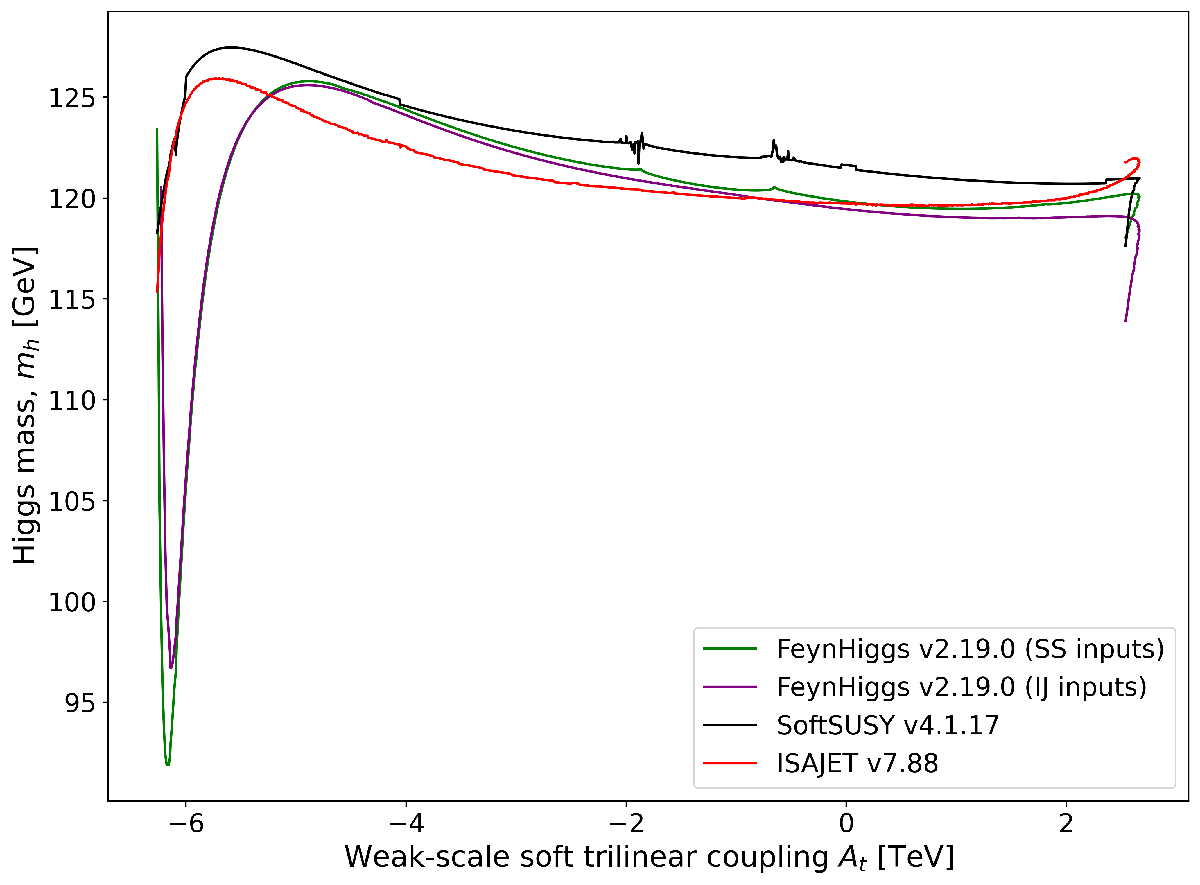}
  \includegraphics[height=0.365\textwidth]{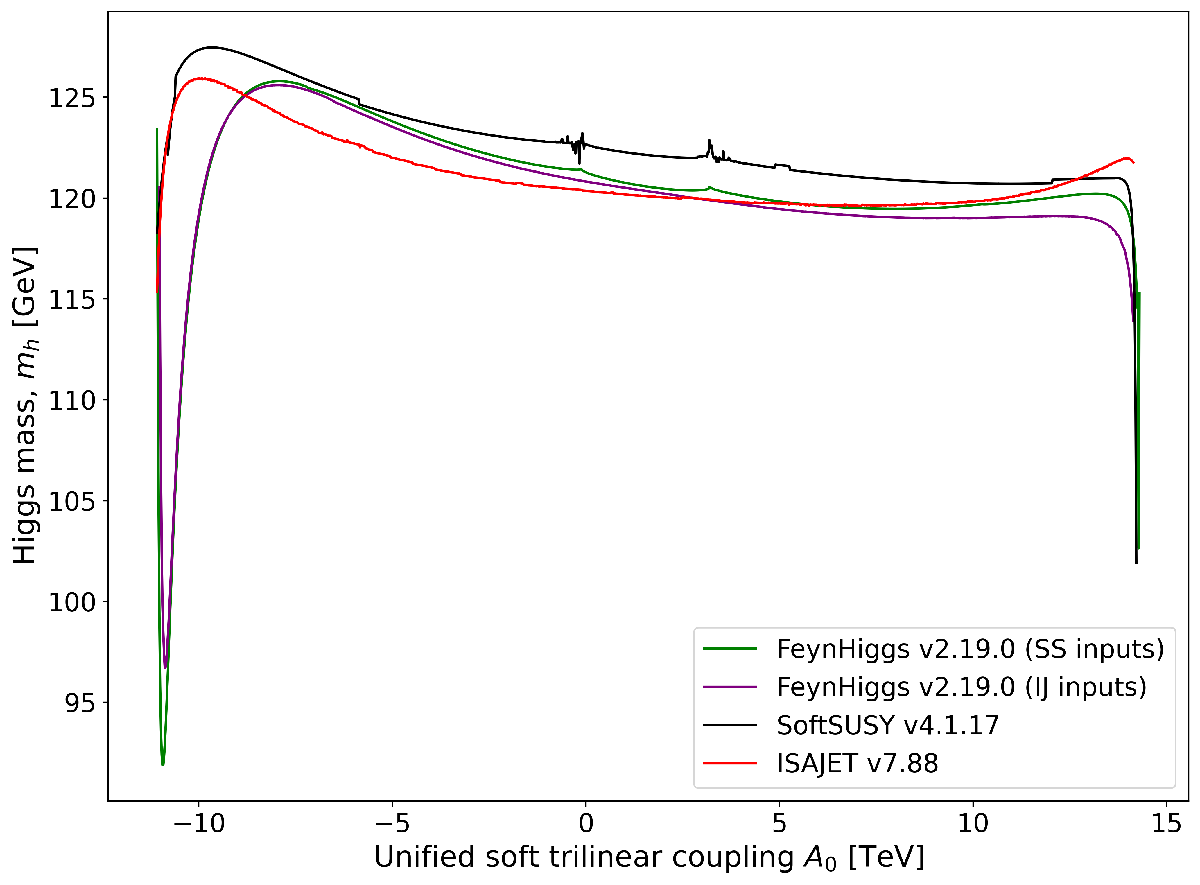}
  \caption{Computed value of $m_h$ vs. {\it a}) $A_t$ and
    {\it b}) $A_0$ in the mSUGRA/CMSSM model for
  $m_0=5$ TeV, $m_{1/2}=1.2$ TeV and $\tan\beta =10$, with $\mu >0$.
  We compare the results from SoftSUSY\cite{Allanach:2001kg},
  FeynHiggs\cite{Bahl:2018qog} using SoftSUSY or Isajet inputs,
  and Isasugra 7.88\cite{Paige:2003mg}.
\label{fig:mh}}
\end{center}
\end{figure}

In Fig. \ref{fig:nat_planes}{\it a}), we plot naturalness favored and unfavored
regions of parameter space in the mSUGRA/CMSSM model $m_0$ vs. $m_{1/2}$
plane where $\tan\beta =10$, $\mu >0$ and $A_0=0$
(spectra generated using SoftSUSY\cite{Allanach:2001kg}).
The latter choice for $A_0$ is traditional in that it displays the FP
region, which otherwise disappears for large $A_0$.
However, it should be remarked here that $A_0=0$ is probably the {\it least}
motivated value for $A_0$ in that in generic SUGRA models  all soft terms are
expected to occur of order $m_{3/2}$ and $m_{3/2}\sim m_0$.
On the phenomenological side, small $A_t$ leads to a near minimum in the
Higgs mass $m_h$ (as shown in Fig. \ref{fig:mh})
whilst $A_t\sim \sqrt{6}m_0$ leads maximal $m_h$ values\cite{Slavich:2020zjv}.
Using high scale parameters $A_0$, the range of $A_t$ doesn't extend to the
$m_h$ maximal value before CCB minima are encountered in the scalar potential
(which forms the endpoints of the plot).
Nowadays, the mSUGRA/CMSSM FP region seems excluded by
\begin{enumerate}
 \item too low a value of $m_h$\cite{Baer:2012mv} and
 \item the LSP DM candidate is of the well-tempered\cite{Arkani-Hamed:2006wnf,Baer:2006te} type which is now
 excluded\cite{Baer:2016ucr,Badziak:2017the,Profumo:2017ntc} by WIMP spin-independent direct
 detection experiments
 such as Xenon\cite{XENON:2018voc} and LZ\cite{LZ:2022ufs}.
\end{enumerate}

%


From Fig. \ref{fig:nat_planes}{\it a}), we see the $\DeltaBG$ contours of
30 and 100 roughly track lines of constant $\mu$ for lower $m_0$
values and lines of $m_0$ for high values\cite{Feng:1999zg}. Meanwhile,
the contour $\DeltaEW =30$ (denoted as blue), follows the low $\mu$
values to much larger $m_0$ whereupon it cuts off due to increasing
top-squark contributions via the $\Sigma_u^u(\tst_{1,2})$.
The conflict of these measures with $\DeltaHS$ is apparent since
$\DeltaHS$ favors light third generation squarks which occur only at
low $m_0$ and low $m_{1/2}$.
All three measures favor the lower corner of
$m_0$ vs. $m_{1/2}$ parameter space, and, when compared to LHC gluino mass limits
($m_{\tg}\agt 2.25$ TeV as shown by the green contour), might lead one
to conclude this model is excluded based on comparisons of
naturalness with LHC constraints. 

A very different picture emerges when one proceeds to the NUHM2 model
as shown in Fig. \ref{fig:nat_planes}{\it b}). In this case,
non-universality of the Higgs soft masses (as expected in gravity-mediation)
allows for low $\mu =200$ GeV throughout the parameter plane.
Also, the large negative $A_0=-1.6 m_0$ term allows for $m_h\sim 125$ GeV
throughout much of the plane (as shown between the yellow contours of
mass $m_h=123$ and $127$ GeV). The large $-A_0$ term helps crunch the
$\DeltaHS$ and $\DeltaBG$ contours into the lower-left region
which actually yields charge-and-color breaking scalar potential minima,
which must be excluded. Meanwhile, the $\DeltaEW$ contours balloon out
to very large $m_0$ and $m_{1/2}$ values, with $\DeltaEW\sim 30$ extending
well beyond present LHC limits on $m_{\tg}$. While the lowest $\DeltaEW$
values are still found in the lower-left corner of $m_0$ vs. $m_{1/2}$ parameter
space, we note that stringy naturalness, which favors a power-law draw to
large soft terms, actually favors the region beyond the LHC $m_{\tg}$
limit\cite{Baer:2019cae}.

\section{Ratios of measures for CMSSM and NUHM2 models}
\label{sec:ratios}

In this Section, we compute the ratios of various naturalness measures
in the CMSSM/mSUGRA and NUHM2 models. Spectra are calculated with
SoftSUSY\cite{Allanach:2001kg} while naturalness measures are computed
with DEW4SLHA\cite{Baer:2021tta}.
The goal here is to quantify potential overestimates of finetuning in
different SUSY models.

\subsection{Ratios of measures for the mSUGRA/CMSSM model}

\subsubsection{Results in $m_0$ vs. $m_{1/2}$ plane}

\begin{figure}[!htbp]
\begin{center}
  \includegraphics[height=0.3\textheight]{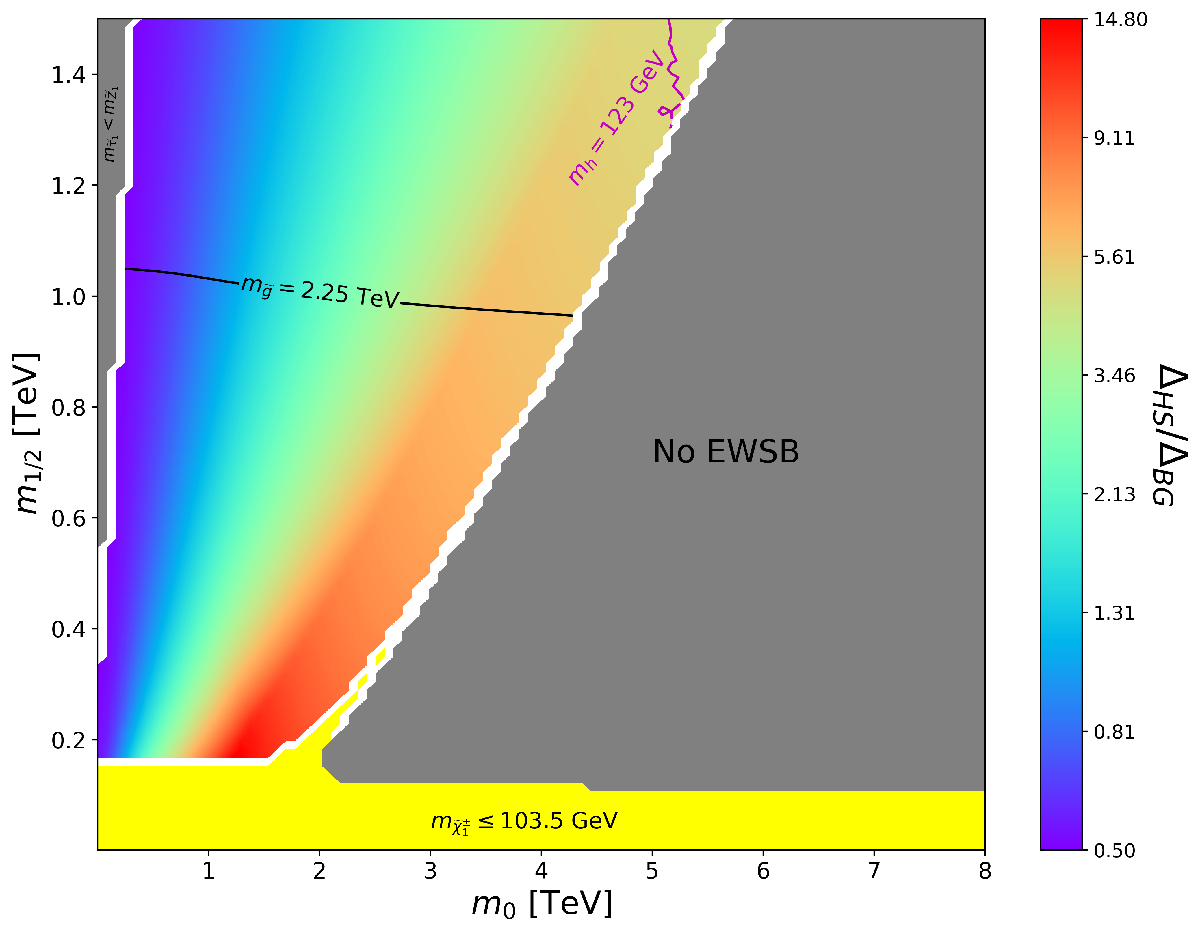}\\
  \includegraphics[height=0.3\textheight]{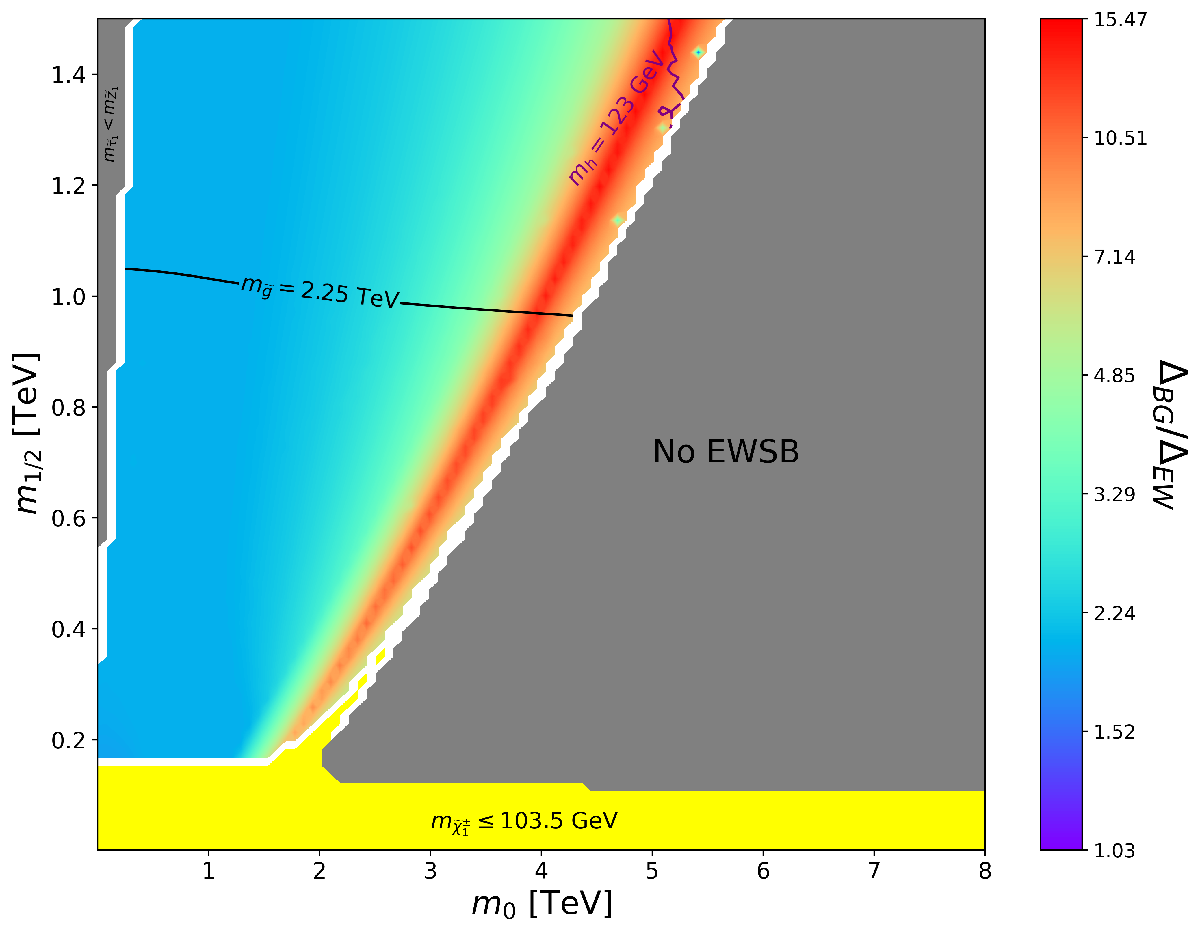}\\
  \includegraphics[height=0.3\textheight]{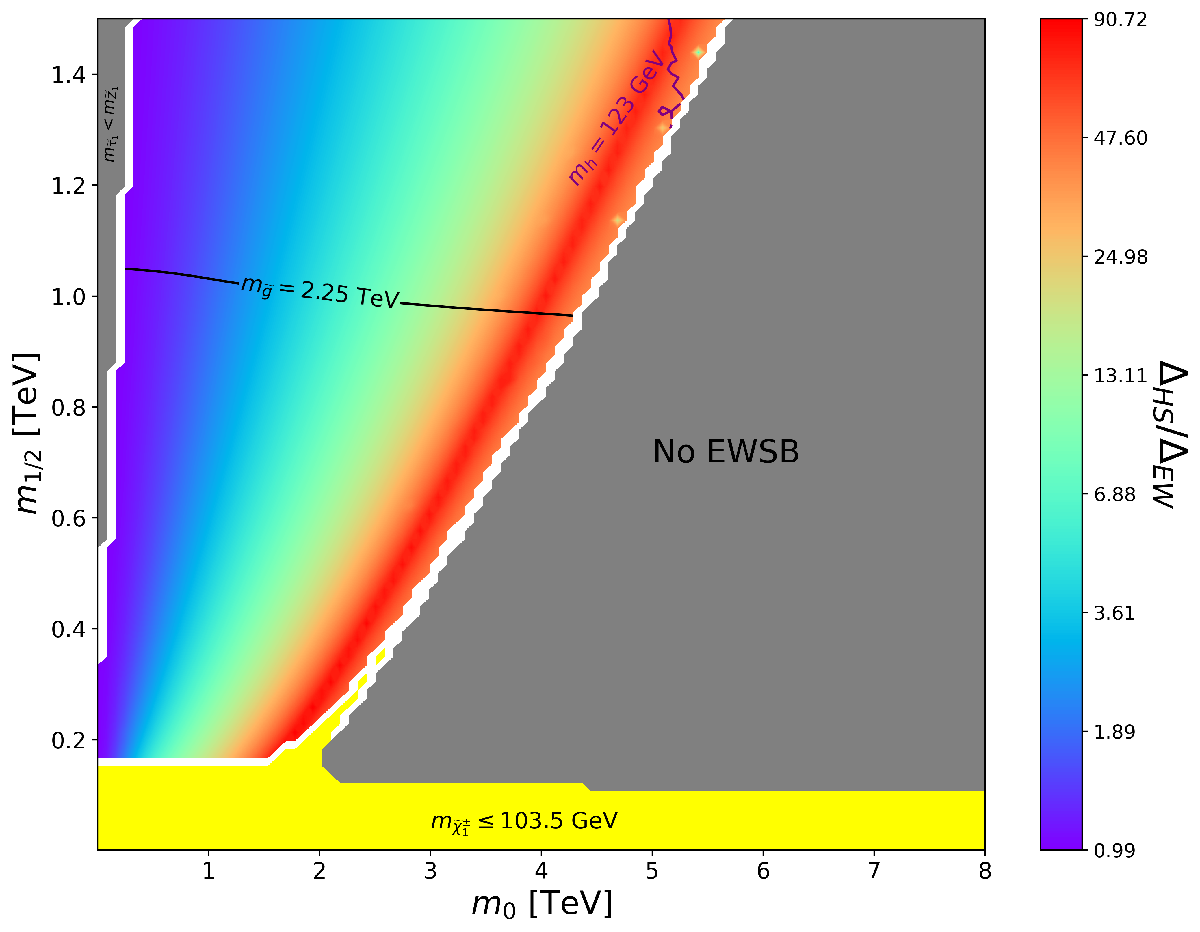}
  \caption{Plot of ratios of naturalness measures in the
    $m_0$ vs. $m_{1/2}$ plane for the CMSSM model for
    for $A_0=0$, $\tan\beta =10$ and $\mu >0$.
    In {\it a}), we plot $\DeltaHS/\DeltaBG$ while in {\it b}) we
    plot $\DeltaBG/\DeltaEW$ and in {\it c}) we plot
    $\DeltaHS/\DeltaEW$.
      The spectrum is calculated using SoftSUSY
  and the naturalness measures with DEW4SLHA.
\label{fig:ratios_cmssm}}
\end{center}
\end{figure}

In Fig. \ref{fig:ratios_cmssm}, we compute the various ratios of
naturalness in the mSUGRA/CMSSM model and display results in the
$m_0$ vs. $m_{1/2}$ plane for $A_0=0$ and $\tan\beta =10$ (such as to include
the HB/FP region. Our first results are shown in frame {\it a}) where
we plot $\DeltaHS/\DeltaBG$. The right-side gray region has no
EWSB while the left-side gray region has a stau LSP.
The lower yellow region has lightest chargino below LEP2 limits of
$m_{\tchi_1}^+<103.5$ GeV. The color-coded ratios are denoted by the scale on
the right-hand-side of the plot and range from 0.5 (purple)
to $\sim 15$ (red).

Starting from the LHS of Fig. \ref{fig:ratios_cmssm}{\it a}), we see that for
low $m_0$ then $\DeltaHS\sim \DeltaBG /2$.
This is because $\DeltaBG$ is
typically dominated by the gluino or $M_3$ contribution which is then canceled
by $\mu^2$ to maintain $m_Z$ at 91.2 GeV. But $\DeltaHS$ is dominated instead
by $\delta m_{H_u}^2$ which is low at low $m_0$.
As $m_0$ increases, then ultimately the two measures are comparable in the
color transition region while for higher $m_0$ values, where the FP-type
cancellation kicks in, then $\DeltaHS$ becomes much larger than
$\DeltaBG$ and reaches nearly a factor of $\sim 5-15$ at the edge of the
``no EWSB'' region. This plot illustrates that the $\DeltaBG$ and $\DeltaHS$
measures are incompatible.

In Fig. \ref{fig:ratios_cmssm}{\it b}), we plot the ratio
$\DeltaBG/\DeltaEW$. In this case, on the left side at low $m_0$
the two measures are comparable, but become as large as a factor $\sim 15$
on the RHS near the edge of the {\it no EWSB} excluded region. In this
region, $\DeltaBG$ is dominated by the gluino/$M_3$ contribution, but in
$\DeltaEW$ this is two-loop suppressed and so $\DeltaBG$ is much larger.

In Fig. \ref{fig:ratios_cmssm}{\it c}), where the ratio $\DeltaHS /\DeltaEW$
is plotted, we see the measures are only comparable on the extreme LHS
but then differ by up to a factor of $\sim 80- 100$ on the RHS in
orange/red region. In this region, top-squarks are in the multi-TeV region
so $\DeltaHS$ is very large while $\DeltaEW$ allows for multi-TeV top
squarks owing to the loop factor in Eq. \ref{eq:Sigmauu}.

\subsubsection{Results from scan over CMSSM parameters}

\begin{figure}[!htb]
\begin{center}
  \includegraphics[height=0.25\textheight]{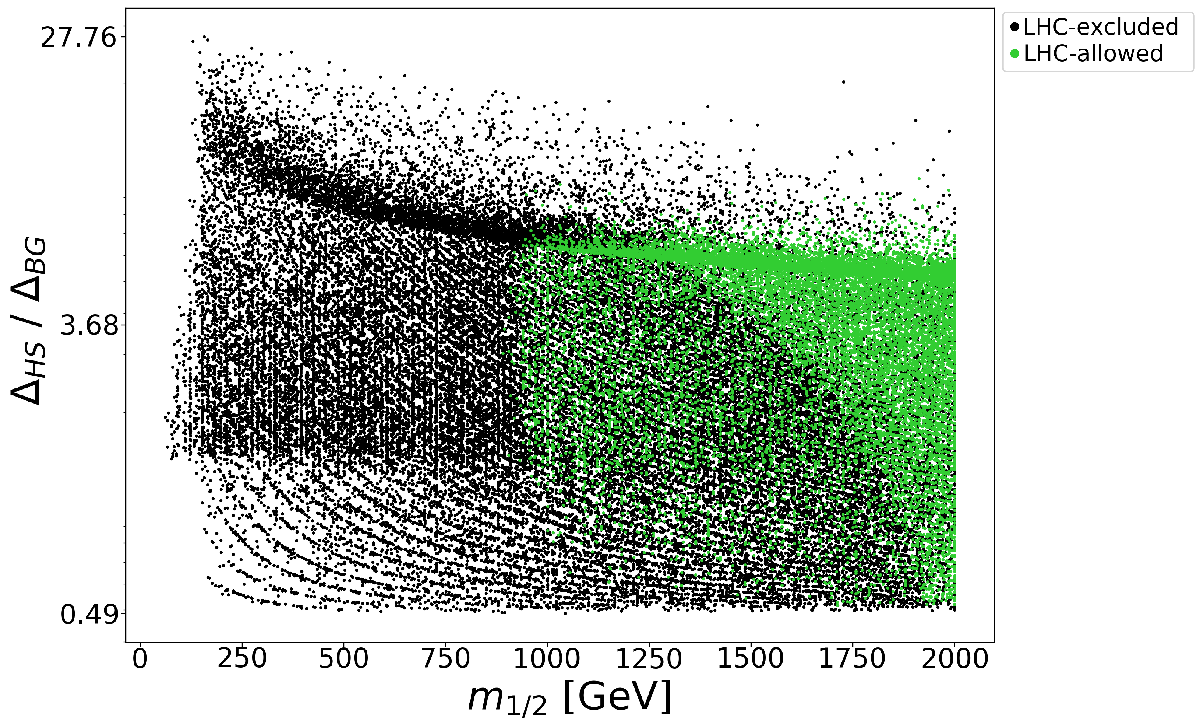}\\
  \includegraphics[height=0.25\textheight]{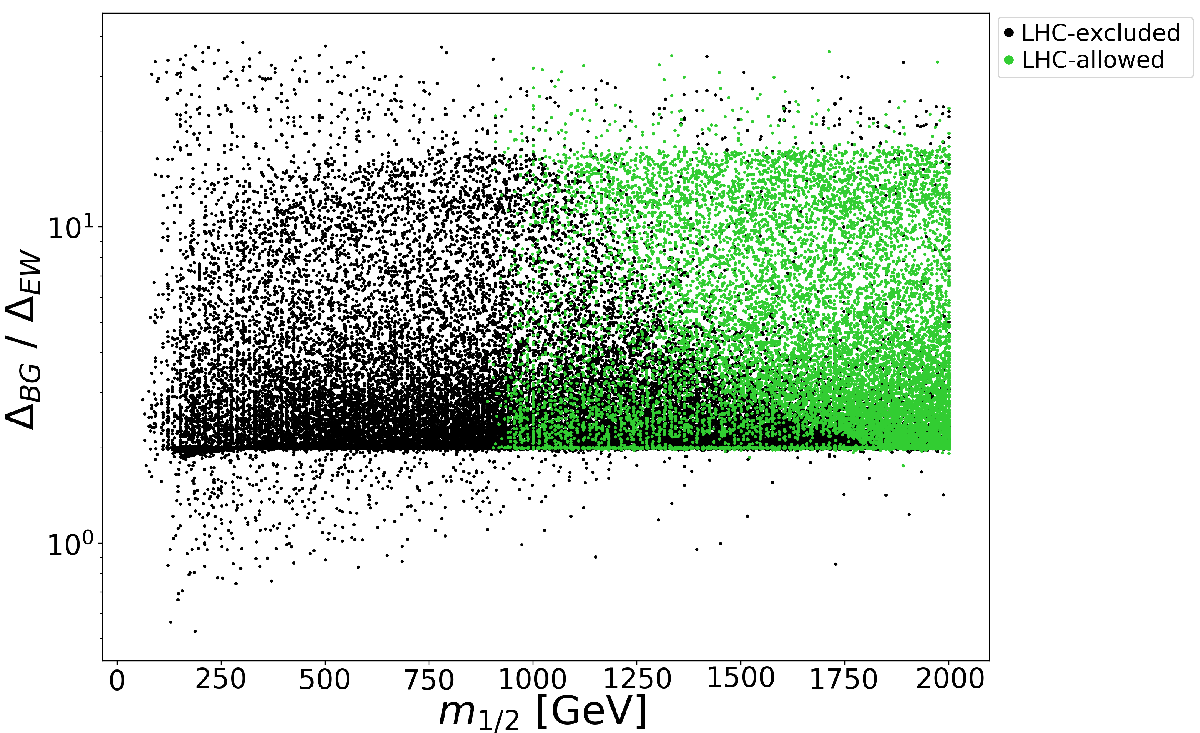}\\
  \includegraphics[height=0.25\textheight]{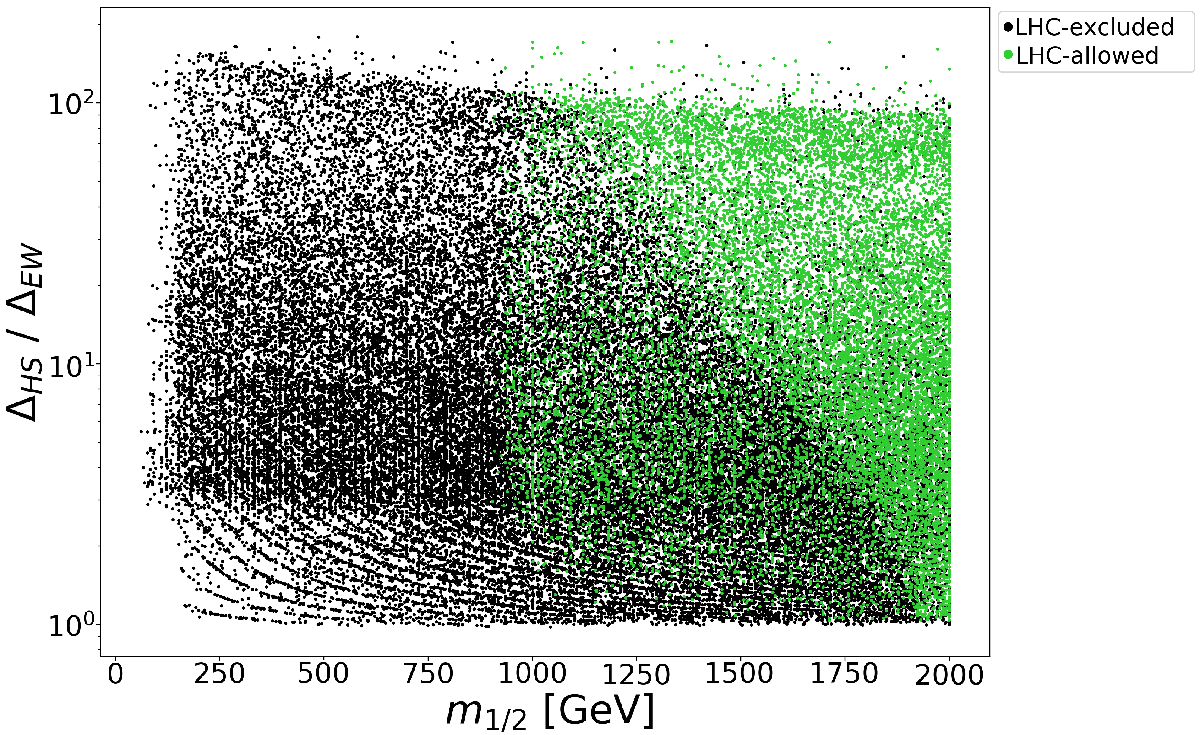}
  \caption{Plot of ratios of naturalness measures vs. $m_{1/2}$ from a scan over CMSSM model parameters. In {\it a}), we plot $\DeltaHS$ vs. $\DeltaBG$ for a general plus a focused scan (at $A_0=0$ to pick up the FP SUSY region)
    while in {\it b}) we plot $\DeltaBG/\DeltaEW$ and in {\it c}) we plot
    $\DeltaHS/\DeltaEW$. The spectrum is calculated using SoftSUSY and the naturalness measures with DEW4SLHA.
\label{fig:ratios_cmssm_scan}}
\end{center}
\end{figure}

Next, we attempt to pick out the maximal ratio of naturalness measures in an attempt to quantify their numerical differences.
First, we scan over CMSSM parameter space
\bi
\item $m_0:\ 0.1-15$ TeV,
  \item $m_{1/2}:\ 0.1-2$ TeV,
  \item $A_0:\ -2.5m_0$ to $+2.5m_0$,
    \item $\tan\beta :\ 3-60$
    \ei
    with $\mu >0$.
    In Fig. \ref{fig:ratios_cmssm_scan}{\it a}), we show the ratio $\DeltaHS/\DeltaBG$
    from the above scan plus a focused scan over the same parameter
    range but with $A_0=0$ so as to pick up the FP region where the ratio is expected to be largest. The green points are LHC-allowed from LHC Run 2. 
In this case, the ratio $\DeltaHS /\DeltaBG$ can be as high as 28 overall, but only 
as high as $\sim 10$ in the LHC-allowed region. These values are somewhat higher than the maximal
ratios obtained from the plane plots.

Similarly, we show in Fig. \ref{fig:ratios_cmssm_scan}{\it b}) the ratio $\DeltaBG/\DeltaEW$ 
which ranges up to 50 (20) in the overall (LHC-allowed) case. In frame {it c}), 
the ratio $\DeltaHS/\DeltaEW$ ranges up to $\sim 200$ for both LHC-allowed and forbidden cases.

\subsection{Ratios of measures for the NUHM2 model}

\subsubsection{Results in $m_0$ vs. $m_{1/2}$ plane}

Next, we compute ratios of naturalness measures for the NUHM2 model.
In this case, we expect much bigger differences between naturalness
measures $\DeltaBG$ and $\DeltaEW$ since for NUHMi models, $m_{H_u}^2$
is now a free parameter, and no longer available to cancel against
other oppositely signed sfermion contributions in Eq. \ref{eq:mZsparam}.
In particular, the FP cancellation between Higgs and third generation
sfermion terms in Eq. \ref{eq:mZsparam} is destroyed when one assumes that
$m_{H_u}^2$ is a free parameter. Furthermore, by adopting natural values
of $\mu\sim m_{weak}$, then this contribution to $\DeltaEW$ is suppressed
and the dominant contribution instead frequently comes from the
(loop-suppressed) $\Sigma_u^u(\tst_{1,2})$ terms.

\begin{figure}[!h]
\begin{center}
  \includegraphics[height=0.25\textheight]{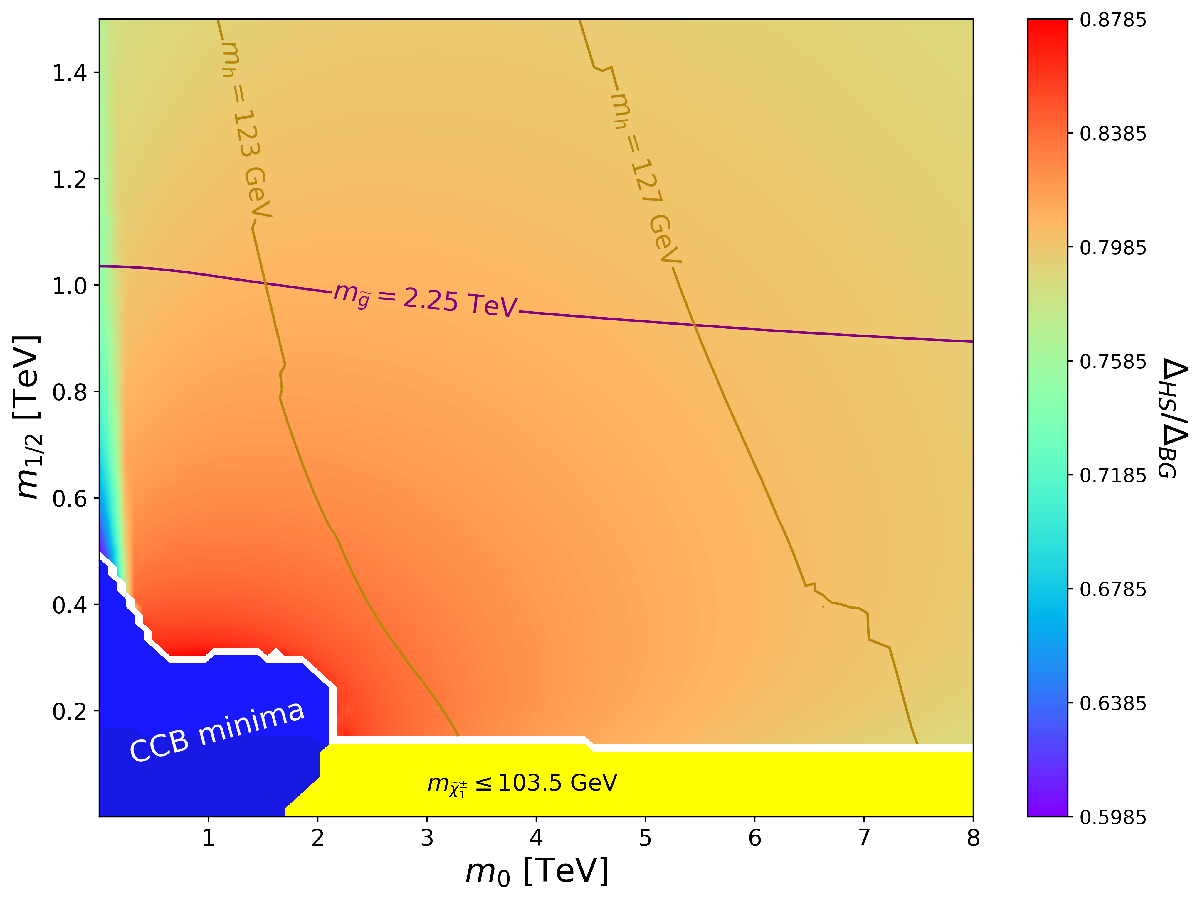}\\
  \includegraphics[height=0.25\textheight]{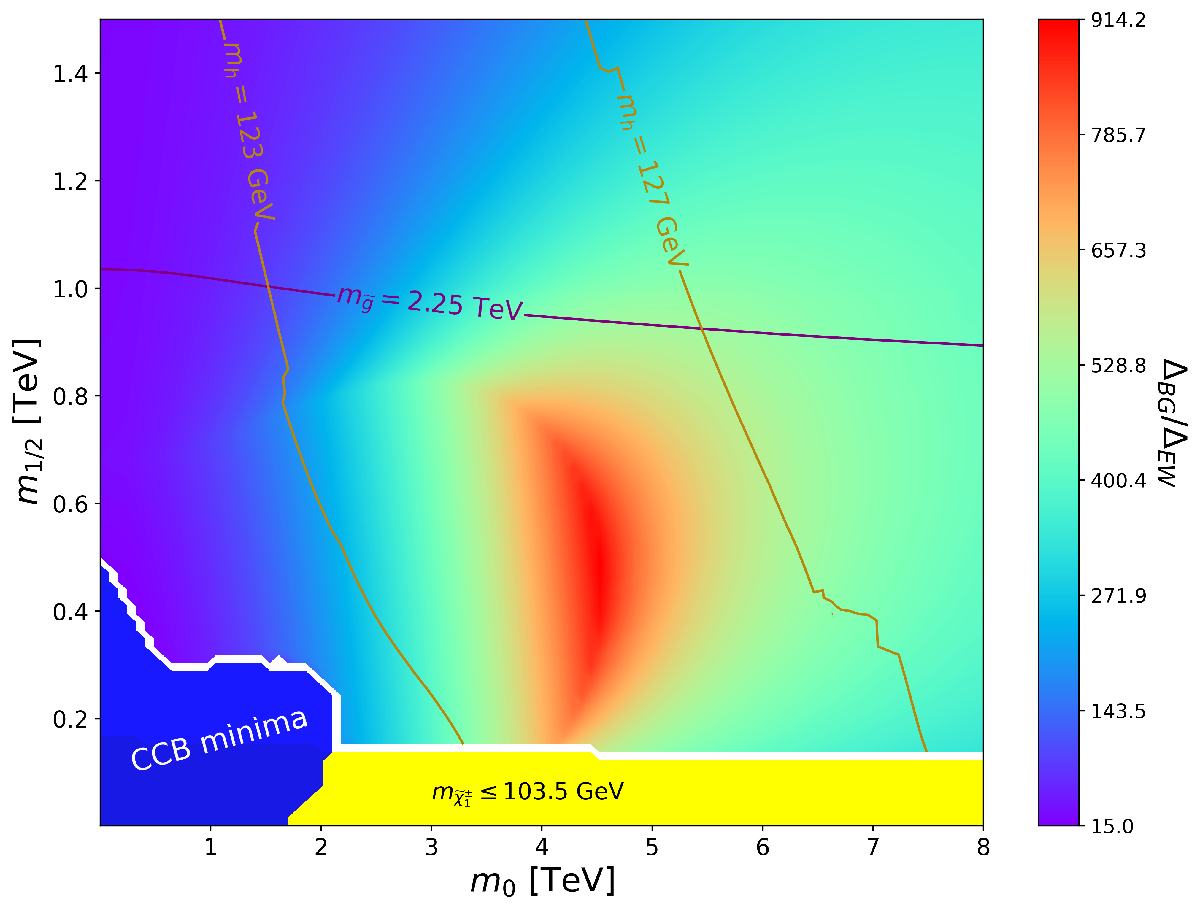}\\
  \includegraphics[height=0.25\textheight]{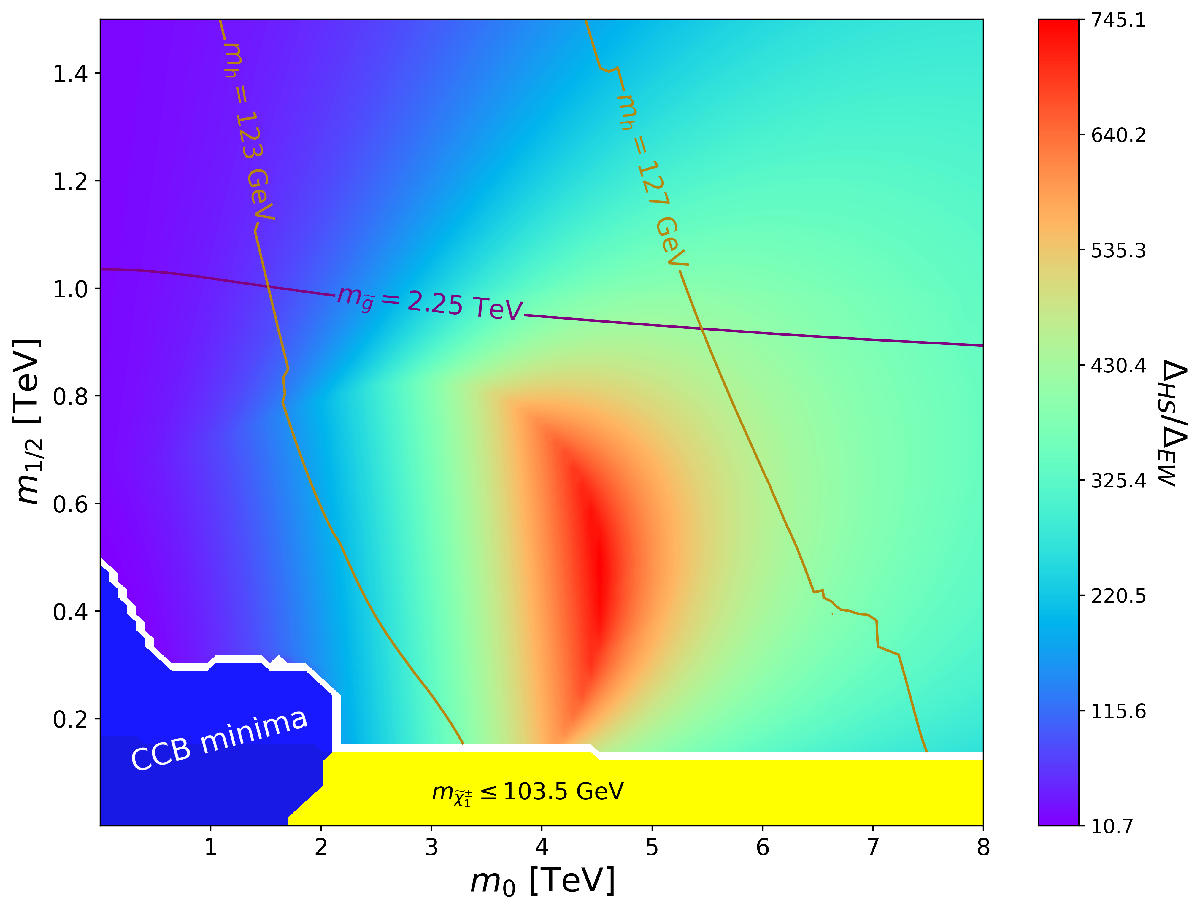}
  \caption{Plot of ratios of naturalness measures in the $m_0$ vs. $m_{1/2}$ plane for the NUHM2 model for $A_0=-1.6m_0$, $\tan\beta =10$ and $\mu =200$ GeV and $m_A=2$ TeV. In {\it a}), we plot $\Delta_{BG}/\DeltaHS$ while in {\it b}) we plot $\Delta_{BG}/\DeltaEW$ and in {\it c}) we plot $\DeltaHS/\DeltaEW$. The spectrum is calculated using SoftSUSY and the naturalness measures with DEW4SLHA.
\label{fig:ratios_nuhm2}}
\end{center}
\end{figure}

In Fig. \ref{fig:ratios_nuhm2}{\it a}), we show color-coded ratios
$\DeltaHS/\DeltaBG$ in the $m_0$ vs. $m_{1/2}$ plane of the NUHM2 model
with $A_0=-1.6m_0$, $\tan\beta =10$, $\mu =200$ GeV and $m_A=2$ TeV.
The lower-left blue shaded region has CCB minima in the scalar potential
owing in part to the large $A_0$ term.
The spectra is generated using SoftSUSY.
The gold contours denote Higgs mass $m_h=123$ GeV (left) and
$m_h=127$ GeV (right), while the purple contour near $m_{1/2}\sim 1$ TeV
denotes the LHC $m_{\tg}=2.25$ TeV limit. On the green-shaded
extreme left side of the plot, $\DeltaBG$ becomes larger than $\DeltaHS$ by a factor $\sim2$.
For the bulk of the plot range (red- and orange- shaded regions), then
$\DeltaHS\sim (0.75-0.85)\DeltaBG$.

In Fig. \ref{fig:ratios_nuhm2}{\it b}), we show the ratio $\DeltaBG/\DeltaEW$
in the same plane as frame {\it a}). In the red-shaded region below the
LHC gluino mass limit, we see that $\DeltaBG /\DeltaEW \sim 900$,
a gross disparity between measures.
Here, we have utilized $m_{H_{u}}$ as the fundamental parameter in place of $m_{H_{u}}^{2}$
for numerical stability purposes--the difference is only a factor of two in the derivative.
In this region, $\DeltaBG$ is dominated by the
$|-1.27\cdot2 m_{H_u}^2|/(m_Z^2/2)$ term in Eq. \ref{eq:mZsparam} which can be
$\sim 11000$ whilst the $\DeltaEW$ measure is dominated by
$\Sigma_u^u (\tst_{1,2})$ which yields $\DeltaEW\sim 10-20$ (very natural).
Throughout the NUHM2 plane, one can be brought to very different conclusions
regarding the naturalness of the NUHM2 model parameter space depending on
which measure one adopts! For the bulk of parameter space above the LHC
gluino bound, then one finds $\DeltaBG\sim (50-400) \DeltaEW$.

In Fig. \ref{fig:ratios_nuhm2}{\it c}), we show the ratio $\DeltaHS /\DeltaEW$.
In this case, the red-shaded region shows that $\DeltaHS\sim 700\DeltaEW$.
Again, one is led to very different conclusions on the naturalness of the
model depending on which measure one chooses. In this case,
in the LHC-allowed region then $\DeltaHS\sim (50-400)\DeltaEW$.

\subsubsection{Results from scan over NUHM2 parameters}

\begin{figure}[!htbp]
\begin{center}
  \includegraphics[height=0.25\textheight]{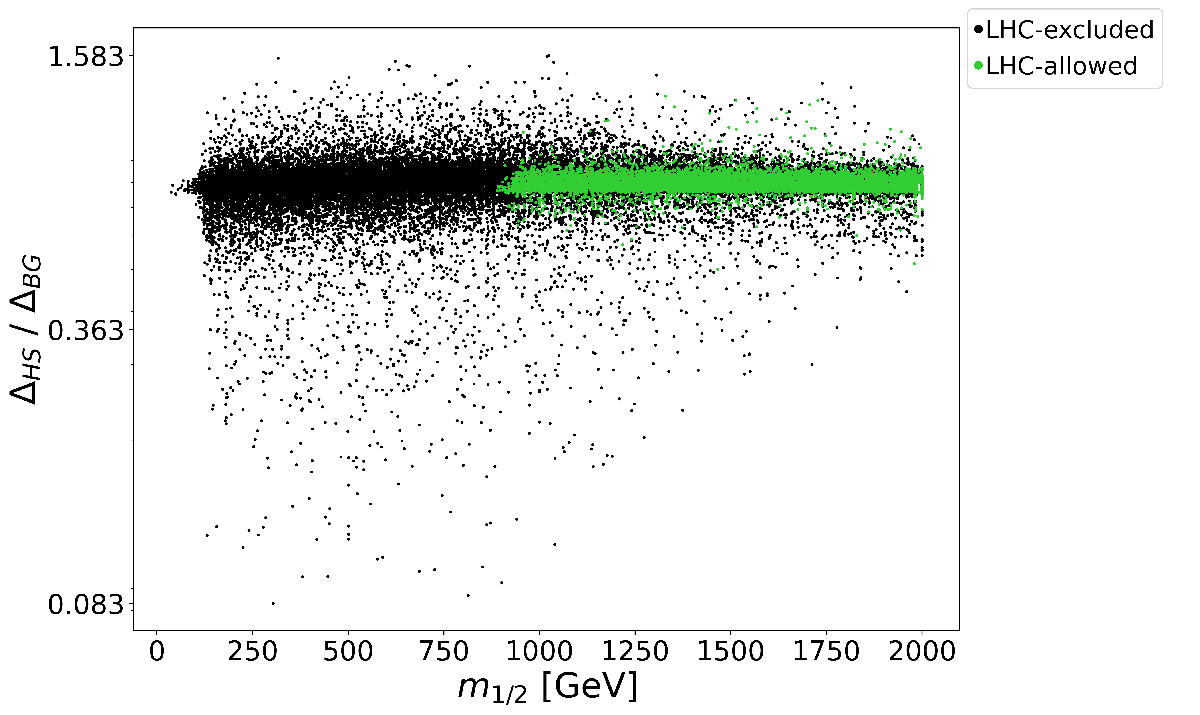}\\
  \includegraphics[height=0.25\textheight]{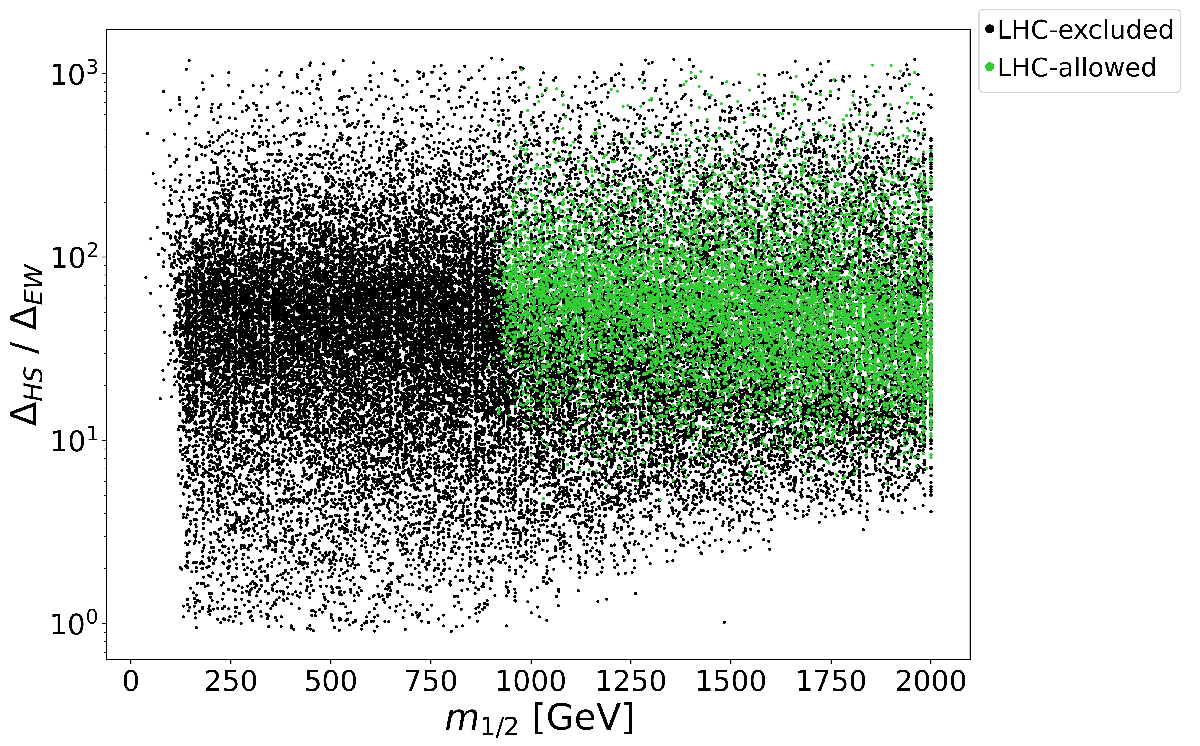}\\
  \includegraphics[height=0.25\textheight]{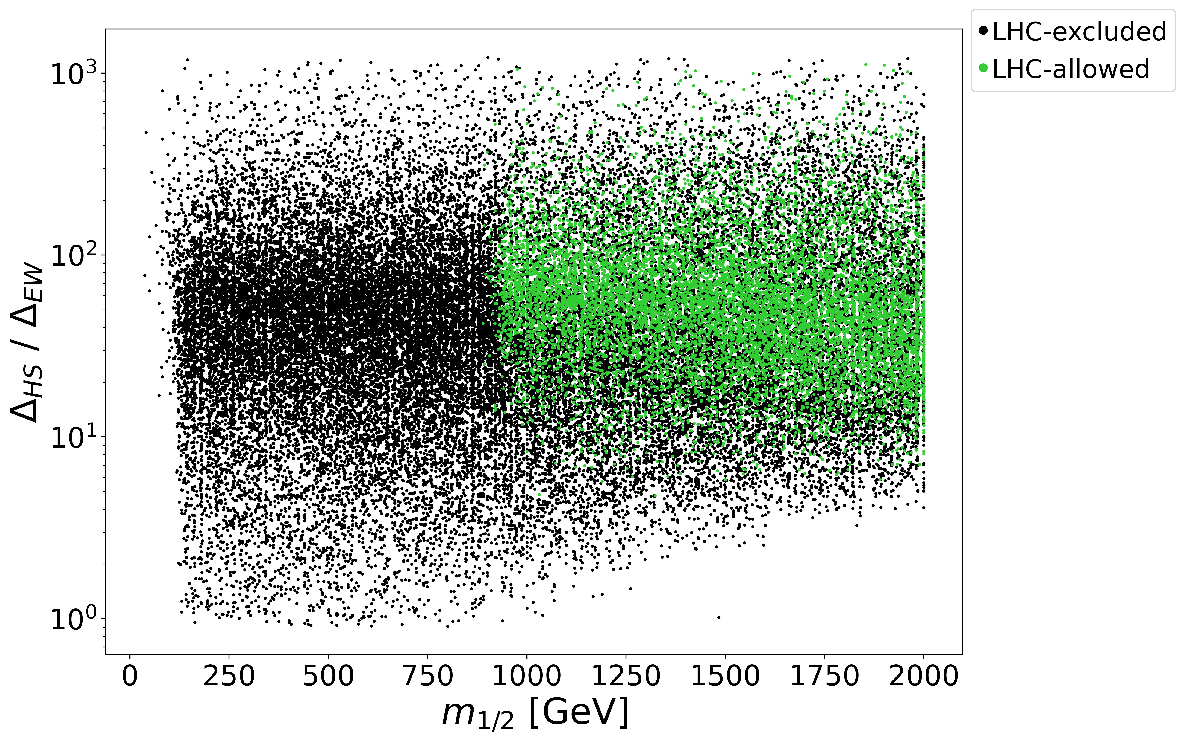}
  \caption{Plot of ratios of naturalness measures vs. $m_{1/2}$ 
    from a scan over NUHM2 model parameters.
        In {\it a}), we plot $\DeltaHS/\DeltaBG$ while in {\it b}) we
    plot $\DeltaBG/\DeltaEW$ and in {\it c}) we plot
    $\DeltaHS/\DeltaEW$.
      The spectrum is calculated using SoftSUSY
  and the naturalness measures with DEW4SLHA.
\label{fig:ratios_nuhm2_scan}}
\end{center}
\end{figure}

Here, we scan over NUHM2 parameter space:
\bi
\item $m_0:\ 0.1-15$ TeV,
  \item $m_{1/2}:\ 0.1-2$ TeV,
  \item $A_0:\ -2.5m_0$ to $+2.5m_0$,
    \item $\tan\beta :\ 3-60$
    \item $\mu :\ 0.1-1$ TeV and
      \item $m_A:\ 0.3-8$ TeV.
      \ei
      From Fig. \ref{fig:ratios_nuhm2_scan}{\it a}), we see the ratio
      $\DeltaHS/\DeltaBG$ ranges from $\sim0.1-1.5$ over the parameter space scanned: the two measures are rather close much of the time in this case, but sometimes $\DeltaBG$ can become a factor of $\sim10$ larger than $\DeltaHS$.
      In Fig. \ref{fig:ratios_nuhm2_scan}{\it b}), instead we plot the
      $\DeltaBG/\DeltaEW$ ratio. While the bulk of parameter points have ratio
      between $\sim1-250$, some few points can range up to $500-1000$. Similar results are obtained for Fig. \ref{fig:ratios_nuhm2_scan}{\it c}) where we plot
        $\DeltaHS/\DeltaEW$ and find ratios ranging again up to over 1000.

\section{Natural generalized anomaly-mediation (nAMSB)}
\label{sec:AMSB}

Anomaly-mediated SUSY breaking (AMSB) models\cite{Randall:1998uk,Giudice:1998xp} are good examples of models where
the soft terms are all correlated and determined by a single parameter, the
gravitino mass $m_{3/2}$.
AMSB models assume a sequestering between the hidden and visible sector
fields such that gravity-mediated soft terms are suppressed;
in such a case, the loop-induced AMSB soft terms,
which depend on the beta functions and anomalous dimensions of the
low energy theory (assumed to be the MSSM) are dominant, and independent of
higher energy physics. At first glance, AMSB SUSY models would seem ruled
out since the AMSB soft terms give rise to tachyonic sleptons.
In the original Randall-Sundrum paper, it is conjectured that additional
bulk soft terms may also be present which can solve the tachyonic slepton
mass problem.

In the so-called {\it minimal} AMSB (mAMSB) model, a universal bulk sfermion
mass $m_0$ is also assumed so that the parameter space of mAMSB is given by
\be
m_0,\ m_{3/2},\ \tan\beta,\ sign(\mu )\ \ \ (mAMSB) .
\ee
Here, the magnitude of a bilinear soft $B$ term is traded for the parameter
$\tan\beta$ and $|\mu |$ is determined from the EW minimization condition.
Famously, the wino-like neutralino turns out to be the LSP.
At present, in light of LHC sparticle and Higgs mass constraints
and direct/indirect wino dark matter constraints,
mAMSB seems ruled out\cite{Cohen:2013ama,Fan:2013faa,Baer:2016ucr}.
The small AMSB $A_t$ terms lead to $m_h\ll 125$ GeV unless sparticle masses $\sim 10-100$ TeV
(highly unnatural) sparticle are assumed\cite{Arbey:2011ab,Baer:2012uya}
and the DM constraints are evaded.

However, in Ref. \cite{Baer:2018hwa} some minor fixes were proposed which
lead to {\it generalized} AMSB models (gAMSB) which allow for naturalness
(nAMSB with $\DeltaEW\alt 30$) and with $m_h\sim 125$ GeV.
The fixes are:
\begin{enumerate}
\item non-universal scalar bulk masses $m_{H_u}\ne m_{H_d}\ne m_0$ and
\item bulk-induced $A_0$ terms.
\item a further option is independent bulk terms for each sfermion generation $m_0(i)$ with $i=1-3$.
\end{enumerate}
As in NUHMi models, the non-universal bulk Higgs soft terms can be traded for
$\mu$ and $m_A$ via scalar potential minimization conditions and the bulk
$A$ terms can be chosen to dial up $m_h\sim 125$ GeV.
Thus, the generalized AMSB parameter space is given by
\be
m_0(i),\ m_{3/2},\ A_0,\ \tan\beta ,\ \mu\ \ {\rm and}\ \ m_A\ \ \ (gAMSB).
\ee
For natural values of $\mu\sim 100-350$ GeV, then in nAMSB the LSP
is instead (usually) higgsino-like, although the wino is still the
lightest of the gauginos. The gAMSB model is what one may expect
in models of {\it charged} SUSY breaking, where the hidden sector
SUSY breaking field $S$ contains some hidden sector charge\cite{Wells:2004di}.
In this case, then usual gravity-mediated gaugino masses are forbidden
since the gauge kinetic function is holomorphic. 
But sfermion masses and $A$-terms, which depend instead on the K\"ahler potential, 
are allowed to obtain gravity-mediated soft terms. The naturalness measure
$\DeltaBG$ is harder to interpret in the AMSB case since $m_{3/2}$ plays a 
more fundamental role than the ad-hoc soft terms. Also, $\DeltaHS$ is problematic in that
the purely AMSB soft terms are famously scale independent, so there is no prescription for which
$\Lambda$ should be used in Eq. \ref{eq:dmHu2_approx}. Alternatively, there is no ambiguity in the
$\DeltaEW$ measure, so we proceed to exhibit its value.

\begin{figure}[!htbp]
\begin{center}
  \includegraphics[height=0.3\textheight]{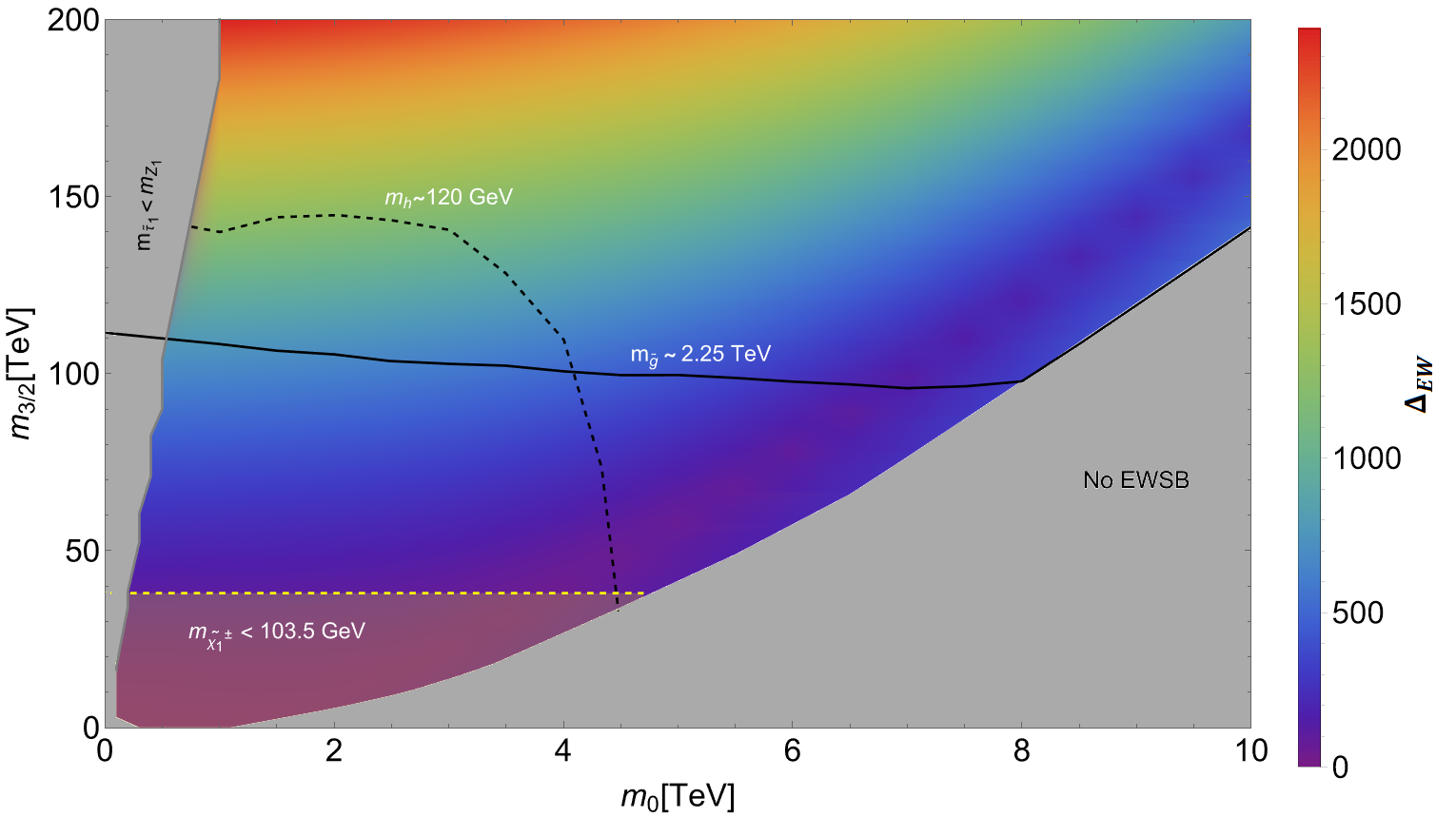}\\
  \includegraphics[height=0.3\textheight]{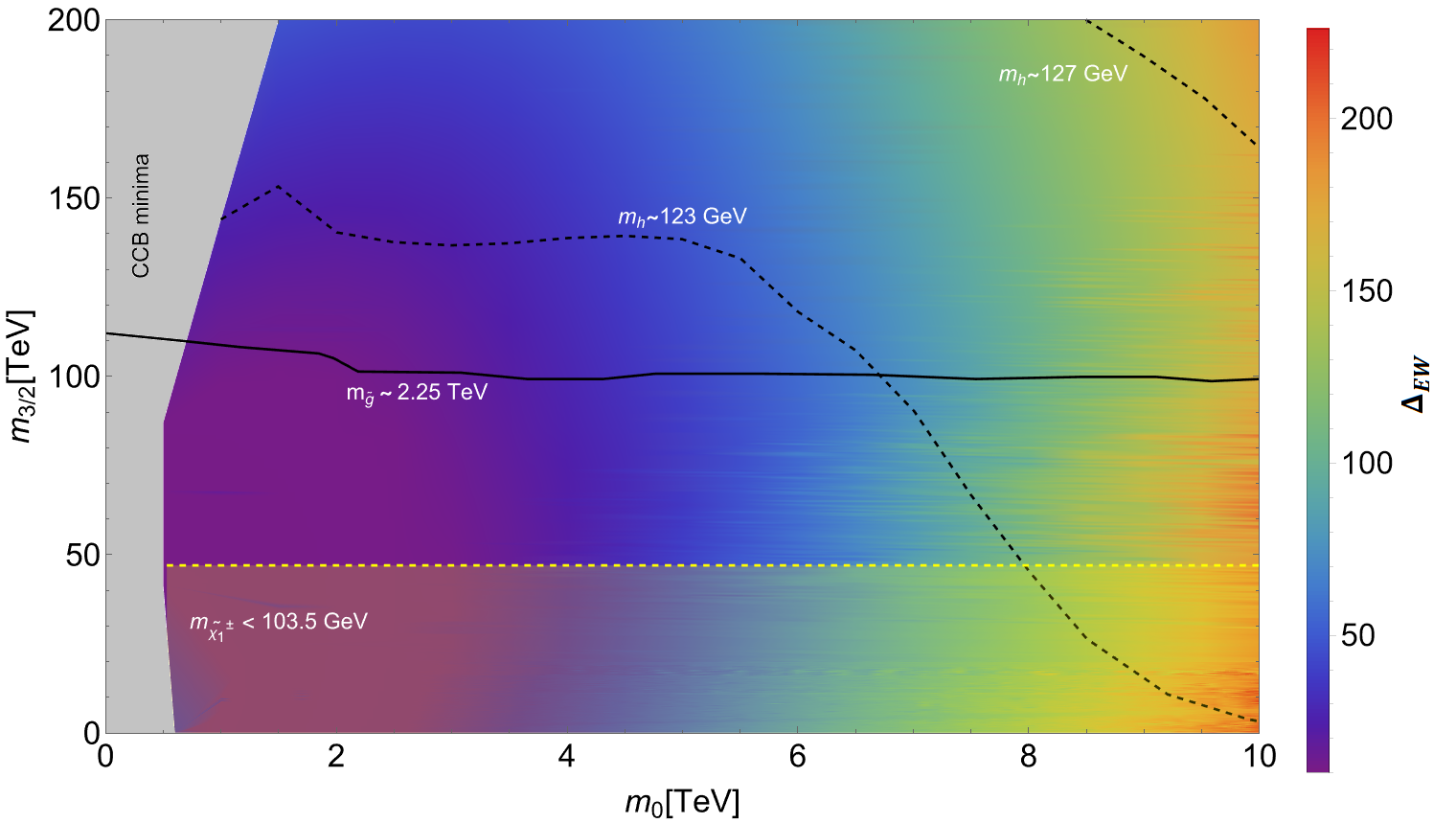}
  \caption{Plot of naturalness measure $\DeltaEW$ in the
    $m_0$ vs. $m_{3/2}$ plane of {\it a}) the
    mAMSB model and {\it b}) the nAMSB model for $\tan\beta =10$.
    For nAMSB, we also require $\mu =200$ GeV and $m_A=2$ TeV.
      The spectrum is calculated using Isasugra.
\label{fig:AMSB_dew}}
\end{center}
\end{figure}

In Fig. \ref{fig:AMSB_dew}, we plot color-coded regions of naturalness
measure $\DeltaEW$ in the $m_0$ vs. $m_{3/2}$ plane for
{\it a}) the mAMSB model and {\it b}) for the nAMSB model.
For both cases, we take $\tan\beta =10$ and $\mu >0$.
For the nAMSB model, we also take $\mu =200$ GeV and $m_A=2$ TeV and $A_0=m_0$.
In the case of frame {\it a}) for mAMSB, the Higgs mass $m_h$ is less
then 123 GeV throughout the entire plane shown (and we do show a contour
of $m_h=120$ GeV via the dotted curve).
The minimal value of $\DeltaEW$ is $\sim 25$ in the purple focus point region
in the LHC-excluded zone, but $\DeltaEW$ can range as high
as over 2000 in the upper-left region. 

In contrast, for frame {\it b}), we see that a large portion of the
plane shown has $m_h\sim 123-127$ GeV.
In addition, $\DeltaEW$ can range as low as $\sim 15$ in the lower-left region,
although this is excluded by LHC.
In the LHC-allowed region, above the $m_{\tg}=2.25$ TeV contour, then
$\DeltaEW$ can be as low as $\sim 20$. Note that the range of color-coded
$\DeltaEW$ values is much smaller in frame {\it b}) than in {\it a}):
in {\it b}), $\DeltaEW$ ranges as high as $\sim 150$ while in {\it a})
it can range beyond 2000.

\section{Natural general mirage mediation (nGMM)}
\label{sec:GMM}

While the AMSB models may seem contrived owing to the requirement of
sequestering and ad hoc bulk soft terms, a perhaps more realistic
alternative is {\it mirage mediation} (MM) models where gravity-mediated and
anomaly-mediated soft terms are comparable. This class of models is
expected to arise from IIB string compactification on an orientifold
with moduli stabilization as in KKLT\cite{Kachru:2003aw}:
the dilaton $S$ and complex structure moduli $U^\beta$ are stabilized by fluxes
and the K\"ahler moduli $T^\alpha$ are stabilized by non-perturbative effects
such as gaugino condensation or instantons. While the $S$ and $U^\beta$ moduli
are expected to gain Kaluza-Klein (KK) scale masses, the $T^\alpha$ moduli
can be much lighter. Moduli stabilization leads to supersymmetric
AdS vacua, but uplifting of the scalar potential via addition of, for instance,
an anti D3 brane at the tip of a Klebanov-Strassler throat can lead to
metastable deSitter vacua. 
A scale of hierarchies $m_T\sim 4\pi^2 m_{3/2}\sim 4\pi^2 m_{soft}$ is expected
to ensue, leading to comparable moduli- and AMSB-contributions to soft masses.

The original KKLT picture assumed a single K\"ahler modulus $T$ and a
simple uplift procedure.
Within the MM model, the soft supersymmetry breaking (SSB) gaugino mass parameters, trilinear SSB
parameters and sfermion mass parameters, all renormalized just below the
unification scale (taken to be $Q=m_{\rm GUT}$), are found to be\cite{Choi:2005ge},
\begin{eqnarray}
M_a&=& M_s\left( l_a \alpha +b_a g_a^2\right),\label{eq:M}\\
A_{ijk}&=& M_s \left( -a_{ijk}\alpha +\gamma_i +\gamma_j +\gamma_k\right),
\label{eq:A}\\
m_i^2 &=& M_s^2\left( c_i\alpha^2 +4\alpha \xi_i -
\dot{\gamma}_i\right) ,\label{eq:m2}
\end{eqnarray}
where $M_s\equiv\frac{m_{3/2}}{16\pi^2}$,
$b_a$ are the gauge $\beta$ function coefficients for gauge group $a$ and 
$g_a$ are the corresponding gauge couplings. The coefficients that
appear in (\ref{eq:M})--(\ref{eq:m2}) are given by
$c_i =1-n_i$, $a_{ijk}=3-n_i-n_j-n_k$ and
$\xi_i=\sum_{j,k}a_{ijk}{y_{ijk}^2 \over 4} - \sum_a l_a g_a^2
C_2^a(f_i).$ 
Finally, $y_{ijk}$ are the superpotential Yukawa couplings,
$C_2^a$ is the quadratic Casimir for the a$^{th}$ gauge group
corresponding to the representation to which the sfermion $\tf_i$ belongs,
$\gamma_i$ is the anomalous dimension and
$\dot{\gamma}_i =8\pi^2\frac{\partial\gamma_i}{\partial \log\mu}$.
Expressions for the last two quantities involving the 
anomalous dimensions can be found in the Appendices of 
Ref's.~\cite{Falkowski:2005ck,Choi:2006xb}.
The quantity $l_a$ is the power of the modulus field entering the gauge kinetic function.
The $n_i$ are modular weights which take on discrete values in the original
construction based on the brane locations of the matter superfields\cite{Choi:2005ge}.

The MM model is then specified by the parameters
\begin{equation}
\ m_{3/2},\ \alpha ,\ \tan\beta ,\ sign(\mu ),\ n_i,\ l_a. 
\label{eq:par1}
\end{equation}
The mass scale for the SSB parameters is dictated by the gravitino mass
$m_{3/2}$. The phenomenological parameter $\alpha$, which could be of
either sign, determines the relative contributions of anomaly mediation
and gravity mediation to the soft terms, and is expected to be $|\alpha|
\sim {\cal O}(1)$. Grand unification implies matter particles within
the same GUT multiplet have common modular weights, and that the $l_a$
are universal. We will assume here that all $l_a=1$ and, for
simplicity, there is a common modular weight for all matter scalars
$c_m$ but we will allow for different modular weights $c_{H_u}$ and
$c_{H_d}$ for each of the two Higgs doublets of the MSSM. Such choices
for the scalar field modular weights are motivated for instance by
$SO(10)$ SUSY GUT models where the MSSM Higgs doublets may live in different
${\bf 10}$-dimensional Higgs reps.

For a variety of discrete parameter choices $n_i$, the various MM models
have all been found to be unnatural when the Higgs mass $m_h$ is restricted to
be $m_h:123-127$ GeV. However, in Ref. \cite{Baer:2016hfa}, it was suggested that to
allow for more realistic compactification schemes
(wherein the K\"ahler moduli may number in the hundreds instead of just one)
and for more diverse uplifting mechanisms, then the discrete valued
parameter choices may be generalized to continuous ones.
This transition to {\it generalized} MM models (GMM)
then allows for natural models with $m_h\sim 125$ GeV.
The parameter space of GMM is given by
\be 
\alpha,\ m_{3/2},\ c_m,\ c_{m3},\ a_3,\ c_{H_u},\ c_{H_d},\
\tan\beta \ \  \ \ (GMM), 
\ee 
where $a_3$ is short for $a_{Q_3H_uU_3}$ (appearing in Eq. \ref{eq:A})
and $c_m$, $c_{m3}$, $c_{H_u}$ and $c_{H_d}$ arise in Eq. \ref{eq:m2}.
Here, we adopt an independent value $c_m$ for the first two
matter-scalar generations whilst the parameter $c_{m3}$ applies to
third generation matter scalars. 
The independent values of $c_{H_u}$ and $c_{H_d}$, which set the
moduli-mediated contribution to the soft Higgs mass-squared soft terms, may
conveniently be traded for weak scale values of $\mu$ and $m_A$ as is
done in the two-parameter non-universal Higgs model (NUHM2)\cite{Baer:2005bu}:
\be
\alpha,\ m_{3/2},\ c_m,\ c_{m3},\ a_3,\ \tan\beta , \mu ,\ m_A
\ \ \ (GMM^\prime ). \label{eq:gmmp}  \ee
This procedure allows for more direct exploration of 
stringy natural SUSY parameter space where most landscape solutions 
require $\mu\sim 100-300$ GeV in anthropically-allowed pocket 
universes\cite{Baer:2019tee}.

Thus, our final formulae for the soft terms are given by
\begin{eqnarray}
M_a&=& \left( \alpha +b_a g_a^2\right)m_{3/2}/16\pi^2,\label{eq:Ma}\\
A_{\tau}&=& \left( -a_3\alpha +\gamma_{L_3} +\gamma_{H_d} +\gamma_{E_3}\right)m_{3/2}/16\pi^2,\\
A_{b}&=& \left( -a_3\alpha +\gamma_{Q_3} +\gamma_{H_d} +\gamma_{D_3}\right)m_{3/2}/16\pi^2,\\
A_{t}&=& \left( -a_3\alpha +\gamma_{Q_3} +\gamma_{H_u} +\gamma_{U_3}\right)m_{3/2}/16\pi^2,\\
m_i^2(1,2) &=& \left( c_m\alpha^2 +4\alpha \xi_i -\dot{\gamma}_i\right)
(m_{3/2}/16\pi^2)^2 ,\label{eq:mi2} \\
m_j^2(3) &=& \left( c_{m3}\alpha^2 +4\alpha \xi_j -\dot{\gamma}_j\right)
(m_{3/2}/16\pi^2)^2 ,\\
m_{H_u}^2 &=& \left( c_{H_u}\alpha^2 +4\alpha \xi_{H_u} -\dot{\gamma}_{H_u}\right)
(m_{3/2}/16\pi^2)^2 ,\\
m_{H_d}^2 &=& \left( c_{H_d}\alpha^2 +4\alpha \xi_{H_d} -\dot{\gamma}_{H_d}\right)
(m_{3/2}/16\pi^2)^2 ,\label{eq:MHd}
\end{eqnarray}
where, for a given value of $\alpha$ and $m_{3/2}$, the values of
$c_{H_u}$ and $c_{H_d}$ are adjusted so as to fulfill the input values
of $\mu$ and $m_A$. 
In the above expressions, the index $i$ runs over first/second 
generation MSSM scalars
$i=Q_{1,2},U_{1,2},D_{1,2},L_{1,2}$ and $E_{1,2}$ while $j$ runs overs
third generation scalars $j=Q_3,U_3,D_3,L_3$ and $E_3$. 
The natural GMM model has been incorporated into the event generator 
program Isajet 7.86\cite{Paige:2003mg}. Here again, there is ambiguity in the
evaluation of $\DeltaBG$ and $\DeltaHS$ while evaluation of $\DeltaEW$ is
unambiguous.

\begin{figure}[!htbp]
\begin{center}
  \includegraphics[height=0.4\textheight]{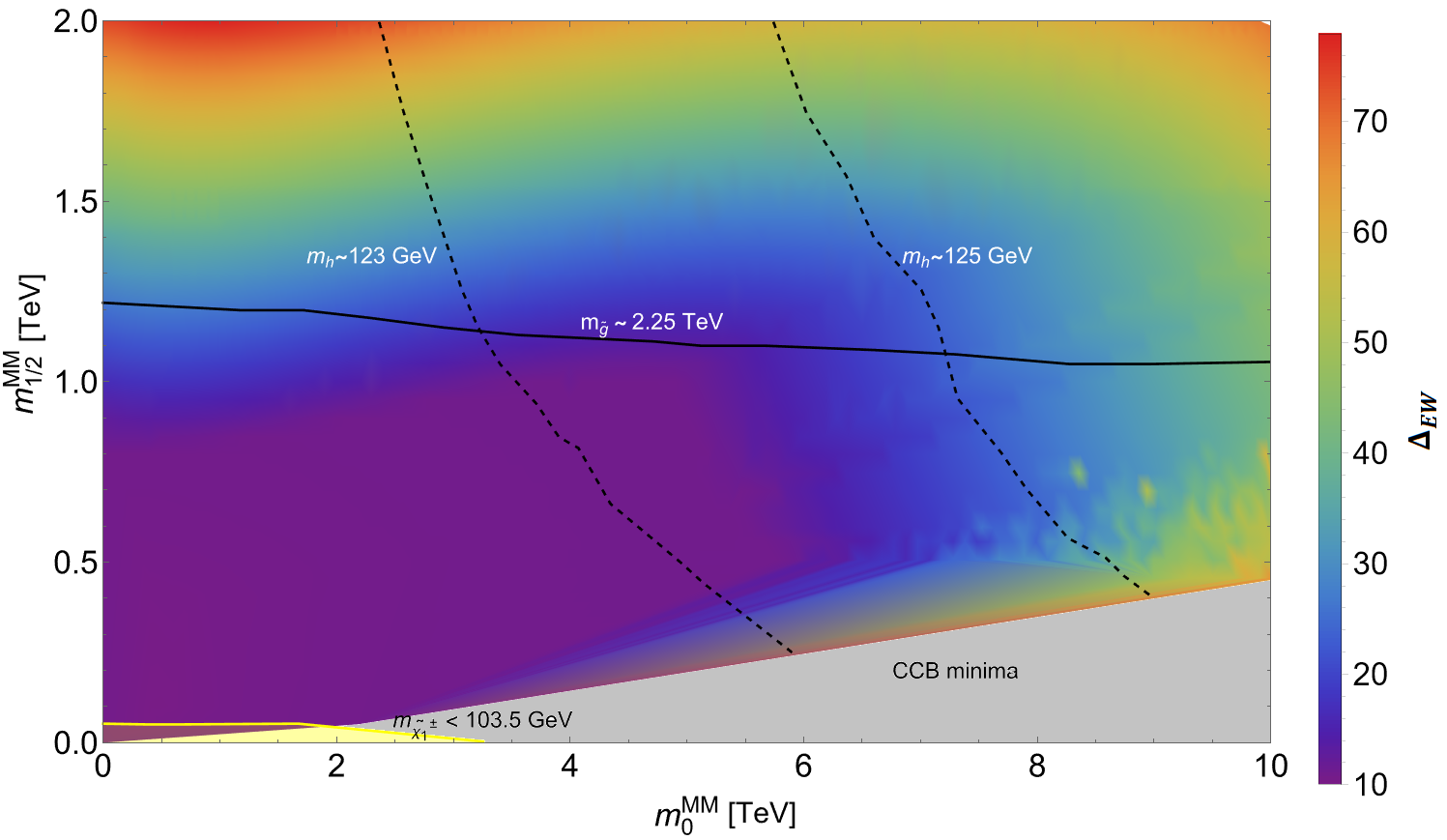}
  \caption{Plot of naturalness measure $\DeltaEW$ in the
    $m_0^{MM}$ vs. $m_{/2}^{MM}$ plane of the generalized mirage-mediation
    model for $\tan\beta =10$ with $c_m=c_{m3}$ and $a_3=1.6\sqrt{c_m}$.
    For GMM$^\prime$, we also require $\mu =200$ GeV and $m_A=2$ TeV.
      The spectrum is calculated using Isasugra.
\label{fig:gmm_dew}}
\end{center}
\end{figure}

In Fig. \ref{fig:gmm_dew}, we show color-coded regions of $\DeltaEW$ for
the GMM$^\prime$ model $m_0^{MM}$ vs. $m_{1/2}^{MM}$ plane for $a_3=1.6\sqrt{c_m}$,
$c_m=c_{m3}$ and with $\mu =200$ GeV and $m_A=2000$ GeV.
Here, $m_0^{MM}$ is defined as $\sqrt{c_m}\alpha m_{3/2}/16\pi^2$ 
and $m_{1/2}^{MM}$ as $\alpha m_{3/2}/16\pi^2$.
We also show contours of $m_h=123$ and $127$ GeV and a contour of
$m_{\tg}=2.25$ TeV. The lower-right region is excluded due to CCB minima,
while the yellow lower-left region has $m_{\tchi_1^+}<103.5$ GeV.
From the plot, we see a vast purple and blue colored region with
$\DeltaEW\sim 10-20$: highly EW natural! Also, much of this region
has a light Higgs scalar with mass $m_h\sim 125$ GeV.
The key signature of GMM models is the fact that the gaugino masses
should unify at scales well below $Q=m_{GUT}$. Thus, if SUSY were
discovered, a high priority issue would be to measure $m_{\tg}$ ($M_3$)
and $m_{\tchi_2^+}$ ($M_2$) to determine the scale $Q$ at which these
values unify. A measurement of the bino mass $M_1$ would also help, but this
may be more difficult than measuring $M_3$ and $M_2$.

\section{Conclusions}
\label{sec:conclude}

In this paper, we have re-examined three finetuning measures which are
widely used in the literature: $\DeltaBG$, $\DeltaHS$ and $\DeltaEW$.
A fourth, stringy naturalness, does not yet admit a quantitative measure
although it may be possible in future work. 
These measures have been quoted vaguely to bolster opinions on future
HE facilities and to set policy for future experiments. Given the situation,
a critical evaluation seems necessary. While naturalness definitions such as
't Hooft naturalness certainly apply to supersymmetry and the big hierarchy
problem,
in that a low scale of SUSY breaking is technically natural in that the
model becomes more (super)symmetric as the SUSY breaking order parameter
$\to 0$, this doesn't apply to the Little Hierarchy which is instead
concerned with the increasing mass gap between the measured value of the weak
scale and the scale of soft SUSY breaking terms (which determine $m_{weak}$
via Eq. \ref{eq:mzs}). For the LHP, we invoke instead the notion of
practical naturalness, where all independent contributions to any observable
should be comparable to or less than the measured value of the observable.

The three naturalness measures are all attempts to measure practical
naturalness.
The $\DeltaEW$ measure is most conservative and unavoidable.
It is model independent within a fixed matter content (such as the MSSM).
It is also unambiguous. Its lessons can be immediately extracted from
Eq. \ref{eq:mzs}: the only superparticles required at the weak scale are the
various higgsinos whose mass derives from the SUSY conserving $\mu$ parameter.
While the value of $\mu$ is frequently tuned in Eq. \ref{eq:mzs} such as to
give the measured value of $m_Z$, its physics origin is rather obscure and
may or may not be directly related to SUSY breaking.\footnote{Twenty solutions
  to the SUSY $\mu$ problem are reviewed in Ref. \cite{Bae:2019dgg}.}
Of the remaining superparticle contributions to the weak scale, all
are suppressed by loop factors times mass-squared factors in Eq. \ref{eq:mzs}
so that the sparticles can lie in the TeV-to-multiTeV range at
little cost to naturalness. In the string landscape picture, in fact, there
is a statistical draw to large soft terms so long as their contributions
to the weak scale are not too large: this then predicts $m_h\sim 125$ GeV with
sparticles typically beyond present reach of LHC\cite{Baer:2017uvn,Baer:2020kwz}.

The traditional $\DeltaBG$ measure which instead famously placed upper
bounds on all sparticles of just a few hundred GeV suffers from the ambiguity
of what to take as free parameters in the log-derivative measure.
While the commonly used SUSY EFTs adopt a variety of ``parameters of ignorance'', it is noted that in more specific models the soft terms are all correlated
(in our universe).
Taking multiple soft parameters as the $p_i$ in $\DeltaBG$ leads to
overestimates of finetuning by factors of up to 500-1000 as compared
to $\DeltaEW$. Also, the measure $\DeltaBG$ is rather complicated to compute,
so we have embedded its numerical evaluation into the publicly available
code DEW4SLHA which computes all three finetuning measures given an input
SUSY Les Houches Accord file. By combining dependent soft terms,
then $\DeltaBG$ reduces to the tree level value of $\DeltaEW$.

The measure $\Delta_{HS}$ which computes $\sim \delta m_{H_u}^2/m_{weak}^2$
is found to overestimate finetuning by artificially splitting
$m_{H_u}^2(weak)$ into $m_{H_u}^{2}(\Lambda )+\delta m_{H_u}^2$ which actually
are not independent contributions to $m_h^2$. In fact, selection of
appropriately broken EW symmetry requires $\delta m_{H_u}^2$ to be large
or else EW symmetry is not broken. This measure, which famously predicts
three third generation squarks below 500 GeV, also overestimates
finetuning by up to three orders of magnitude. By combining the dependent terms
$m_{H_u}^2(\Lambda )$ with $\delta m_{H_u}^2$, then $\DeltaHS$ reduces to
$\DeltaEW$ according to Eq. \ref{eq:mzsHS}.

Our ultimate conclusion is that the so-called {\it naturalness crisis}\cite{Lykken:2014bca,Dine:2015xga}
which arose from non-observation of SUSY particles at LHC is not a crisis
at all, but is based on faulty estimates of finetuning by the $\DeltaBG$
and $\DeltaHS$ measures (which are actually inconsistent with each other).
The more conservative measure $\DeltaEW$ rules out old favorites
such as the CMSSM/mSUGRA model based on naturalness, but allows for plenty
of natural parameter space in models like NUHMi, nAMSB and nGMM
(and of course less theoretically constrained exploratory constructs like pMSSM).
In fact, the naturalness-allowed and LHC-allowed parameter space regions
are precisely those which seem most prevalent from rather
general considerations of the string landscape. In this light,
the above natural SUSY models maintain a high degree of motivation,
and are perhaps even more highly motivated than pre-LHC times due to
the emergence of the string landscape. Thus, policy decisions for future
HEP facilities, especially future accelerators, should bear this resolution
in mind in that it may be that we just need a much more energetic collider
for the discovery of superpartners.

\section*{Acknowledgements} 

This material is based upon work supported by the U.S. Department of Energy, 
Office of Science, Office of Basic Energy Sciences Energy Frontier Research 
Centers program under Award Number DE-SC-0009956 and U.S. Department of Energy 
Grant DE-SC-0017647. 




\bibliography{nat.bib}
\bibliographystyle{elsarticle-num}

\end{document}